\documentclass{raa_twocolumn}
\usepackage{graphicx,times}  
\usepackage{natbib}
\usepackage{amssymb,amsmath}
\bibpunct{(}{)}{;}{a}{}{,}
\usepackage[morefloats = 38]{morefloats}
\usepackage[figuresright]{rotating}
\usepackage[flushleft]{threeparttable}
\usepackage{multirow}

\begin{document}

   \title{A Fermi-LAT Study of Globular Cluster Dynamical Evolution in the Milky Way: Millisecond Pulsars as the Probe
%\footnotetext{$*$ Supported by the National Natural Science Foundation of China.}
}
%   \subtitle{I. Place Your Subtitle Here}

   \volnopage{Vol.0 (200x) No.0, 000--000}      %%preserved for Editor. DOn't remove!
   \setcounter{page}{1}          %%starting page, preserved for Editor. DOn't remove!

   \author{Li Feng
      \inst{1}
   \and  Zhongqun Cheng
      \inst{2,3}   
    \and  Wei Wang
      \inst{2,3}     
   \and  Zhiyuan Li
      \inst{1,4}
     \and Yang Chen
        \inst{1,4}
   }
%% For single author or all the authors from an institute, use "{}" only

\institute{School of Astronomy and Space Science, Nanjing University, Nanjing 210023, China\\
\and 
Department of Astronomy, School of Physics and Technology, Wuhan University, Wuhan 430072, China; chengzq@whu.edu.cn\\
\and 
WHU-NAOC Joint Center for Astronomy, Wuhan University, Wuhan 430072, China\\
\and 
Key Laboratory of Modern Astronomy and Astrophysics, Nanjing University, China\\
  }
   
          
% \date{Received~~2009 month day; accepted~~2009~~month day}

\abstract{Using archival {\it Fermi}-LAT data with a time span of $\sim12$ years, we study the population of Millisecond Pulsars (MSPs) in Globular Clusters (GlCs) and investigate their dependence on cluster dynamical evolution in the Milky Way Galaxy. 
We show that the $\gamma$-ray luminosity ($L_{\gamma}$) and emissivity (i.e., $\epsilon_{\gamma}=L_{\gamma}/M$, with $M$ the cluster mass) are good indicators of the population and abundance of MSPs in GlCs, and they are highly dependent on the dynamical evolution history of the host clusters. Specifically speaking, the dynamically older GlCs with more compact structures are more likely to have larger $L_{\gamma}$ and $\epsilon_{\gamma}$, and these trends can be summarized as strong correlations with cluster stellar encounter rate $\Gamma$ and the specific encounter rate ($\Lambda=\Gamma/M$), with $L_{\gamma}\propto \Gamma^{0.70\pm0.11}$ and $\epsilon_{\gamma}\propto \Lambda^{0.73\pm0.13}$ for dynamically normal GlCs. However, as GlCs evolve into deep core collapse, these trends are found to be reversed, implying that strong encounters may have lead to the disruption of Low-Mass X-ray Binaries (LMXBs) and ejection of MSPs from core-collapsed Systems. Besides, the GlCs are found to exhibit larger $\epsilon_{\gamma}$ with increasing stellar mass function slope ($\epsilon_{\gamma}\propto 10^{(0.57\pm0.1)\alpha}$), decreasing tidal radius ($\epsilon_{\gamma}\propto R_t^{-1.0\pm0.22}$) and distances from the Galactic Center (GC, $\epsilon_{\gamma}\propto R_{gc}^{-1.13\pm0.21}$). These correlations indicate that, as GlCs losing kinetic energy and spiral in towards GC, tidal stripping and mass segregation have a preference in leading to the loss of normal stars from GlCs, while MSPs are more likely to concentrate to cluster center and be deposited into the GC. Moreover, we gauge $\epsilon_{\gamma}$ of GlCs is $\sim10-1000$ times larger than the Galactic bulge, the latter is thought to reside thousands of unresolved MSPs and may responsible for the GC $\gamma$-ray excess, which support that GlCs are generous contributors to the population of MSPs in the GC.}

   \authorrunning{L. Feng, Z.-Q. Cheng, W. Wang, Z. -Y. Li, \& Y. Chen}            %author_head in even pages
   \titlerunning{A Fermi-LAT Study of Globular Cluster Dynamical Evolution in Milky Way Galaxy}  % title_head in odd pages
    \maketitle
%% The author head (on even pages) and the title head (on odd pages) will be
%% automatically extracted from \author{} and \title{}. Whenever the title is too long,
%% you will be asked to supply a shorter one by inserting either \authorrunning{} or
%% \titlerunning{} before \maketitle. Anyway, you can specify your own heads.
%%
%%
%% Note: In the following text body of your manuscript, please note several differences from
%%       other major journals:
%% (1) \subsection{Please Capitalize the First Letter of Each Notional Word in Subsection Title}
%% (2) Please Capitalize the First Letter of Each Notional Word in all tables' captions

%
%_______________________________________________ sections below
%


\section{Introduction} \label{sec:intro}
Globular Clusters (GlCs) are self-gravitating systems that evolve with dynamical timescale (i.e., two-body relaxation timescale) much smaller than their age \citep{Heggie2003}, and the binary burning process (i.e., extraction of binary binding energy via dynamical encounters) is the ``heating mechanism'' that supports the cluster against gravother-thermal collapse \citep{Fregeau2003}, which also makes GlCs promising breeding grounds for exotic objects, such as blue straggler stars \citep{Fregeau2004, Chatterjee2013}, coronal active binaries (ABs), cataclysmic variables (CVs; \citealp{Ivanova2006, Shara2006, Belloni2016, Belloni2017, Belloni2019}), low-mass X-ray binaries (LMXBs; \citealp{Rasio2000, Ivanova2008, Kremer2018}), millisecond pulsars (MSPs; the offspring of LMXBs; \citealp{Ye2019, Kremer2020a, Ye2022}), and gravitational-wave sources made up of compact objects \citep{Rodriguez2015, Rodriguez2016, Askar2017, Clausen2013, Ye2020, Arcasedda2020, Kremer2021}.

Depend on their distances from the Galactic Center (GC), the dynamical evolution of GlCs is also subjected to the gravitational tidal field of the Milky Way Galaxy \citep{Baumgardt2003}. The expanding cluster halo could be truncated by the external tidal field, and tidal stripping may lead to the loss of stars from GlCs, enhance the energy outflow of GlCs and thereby accelerating the dynamical evolution of the cluster \citep{Gnedin1997, Gnedin1999}. Besides, dynamical friction between GlCs and Galacitc background stars may lead to the spiral in of GlCs toward the deep potential well of the Galaxy \citep{Tremaine1975, Moreno2022}, where tidal stripping would eventually lead to the fully dissolution of the clusters. GlCs therefore can take their binary burning products into the Inner Galaxy \citep{Fragione2018a, Arcasedda2018b}, and contribute the cluster mass to the growing Galactic nuclei \citep{Antonini2012, Antonini2013, Gnedin2014}.

Thanks to their exclusive formation process and emission properties, the binary burning products are proved to be superb tracing particles of GlC stellar dynamical interactions and cluster evolution. For example, compared with normal stars, the emission of LMXBs, MSPs, CVs, ABs and BSS are found to be much more luminous either in optical, X-ray, $\gamma$-ray or radio band and can be detected and picked out from the dense core of GlCs, making them ideal tracers for studying stellar dynamical encounters \citep{Verbunt1987, Verbunt2003, Pooley2003, Pooley2006}, two-body relaxation and mass segregation effect of GlCs \citep{Ferraro2012, Cheng2019a, Cheng2019b}. The cumulative cluster X-ray luminosity $L_{\rm X}$ and emissivity $\epsilon_{\rm X}$ ($L_{\rm X}$ per unit stellar mass), proxies of population and abundance of weak X-ray sources (mainly CVs and ABs) in GlCs \citep{Cheng2018a, Heinke2020}, are also shown to be effective indicators for diagnosing cluster binary burning process \citep{Cheng2018b, Cheng2020b}. 

Among all the binary burning products in GlCs, MSPs are outstanding for many reasons. First, compared with BSS, CVs and ABs that can be formed through both primordial binary evolution and dynamical formation channels, it is widely recognized that LMXBs, the progenitors of MSPs, are dynamically formed in GlCs \citep{Verbunt1987, Verbunt2003, Pooley2003}, and the abundances (number per unit stellar mass) of LMXBs and MSPs are more than $\sim 100$ times higher in GlCs than in the Galactic field \citep{Clark1975, Katz1975, Camilo2005, Ransom2008}. Besides, unlike other tracers that could be contaminated by fore- and background sources, MSPs are mainly detected via either radio or $\gamma$-ray observations, in which contamination is negligible. Moreover, the lifetime of MSPs is very long ($\sim 10^{10}\, {\rm yr}$) and their luminosities are observed to be very stable. Taking these aspects together, it is safe to say that MSPs are the best probe of cluster dynamical evolution, especially in tracing the tidal dissolution of GlCs in the Milky Way Galaxy.

In $\gamma$-ray astronomy, both MSPs and GlCs are recognized as important $\gamma$-ray emitters in the Galaxy \citep{Abdo2010, Tam2011}, and $\sim 30$ GlCs have been identified as point sources in the fourth Fermi Large Area Telescope catalog (4FGL, \citealp{Abdollahi2020}). 
The $\gamma$-ray emission of GlCs is widely assumed to originate from MSPs that reside in the clusters \citep{Cheng2010, Bednarek2013}, and the detection of pulsed gamma-ray emission from two GlCs has further strengthened the case for this connection \citep{Freire2011, Wu2013}. In concert with LMXBs, a strong correlation between $\gamma$-ray luminosity $L_{\gamma}$ and the stellar encounter rate $\Gamma$ has been established among GlCs \citep{Abdo2010, Hui2011, Zhang2016, Tam2016, Hooper2016, Lloyd2018, deMenezes2019}, which lend a support to the dynamical origin of MSPs in GlCs.

On the other hand, the {\it Fermi} Large Area Telescope (Fermi-LAT) has discovered an unexpected $\gamma$-ray excess around the GC, peaking at a few GeV, with an approximately spherical density profile $\propto r^{-2.4}$ and extends to $10\degr -20\degr$ ($1.5-3$ kpc) from the GC \citep{goodenough2009, Hooper2011a, Hooper2011b, DiMauro2021}. The nature of this ``Galactic Center Excess'' (GCE) is still not clear, and many promising models have been proposed over the past decade, including dark matter annihilation \citep{Abazajian2012, Daylan2016}, emission of thousands of unresolved MSPs \citep{Abazajian2011, Yuan2014, Bartels2016}, or emission from cosmic rays injected at the GC \citep{Carlson2014, Petrovic2014, Cholis2015a, Gaggero2015}. In the MSP scenario, Galactic GlCs fallen into the GC are expected to deposit their population of MSPs, which are then inherited by the Galactic nuclear star cluster and nuclear bulge, and contribute to the observed $\gamma$-ray excess \citep{Bednarek2013, Brandt2015, Fragione2018a, Abbate2018, Arcasedda2018b}.
 
In the present work, we perform a $\gamma$-ray study of the Galactic GlCs based on the archival {\it Fermi} data. 
Unlike previous works that only focus on the cumulative GlC $\gamma$-ray luminosity $L_{\gamma}$ \citep{Abdo2010, Hooper2016, deMenezes2019}, we also measure the $\gamma$-ray emissivity $\epsilon_{\gamma}$ ($L_{\gamma}$ per unit stellar mass) of the GlCs, and explore their dependence on various cluster parameters. 
The cumulative cluster $\gamma$-ray luminosity offers us a chance to study the population of MSPs at the Inner Galaxy, where single MSP is hard to be detected via radio or $\gamma$-ray observation. On the other hand, as demonstrated by \citet{Cheng2018a} and \citet{Heinke2020} in X-ray band, the emissivity has been proved to be a reliable indicator of exotic objects abundance in GlCs, which is insensitive to the luminosity function (LF) and can be applied to a large cluster sample in a highly uniform fashion. 
More importantly, the derived GlC $\gamma$-ray emissivity can be directly compared to the stellar $\gamma$-ray emissivity of the Galactic nuclear star cluster and nuclear bulge, which are crucial to evaluating the relative abundance of putative MSPs in these environments, thus estimating the possible stellar origins and contribution to the GCE by dissolved GlCs.

The limitation of our approach is that we assume the measured cluster $\gamma$-ray luminosity is a good proxy of the population of hosted MSPs, that may not be true for all clusters, especially for GlCs contains few MSPs. However, as we will show below, although the uncertainty of small counts of MSPs in individual GlCs may introduce large scatter to our cluster sample\footnote{In fact, the scatter of derived correlations in this paper is about an order of magnitude larger.}, but are unlikely to create the correlations and trends observed in this work.

The paper is organized as follows. Section 2 describes the data reduction and analysis that lead to the detection and the measurement of the $\gamma$-ray luminosity $L_{\gamma}$ and emissivity $\epsilon_{\gamma}$ of individual GlCs. Section 3 explores the dependence of $L_{\gamma}$ and $\epsilon_{\gamma}$ on various cluster physical properties. A discussion and a summary of our results are presented in Sections 4 and 5, respectively. Throughout this work we quote $1\sigma$ errors, unless otherwise stated.

\section{$\gamma$-Ray Data Analysis}
\subsection{Data Reduction and Analysis}
We analyzed the archival {\it Fermi}-LAT data of the 157 GlCs presented in the catalog of \citet{Harris1996}. The {\it Fermi} data was observed from 2008-08-08 to 2020-11-08 (MET: 239846401--626486405), with a time span of $\sim12$ years. We used the {\it Fermi} tools release 2.0 to analyse the data, with the energy band was restricted to 100MeV-300GeV and divided into 15 logarithmically spaced energy bins. We selected the events with source class (evclass=128, evtype=3) and filter the data with $\rm DATA QUAL > 0$, $\rm LAT CONFIG == 1$. A zenith angle cut of $<$90$\degr$ and a satellite rocking angle cut of $<52\degr$ was applied to avoid contamination from the Earth limb. 

The region of interest (ROI) of each target was restricted to a $14\degr\times14 \degr$ rectangular box centred on the optical center of the GlCs. We used the Make4FGLxml.py tool and the 4FGL DR2 catalog to generate the background source list within the ROI. For diffuse background modelling, we adopted the most recent Galactic interstellar emission model gll\_iem\_v07.fits and the isotropic spectral template iso\_P8R3\_SOURCE\_V3\_v1.txt. The instrument response function (IRF) was set as P8R3\_SOURCE\_V3. 

All targets were investigated by means of binned likelihood analysis ({\it gtlike} tool -- DRMNFB, NEWMENUIT algorithm). To search for the $\gamma$-ray emission from the GlCs, we added a putative point source at the optical center of the cluster, and used the {\it gttsmap} tool to derive the TS maps. From the TS maps, one can visually inspect the detections of unkonwn sources located within the ROI. We then removed the new sources by adding them to the background source lists and refit the data again. The spectral models of GCs were set as either a power-law (PL), a PL plus exponential cut-off (PL+Expcut), or a logparabola (LP) model. During the fitting, all the GlC parameters (i.e., coordinates, photon index, cut off energy and normlization) were set as free, whereas for diffuse backgrounds and point sources located within $5\degr$ of the ROI, only the normalization parameter was left free to vary. To quantify the significance of the detection, a test statistic (TS) was calculated, defined as $TS=2({\rm log} L_1- {\rm log} L_0)$, where $L_1$ ($L_0$) is the maximum likelihood value of the model with (without) the putative source. The chosen criteria for detection was $TS>25$, corresponding to a significance slightly above $4\sigma$.

\subsection{Data Analysis Results}

\begin{figure*}
\centerline{
\includegraphics[width=1.0\textwidth]{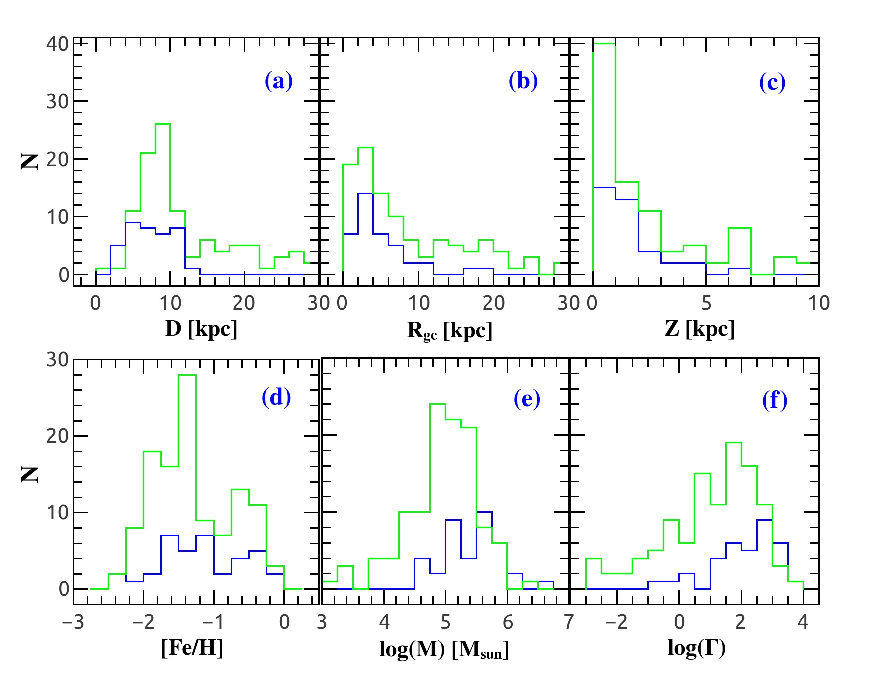}}
\caption{Histogram distributions of GlCs as a function of cluster parameters: distance from the Earth (a), distance from the Galactic center (b), distance from the Galactic disk (c), cluster metallicity (d), cluster mass (e) and the stellar encounter rate (f). The blue lines represent the $\gamma$-ray bright GlCs, while clusters without $\gamma$-ray detection are shown as green lines.}
\label{fig:detection}
\end{figure*}

For the 157 GlCs tabulated in the catalog of \citet{Harris1996}, about $25\%$ (39/157) of them are found to be $\gamma$-ray bright ($TS>25$) in this work, with 2 clusters, HP 1 ($TS=44.3$) and Terzan 9 ($TS=42.1$), are first identified as $\gamma$-ray emitters. 
Cross-checking the 39 clusters with the 4FGL-DR3 catalog \citep{Abdollahi2022}, 5 clusters, NGC 362 ($TS=16$) and NGC 6304 ($TS=21.7$), are found to have a smaller significance than our detection threshold. Besides, Liller 1 was found to be $\gamma$-ray bright by \citet{Tam2011}, while our data fitting suggests that the detection is marginal ($TS=21.6$). All the 39 GlCs are located within a distance of $D=15$ kpc from the Earth (Fig-\ref{fig:detection}(a)), and there is no evident observation bias for GlCs near the GC (Fig-\ref{fig:detection}(b)) and the Galactic disk (Fig-\ref{fig:detection}(c)), or significant dependence on cluster metallicity (Fig-\ref{fig:detection}(d)). However, it seems that GlCs are more likely to be identified as $\gamma$-ray emitters, provided that they are more massive (Fig-\ref{fig:detection}(e)) or have larger stellar encounter rate (Fig-\ref{fig:detection}(f)). Interestingly, the $\gamma$-ray detection rate of core-collapsed GlCs (12/29) is found to be $\sim2$ times higher than the dynamically normal GlCs (27/128), which suggest that the dynamical evolution history is also an important factor influence the $\gamma$-ray luminosity of GlCs.

The $\gamma$-ray data analysis results are summarized in Table-\ref{tab:results}, which is segmented into two panels according to the core collapse flag in the catalog of \citet{Harris1996}. The likelihood fit parameters arranged from left to right are: cluster name, spectral models, test statistic values, PL photon index, exponential cut-off energy $E_{c}$ and measured GlC energy flux $f_{\gamma}$.   Adopting the cluster distance ($D$) presented in the catalog of \citet{Baumgardt2018}, we also calculated the cumulative GlC $\gamma$-ray luminosity with function $L_{\gamma}=4\pi D^2 f_{\gamma}$. We define the GlC $\gamma$-ray emissivity as $\epsilon_{\gamma}=L_{\gamma}/M$, with $M$ being the cluster mass given in the catalog of \citet{Baumgardt2018}. $L_{\gamma}$ and $\epsilon_{\gamma}$ are listed in Columns (7) and (8) of Table-\ref{tab:results}. The errors are given in $1\sigma$ standard deviation level.

\begin{table*}
\centering
\caption{$\gamma$-ray Data Analysis Results of the 27 dynamically normal GlCs (upper panel) and 12 core-collapsed GlCs (bottom panel) detected in this work. Columns 1-2: Name of GlCs and Spectral models. Columns 3-4: The test statistic value and the power-law phonton index. Columns 5-7: Cut-off energy (in units of GeV), GlC $\gamma$-ray flux (in units of $10^{-12}\, {\rm erg\,cm^{-2}\,s^{-1}}$) and luminosity (in units of $10^{33}\, {\rm erg\,s^{-1}}$). Columns 8-9: GlC $\gamma$-ray emissivity (in units of $10^{28}\, {\rm erg\,s^{-1}}\, M^{-1}_{}\odot $) and the random detection probability.}
\label{tab:results}
\begin{tabular}{llllllllr}
\hline
\multicolumn{1}{l}{Name}  &\multicolumn{1}{c}{Model} & \multicolumn{1}{c}{TS} & \multicolumn{1}{c}{Photon Index} & \multicolumn{1}{c}{$E_{c}$} & \multicolumn{1}{c}{$f_{\gamma}$} & \multicolumn{1}{c}{$L_{\gamma}$} & \multicolumn{1}{c}{$\epsilon_{\gamma}$} & \multicolumn{1}{c}{$P_{\rm random}$} \\
\multicolumn{1}{c}{(1)} & \multicolumn{1}{c}{(2)} & \multicolumn{1}{c}{(3)} & \multicolumn{1}{c}{(4)} & \multicolumn{1}{c}{(5)} & \multicolumn{1}{c}{(6)} & \multicolumn{1}{c}{(7)} & \multicolumn{1}{c}{(8)} & \multicolumn{1}{c}{(9)}\\
\hline
\multicolumn{9}{c}{Dynamically Normal GlCs}\\
\hline
NGC 104	   & PL+Expcut & 5557.3 & 1.39$\pm$0.07 & 2.82$\pm$0.30 & 26.9$\pm$0.7  & 65.9$\pm$1.8  & 7.37$\pm$0.21 & $16.9\%$ \\
NGC 1851   & PL+Expcut & 42.8   & 1.46$\pm$0.24 & 1.71$\pm$0.32 & 1.25$\pm$0.32 & 18.2$\pm$4.9  & 5.71$\pm$1.53 & $0.5\%$\\
NGC 2808   & PL+Expcut & 102.5  & 1.22$\pm$0.37 & 2.29$\pm$1.00 & 2.99$\pm$0.47 & 36.3$\pm$5.7  & 4.20$\pm$0.66 & $0.8\%$\\
NGC 5139   & PL+Expcut & 977.3  & 1.20$\pm$0.17 & 1.98$\pm$0.37 & 11.6$\pm$0.8  & 41.0$\pm$2.8  & 1.13$\pm$0.13 & $34.5\%$\\
NGC 5286   & PL	     & 25.4   & 2.67$\pm$0.19 & --            & 3.11$\pm$0.67 & 46.0$\pm$9.9  & 13.0$\pm$2.8  & $0.8\%$\\
NGC 5904   & PL+Expcut & 46.1   & 0.93$\pm$0.7  & 2.22$\pm$1.30 & 1.18$\pm$0.26 & 7.92$\pm$1.77 & 2.01$\pm$0.45 & $4.8\%$\\
NGC 6093   & PL+Expcut & 105.6  & 1.94$\pm$0.2  & 6.29$\pm$2.10 & 3.84$\pm$0.72 & 49.3$\pm$9.3  & 14.6$\pm$2.8  & $0.5\%$\\
NGC 6139   & PL+Expcut & 63.5   & 1.94$\pm$0.25 & 4.09$\pm$2.24 & 5.05$\pm$0.90 & 61.1$\pm$10.8 & 18.9$\pm$3.9  & $3.8\%$\\
NGC 6218   & PL+Expcut & 30.3   & 3.5$\pm$0.01  & 0.10$\pm$0.00 & 2.54$\pm$0.58 & 7.96$\pm$1.82 & 7.44$\pm$1.72 & $2.5\%$\\
NGC 6254   & LP	     & 57.4   & --            & --            & 1.98$\pm$0.30 & 5.43$\pm$1.01 & 2.65$\pm$0.50 & $2.9\%$\\
NGC 6316   & PL+Expcut & 309.0  & 1.93$\pm$0.14 & 5.32$\pm$1.87 & 12.8$\pm$1.1  & 190.9$\pm$16.3& 60.0$\pm$9.0  & $3.0\%$\\
NGC 6341   & PL+Expcut & 59.9   & 1.89$\pm$0.62 & 2.93$\pm$2.56 & 2.48$\pm$0.36 & 21.5$\pm$3.1  & 6.11$\pm$0.88 & $2.3\%$\\
NGC 6388   & PL+Expcut & 1266.5 & 1.46$\pm$0.11 & 2.45$\pm$0.39 & 19.2$\pm$0.4  & 287.4$\pm$6.1 & 23.0$\pm$0.5  & $1.9\%$\\
NGC 6402   & PL+Expcut & 56.0   & 1.51$\pm$0.38 & 3.71$\pm$1.89 & 2.80$\pm$0.62 & 28.1$\pm$6.2  & 4.74$\pm$1.07 & $0.7\%$\\
NGC 6440   & PL+Expcut & 321.6  & 2.09$\pm$0.12 & 8.78$\pm$3.29 & 20.3$\pm$1.3  & 165.8$\pm$10.5& 33.9$\pm$3.9  & $1.7\%$\\
NGC 6441   & PL+Expcut & 551.1  & 1.71$\pm$0.09 & 3.03$\pm$0.34 & 21.6$\pm$1.1  & 420.0$\pm$21.0& 31.8$\pm$1.6  & $3.0\%$\\
NGC 6528   & PL+Expcut & 40.9   & 1.29$\pm$0.26 & 1.55$\pm$0.31 & 3.52$\pm$0.89 & 25.9$\pm$6.6  & 45.7$\pm$12.7 & $1.1\%$\\
NGC 6626   & PL+Expcut & 1102.2 & 1.51$\pm$0.05 & 1.16$\pm$0.06 & 21.1$\pm$1.1  & 73.0$\pm$3.8  & 24.4$\pm$2.0  & $6.2\%$\\
NGC 6637   & LP	     & 29.1   & --            & --            & 1.49$\pm$0.35 & 14.2$\pm$3.3  & 9.14$\pm$2.15 & $2.3\%$\\	
NGC 6652   & PL+Expcut & 158.7  & 1.49$\pm$0.4  & 2.21$\pm$1.39 & 4.84$\pm$0.60 & 52.0$\pm$6.4  & 108$\pm$20.6  & $1.2\%$\\
NGC 6656   & PL+Expcut & 100.5  & 0.83$\pm$0.29 & 1.05$\pm$0.13 & 4.17$\pm$0.47 & 5.41$\pm$0.85 & 1.14$\pm$0.18 & $40.1\%$\\
NGC 6717   & PL+Expcut & 86.1   & 0.63$\pm$0.52 & 1.82$\pm$0.70 & 2.50$\pm$0.43 & 17.0$\pm$2.9  & 47.4$\pm$13.6 & $1.8\%$\\
NGC 6838   & PL        & 31.8   & 2.71$\pm$0.2  & --            & 3.92$\pm$0.79 & 7.53$\pm$1.51 & 16.5$\pm$3.4  & $1.5\%$\\
2MASS-GC01 & PL+Expcut & 92.8   & 1.87$\pm$0.19 & 3.55$\pm$1.29 & 19.6$\pm$2.1  & 26.7$\pm$2.9  & 76.3$\pm$8.3  & $1.8\%$\\
Glimpse 01 & PL+Expcut & 760.2  & 1.65$\pm$0.16 & 4.13$\pm$0.83 & 78.2$\pm$2.7  & 108.5$\pm$3.8 & 36.2$\pm$3.8  & $6.3\%$\\
Glimpse 02 & PL+Expcut & 138.1  & 1.47$\pm$0.13 & 1.62$\pm$0.18 & 30.1$\pm$2.4  & 76.4$\pm$6.0  &	--            & $10.1\%$\\
Terzan 5   & PL+Expcut & 5290.6 & 1.74$\pm$0.03 & 3.98$\pm$0.18 & 125.0$\pm$2.2 & 657.3$\pm$11.5& 70.3$\pm$5.3  & $2.9\%$\\
\hline
\multicolumn{9}{c}{Core-Collapsed GlCs}\\
\hline
NGC 1904   & PL	     & 53.9   & 2.54$\pm$0.16 & --            & 2.5$\pm$0.4   & 52.3$\pm$8.5  & 37.7$\pm$6.8  & $0.6\%$\\
NGC 6266   & PL+Expcut & 1332.0 & 1.45$\pm$0.11 & 2.75$\pm$0.44 & 19.4$\pm$0.8  & 95.6$\pm$3.9  & 15.7$\pm$0.7  & $5.2\%$\\
NGC 6397   & PL+Expcut & 44.8   & 2.15$\pm$0.55 & 3.89$\pm$4.09 & 4.67$\pm$0.68 & 3.45$\pm$0.50 & 3.57$\pm$0.52 & $6.4\%$\\
NGC 6541   & PL+Expcut & 136.1  & 1.61$\pm$0.34 & 2.89$\pm$1.65 & 4.51$\pm$0.58 & 31.3$\pm$4.1  & 10.7$\pm$1.4  & $4.1\%$\\
NGC 6624   & PL+Expcut & 812.1  & 1.15$\pm$0.22 & 1.22$\pm$0.44 & 13.10$\pm$1.1 & 101.1$\pm$8.2 & 64.8$\pm$5.5  & $18.1\%$\\
NGC 6723   & PL+Expcut & 29.8   & 1.54$\pm$0.43 & 4.71$\pm$5.73 & 1.75$\pm$0.48 & 14.4$\pm$3.9  & 8.11$\pm$2.28 & $1.7\%$\\
NGC 6752   & PL+Expcut & 193.1  & 0.59$\pm$0.43 & 1.06$\pm$0.28 & 3.26$\pm$0.39 & 6.64$\pm$0.79 & 2.41$\pm$0.29 & $32.1\%$\\
NGC 7078   & PL+Expcut & 103.0  & 1.98$\pm$0.4  & 1.55$\pm$0.92 & 3.64$\pm$0.53 & 50.1$\pm$7.2  & 7.91$\pm$1.14 & $7.0\%$\\
HP1 	   & PL+Expcut & 44.3   & 0.48$\pm$0.24 & 1.55$\pm$0.21 & 5.62$\pm$0.98 & 33.0$\pm$5.8  & 26.6$\pm$5.9  & $5.6\%$\\
Terzan 1   & LP        & 78.0   & --            & --            & 6.34$\pm$1.05 & 24.5$\pm$4.1  & 16.3$\pm$3.8  & $10.6\%$\\
Terzan 2   & PL+Expcut & 58.4   & 0.55$\pm$0.22 & 1.55$\pm$0.2  & 10.4$\pm$1.5  & 75.0$\pm$10.4 & 55.1$\pm$12.7 & $5.5\%$\\
Terzan 9   & PL+Expcut & 42.1   & 1.47$\pm$0.35 & 1.55$\pm$0.26 & 6.28$\pm$3.94 & 25.1$\pm$15.7 & 20.9$\pm$13.3 & $6.0\%$\\
\hline 
\end{tabular}
\end{table*}

To check the spatial consistency between the $\gamma$-ray emission and the optical centers of GlCs, we plot in Figure-\ref{fig:tsmaps} the TS maps of the 39 clusters. For each GlC, the image was restricted to a $5\degr\times5 \degr$ box, with the central green circle indicates the tidal radius of the cluster. All maps have evidence for good coincidence between the $\gamma$-ray emission and the GlC centroids, except for NGC 1904, where the offset between the optical center and the peak of the $\gamma$-ray emission (cayan ellipse) is $\sim 0.3\degr$. Besides the coincidence of spatial coordinates, it is possible that the $\gamma$-ray emission is associated with a fore/background source rather than the cluster. We estimated this probability with function $P_{\rm random}=\pi R_{t}^{2} N/S_{\rm ROI}$, where $R_{t}$ is the GlC tidal radius, $S_{\rm ROI}$ is the area of the ROI, and $N$ is the number of $\gamma$-ray sources detected within $S_{\rm ROI}$. The values of $P_{\rm random}$ are presented in the last column of Table-\ref{tab:results}. In most clusters, the $P_{\rm random}$ value is less than $\sim 10\%$, the exceptions are NGC 104 ($16.9\%$), NGC 5139 ($34.5\%$), NGC 6624 ($18.1\%$), NGC 6752 ($32.1\%$) and NGC 6656 ($40.1\%$), where the random detection probability is over ten percent. However, as illustrated in Figure-\ref{fig:tsmaps}, $R_{t}$ of these clusters are also found to be very large, thus the larger $P_{\rm random}$ is more likely to result from the larger tidal radius rather than being associated with a fore/background source. 

\begin{figure*}
\leftline{
\includegraphics[width=0.25\textwidth]{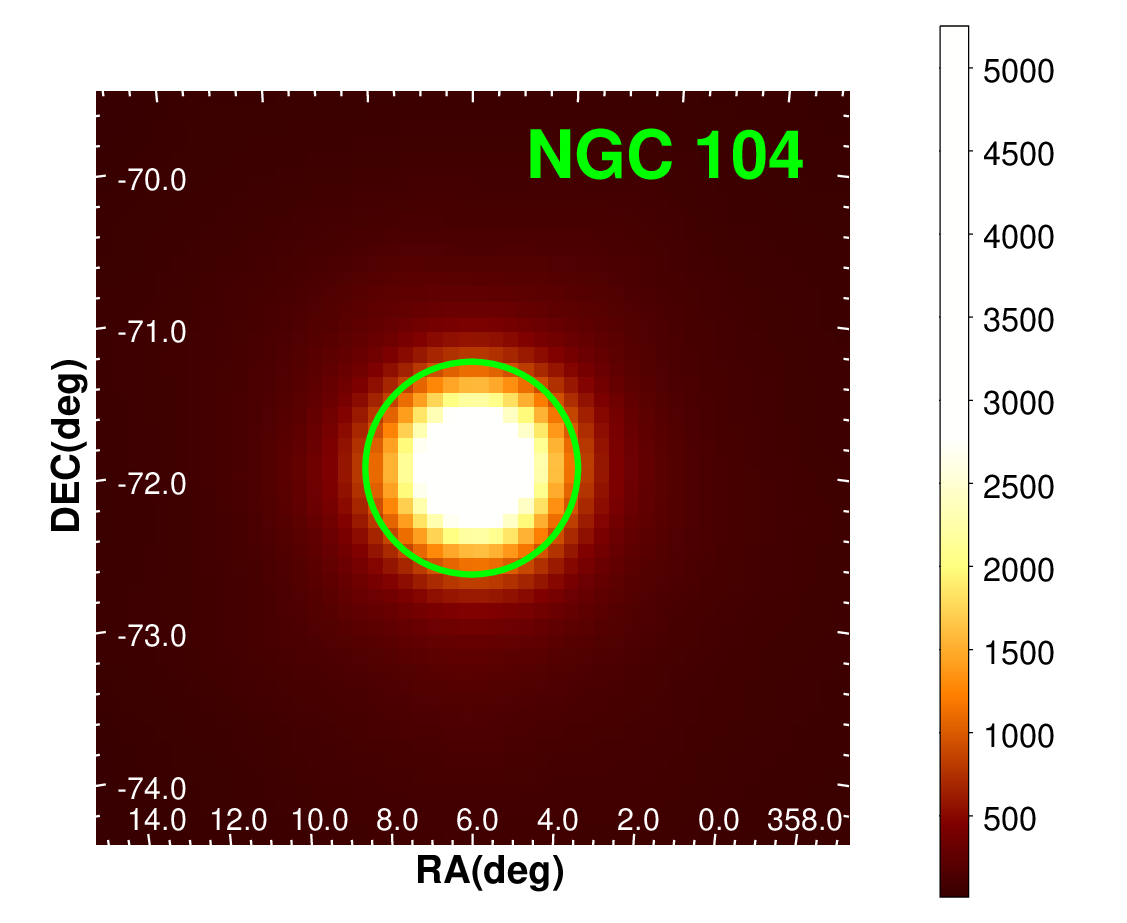}
\includegraphics[width=0.25\textwidth]{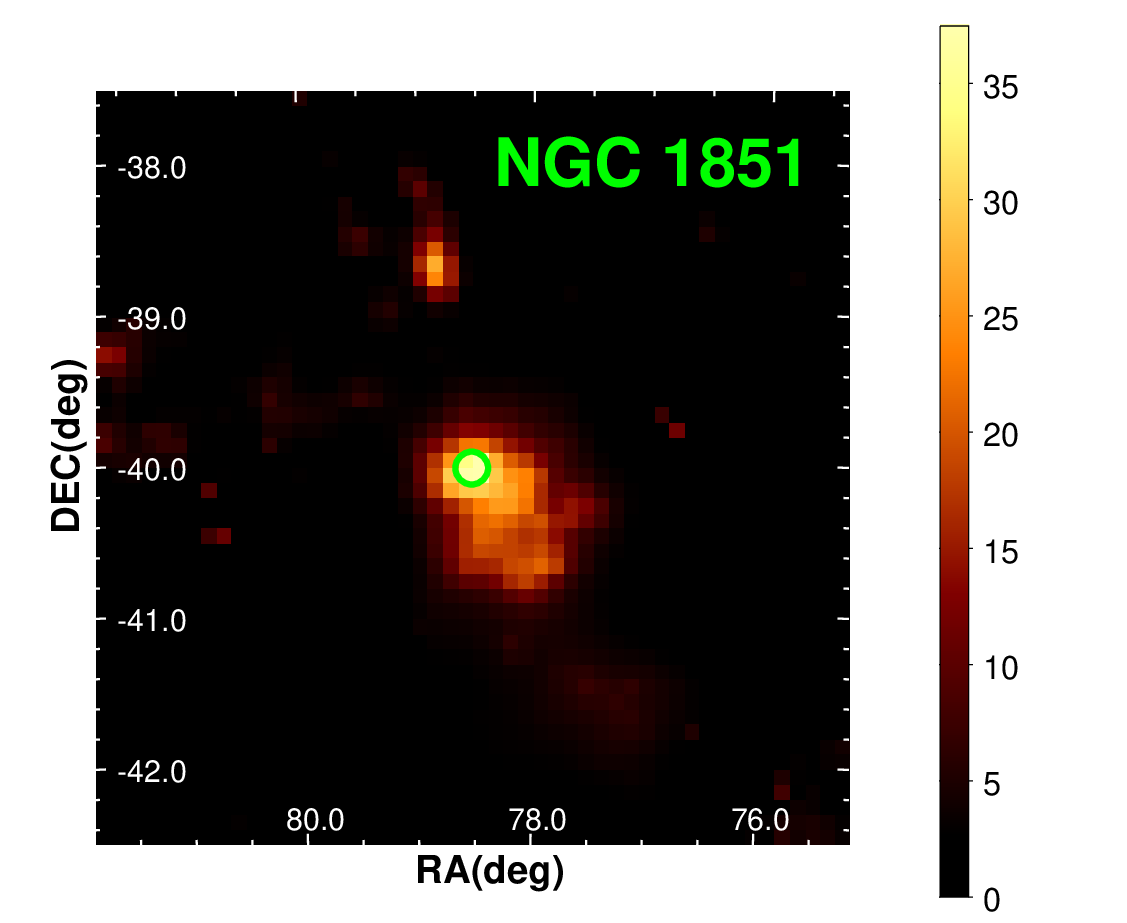}
\includegraphics[width=0.25\textwidth]{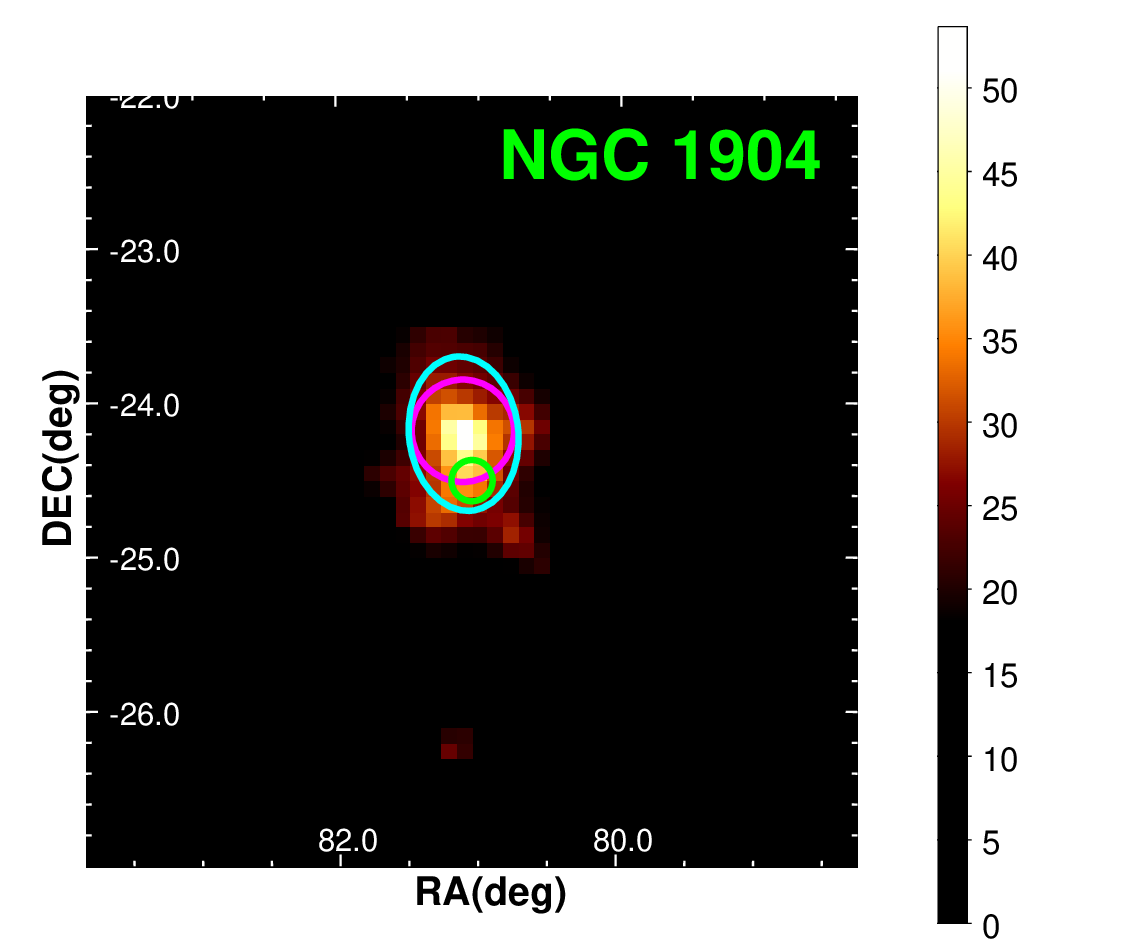}
\includegraphics[width=0.25\textwidth]{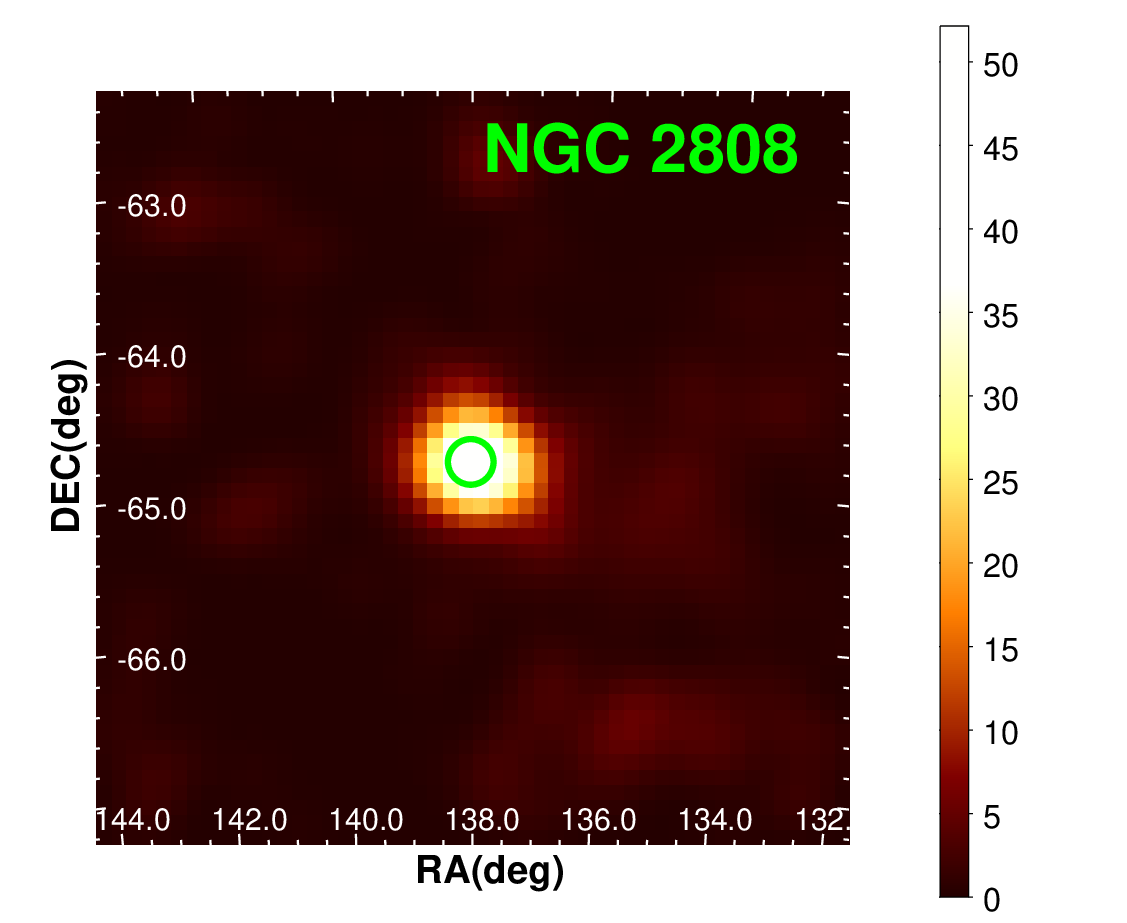}
}
\leftline{
\includegraphics[width=0.25\textwidth]{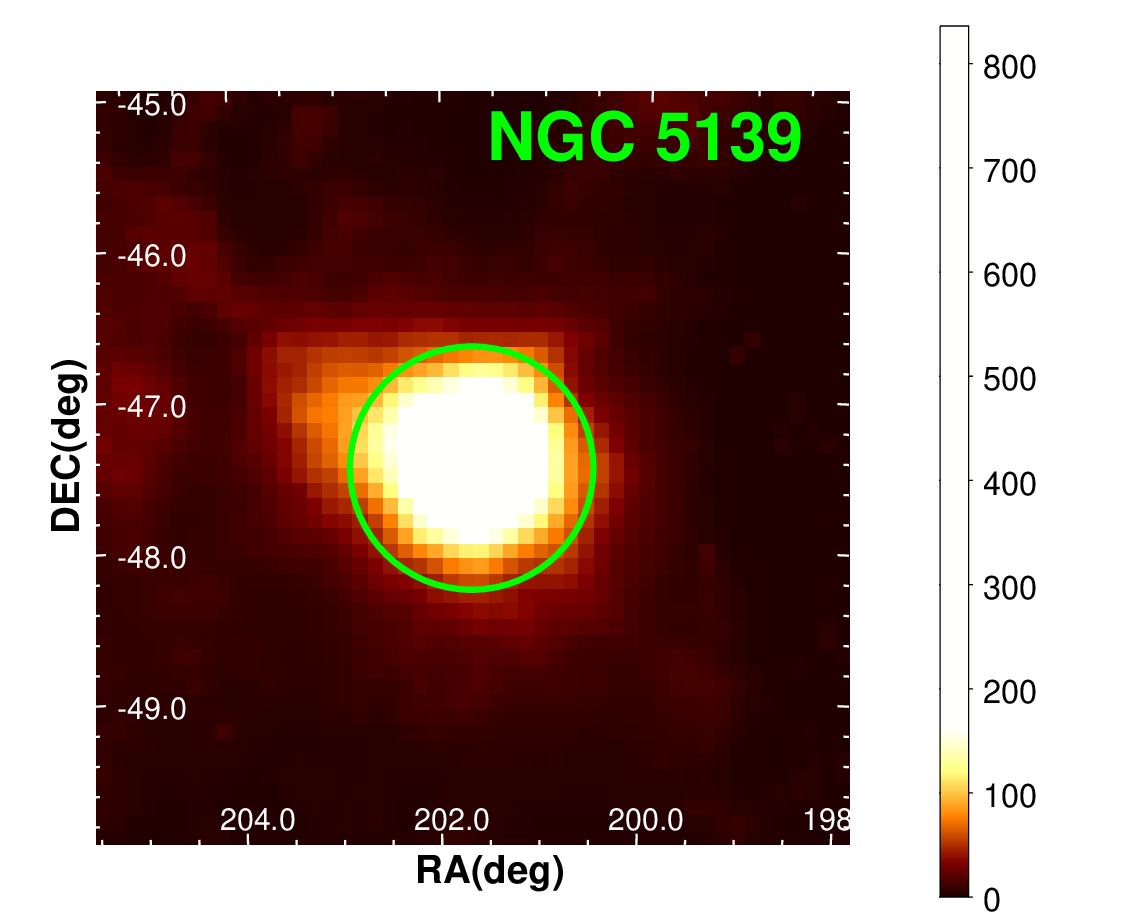}
\includegraphics[width=0.25\textwidth]{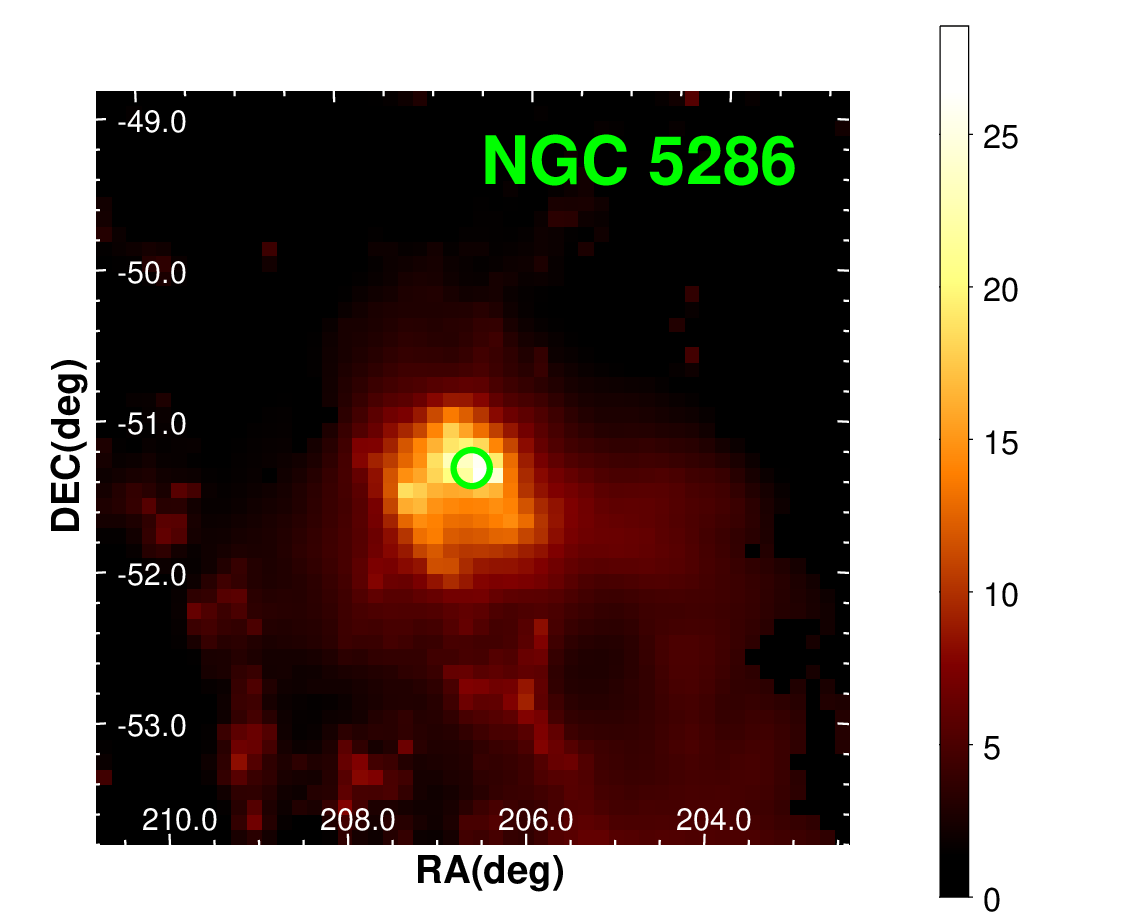}
\includegraphics[width=0.25\textwidth]{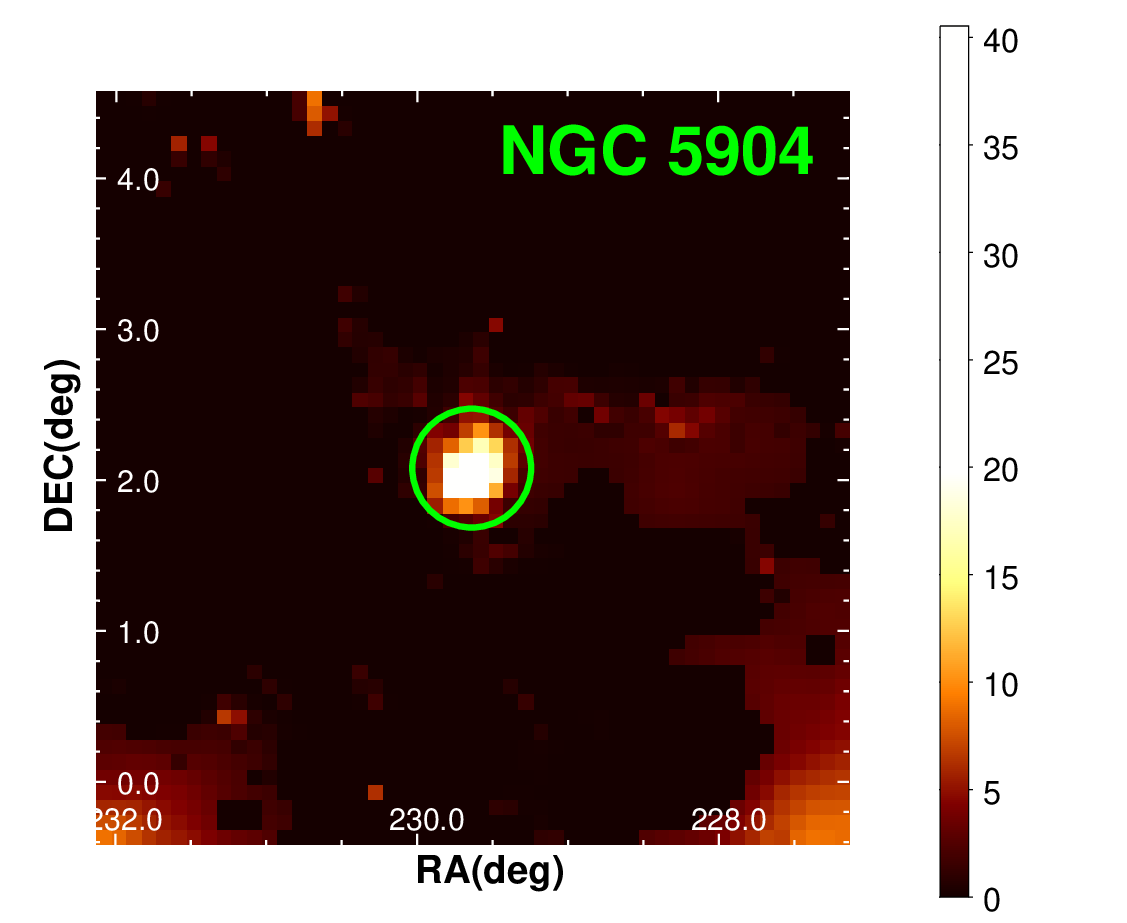}
\includegraphics[width=0.25\textwidth]{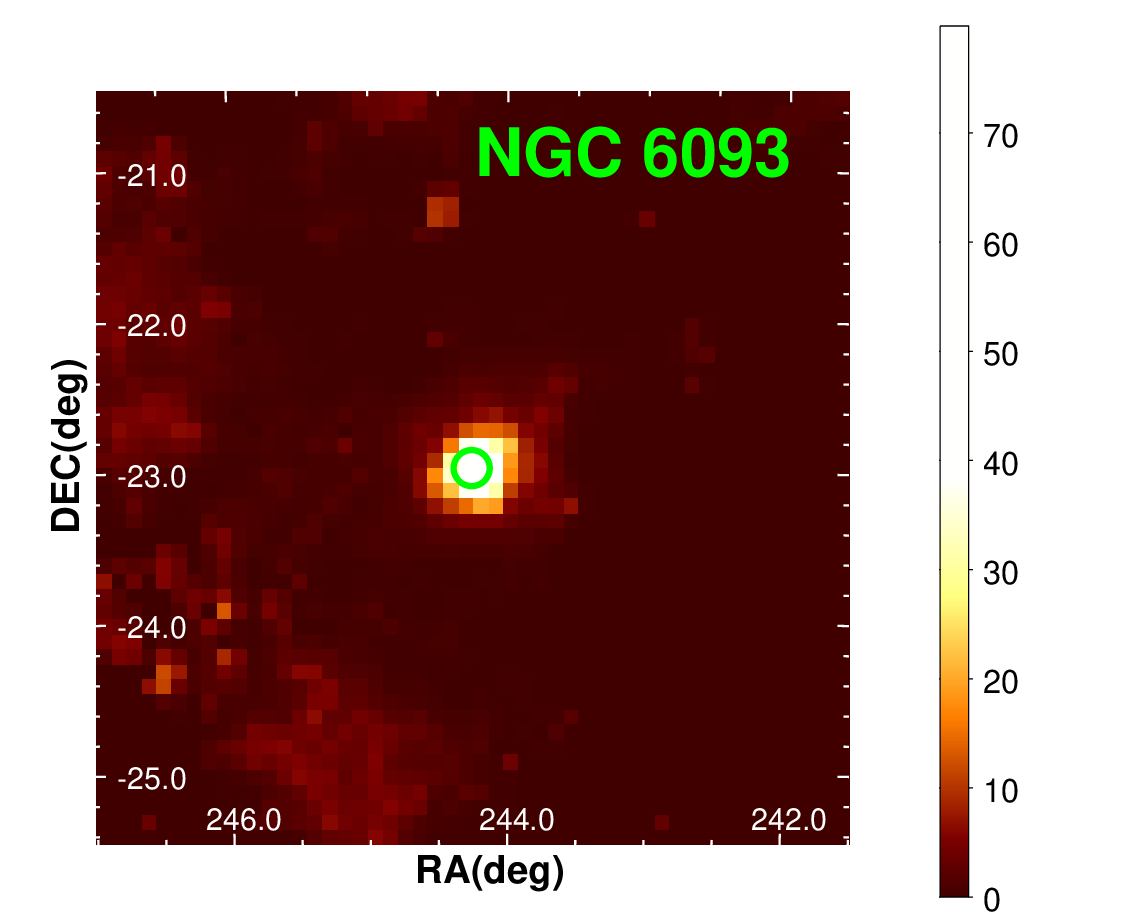}
}
\leftline{
\includegraphics[width=0.25\textwidth]{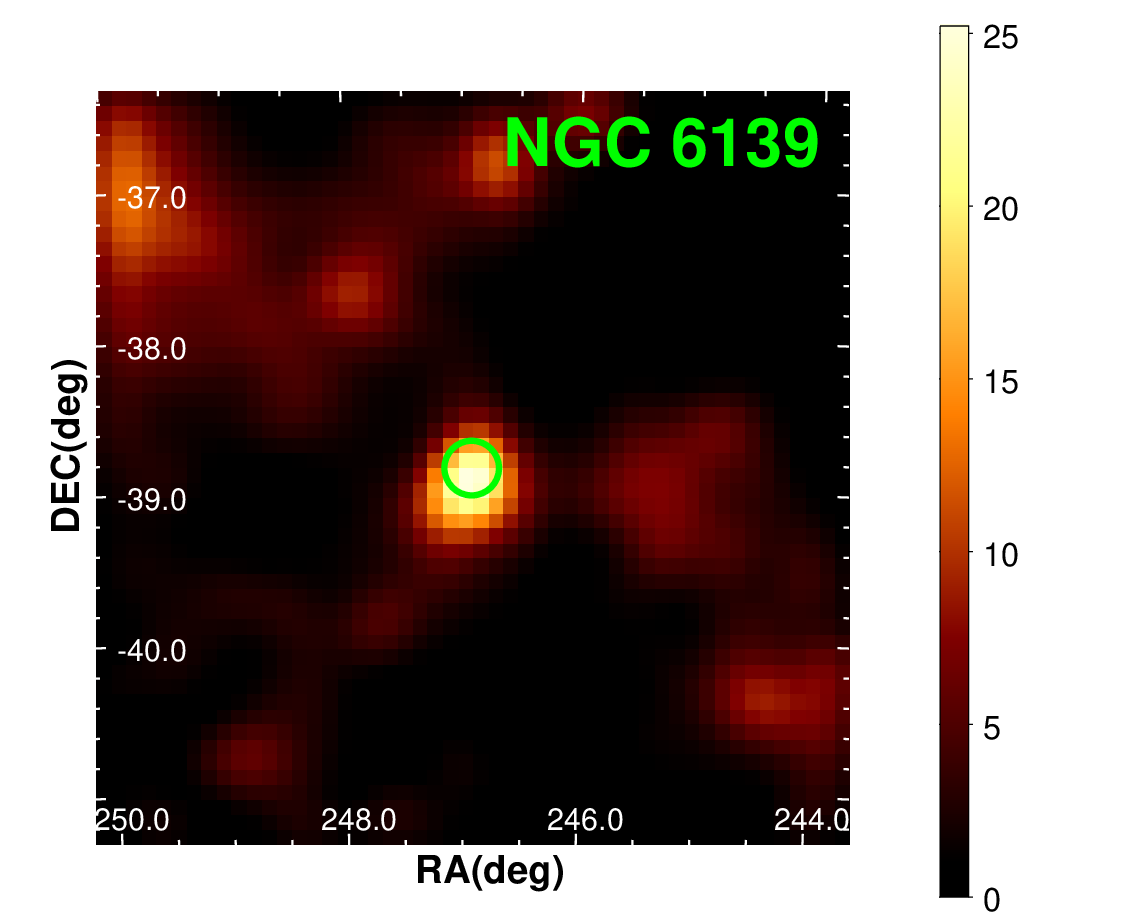}
\includegraphics[width=0.25\textwidth]{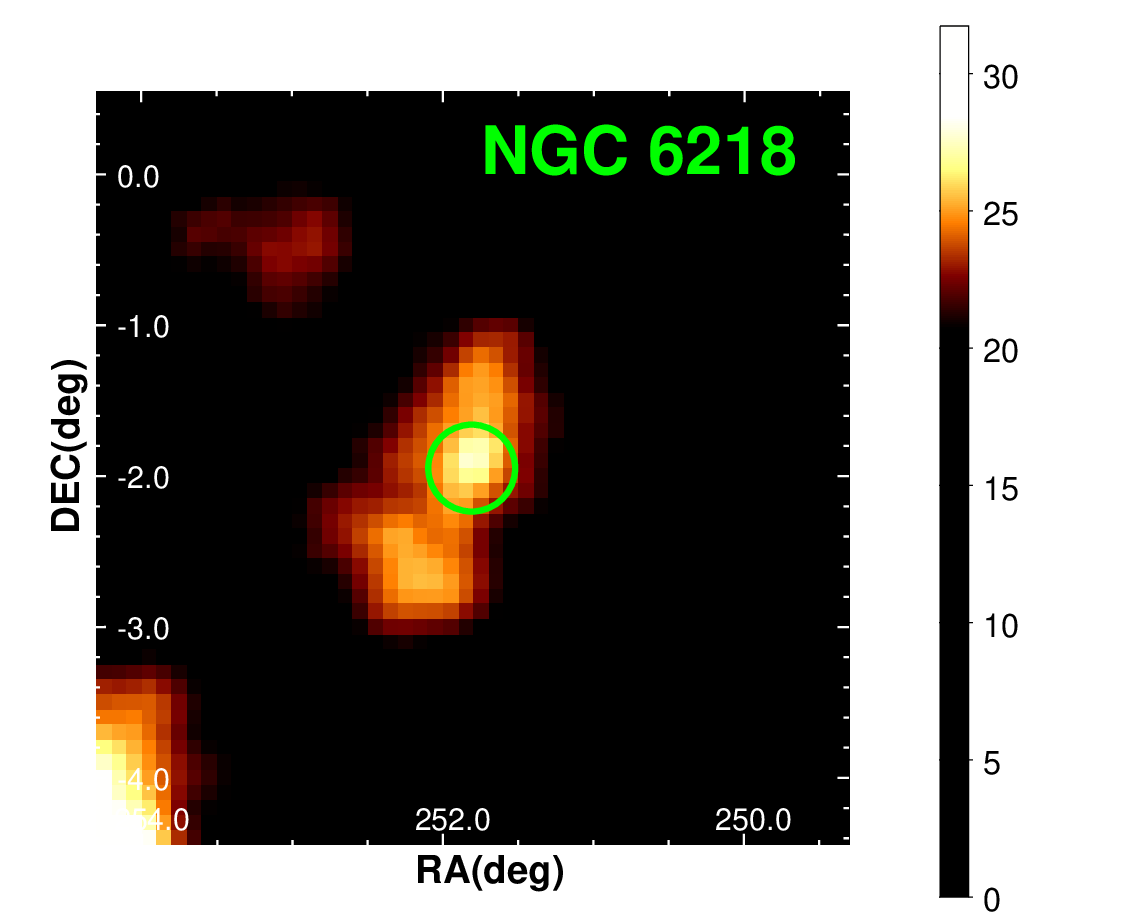}
\includegraphics[width=0.25\textwidth]{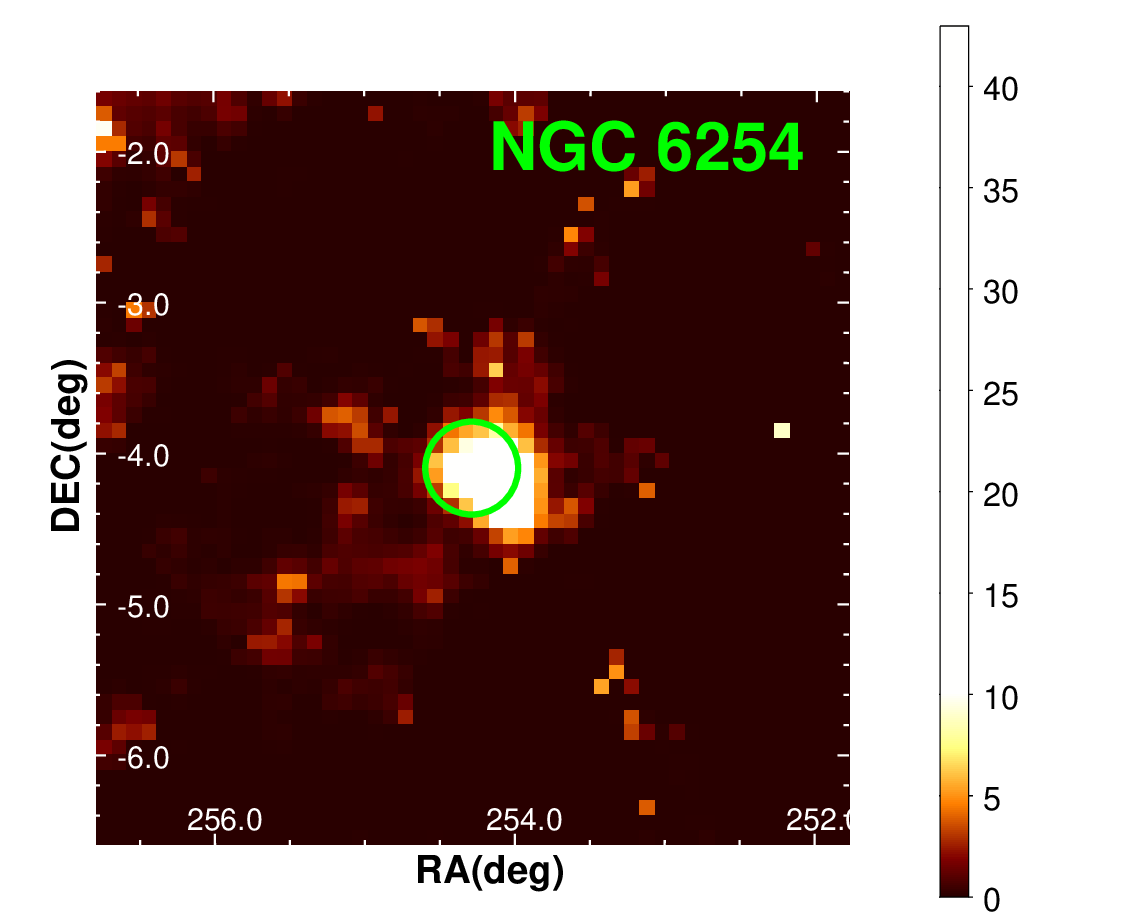}
\includegraphics[width=0.25\textwidth]{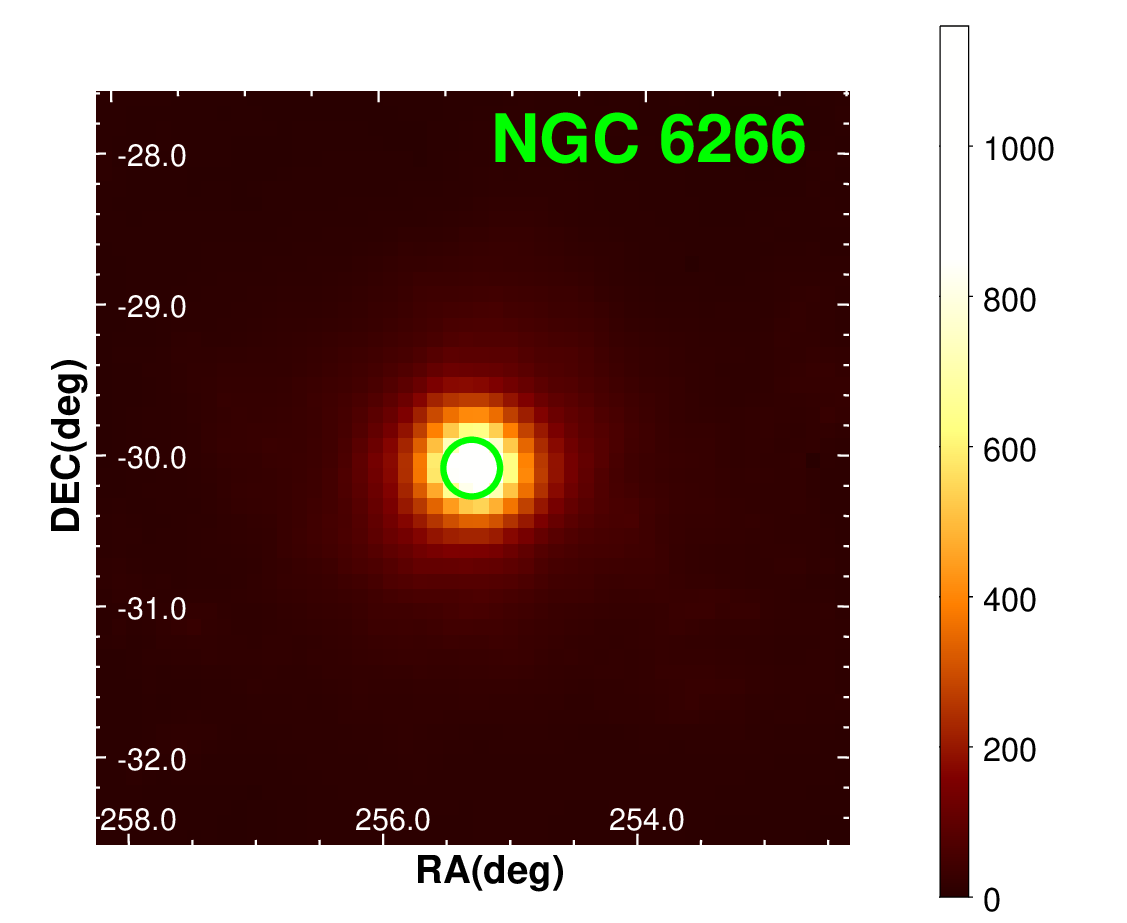}
}
\leftline{
\includegraphics[width=0.25\textwidth]{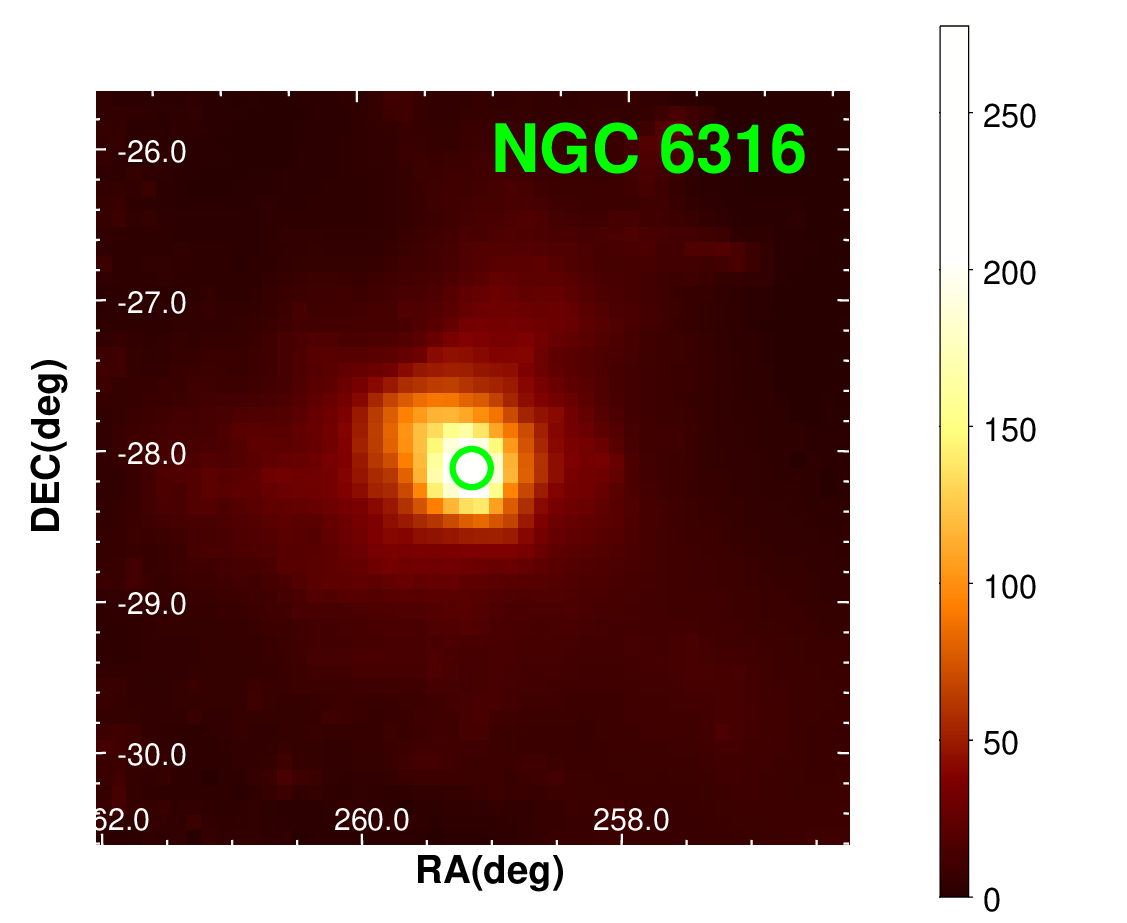}
\includegraphics[width=0.25\textwidth]{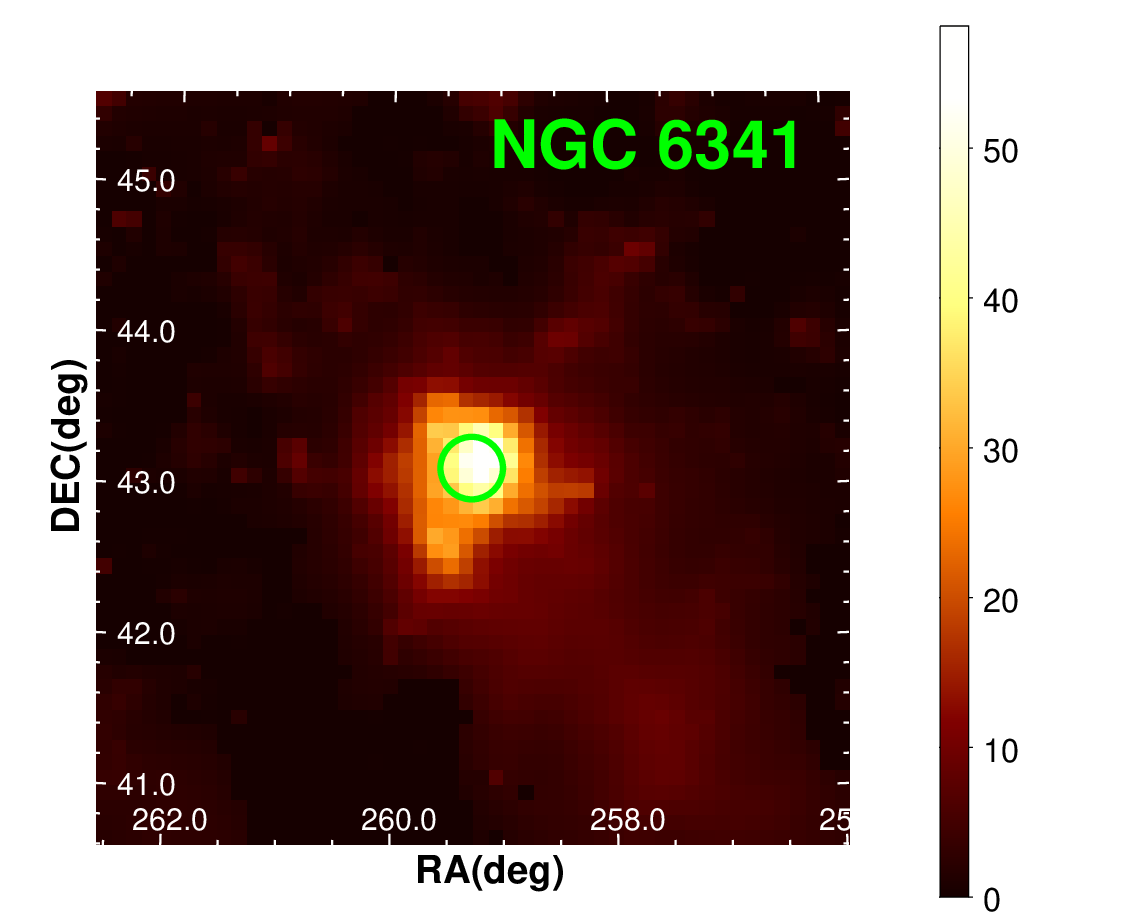}
\includegraphics[width=0.25\textwidth]{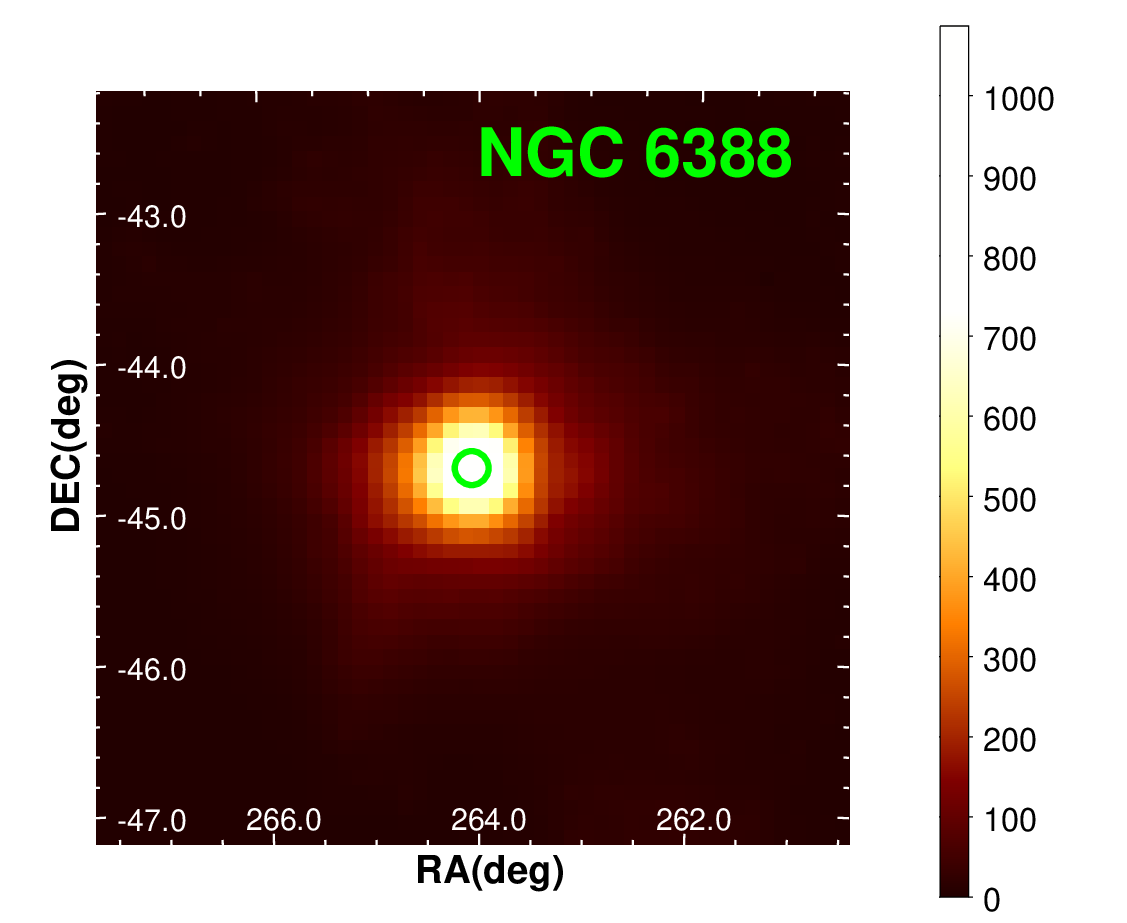}
\includegraphics[width=0.25\textwidth]{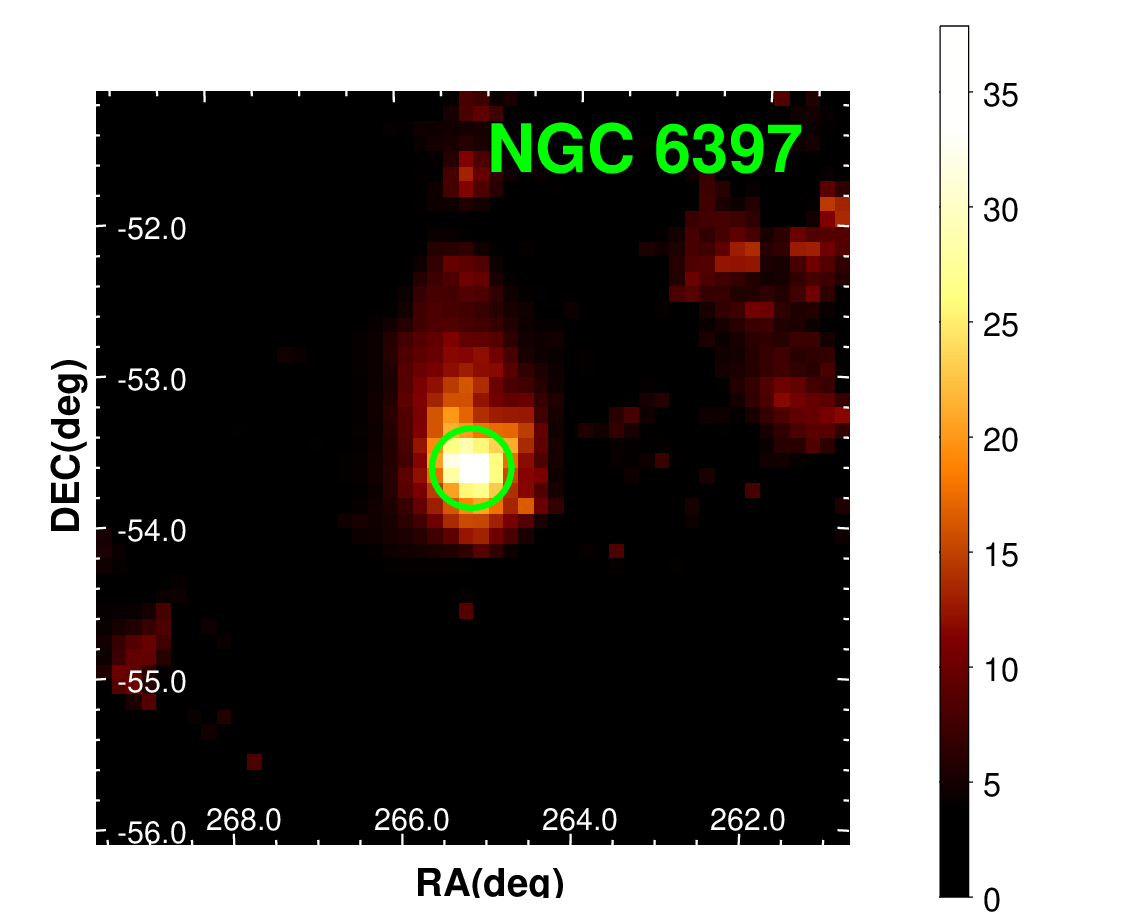}
}
\leftline{
\includegraphics[width=0.25\textwidth]{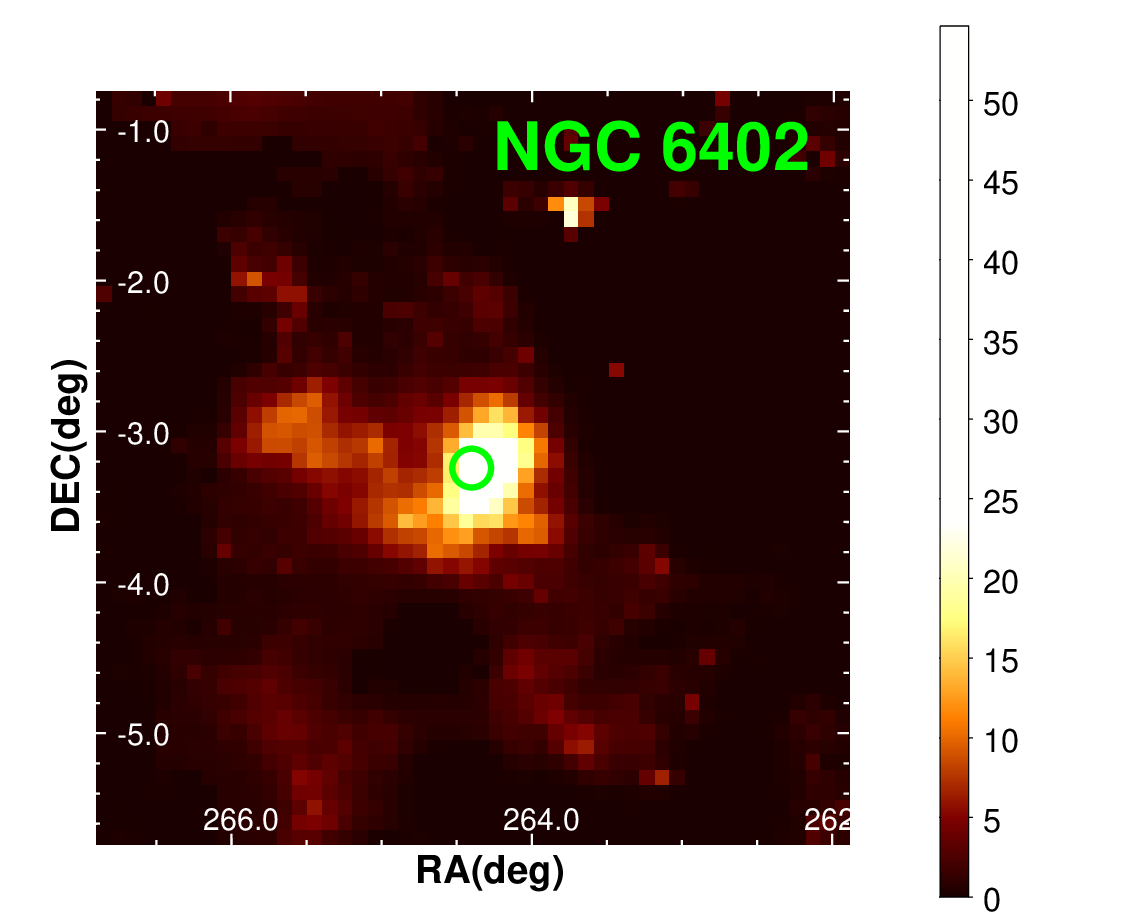}
\includegraphics[width=0.25\textwidth]{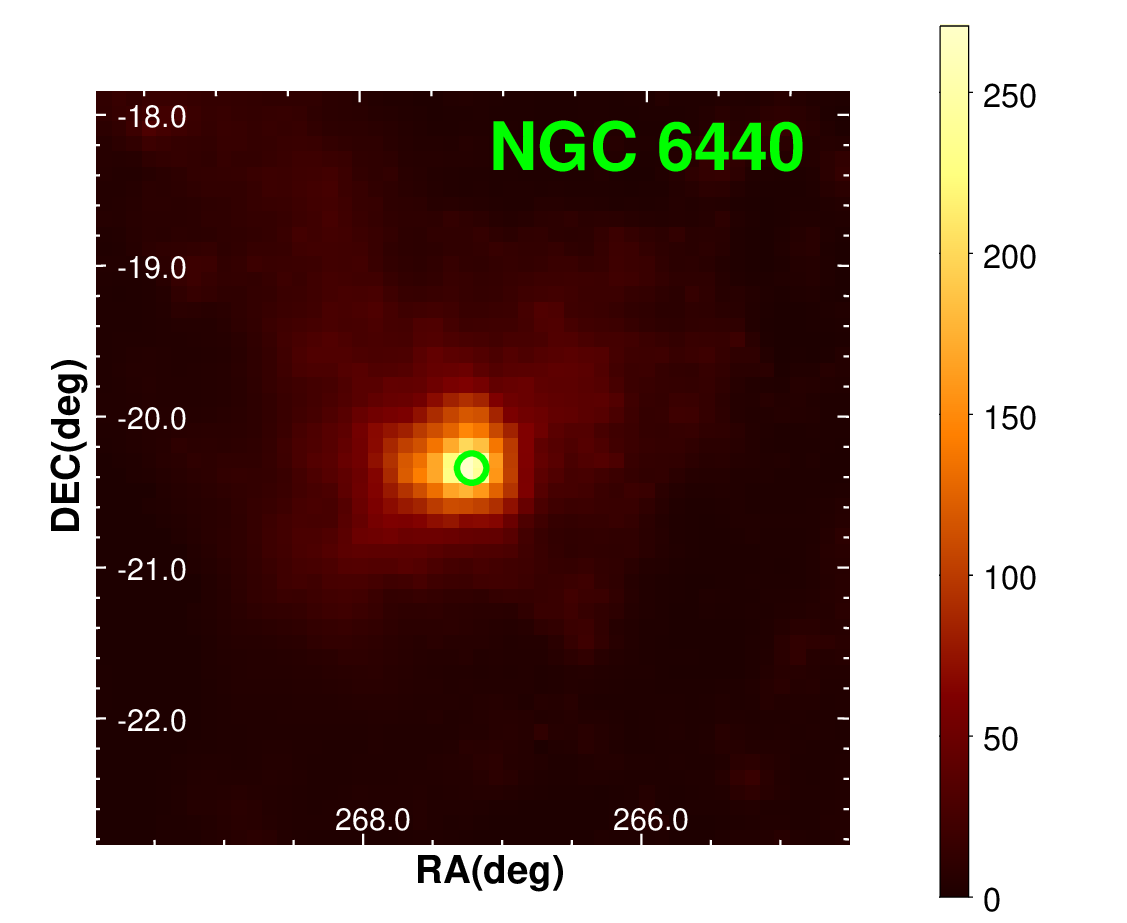}
\includegraphics[width=0.25\textwidth]{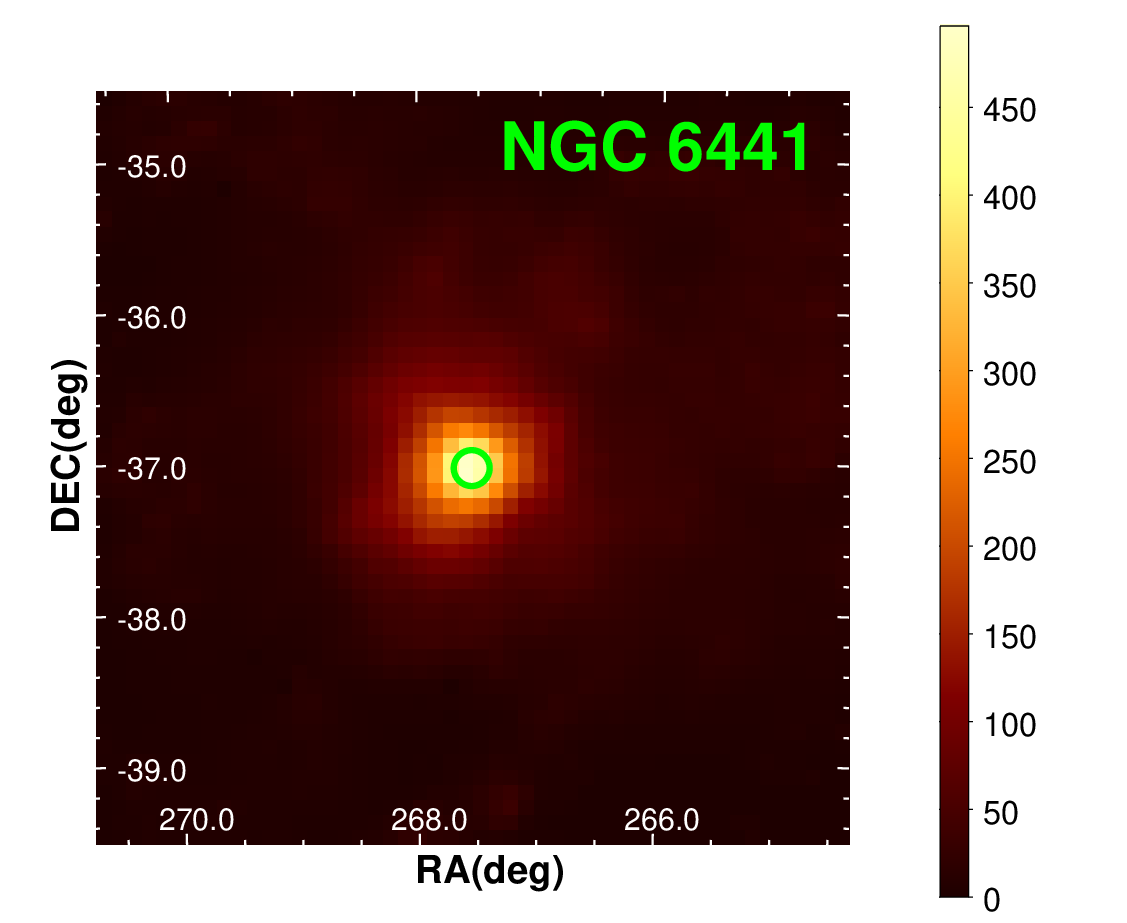}
\includegraphics[width=0.25\textwidth]{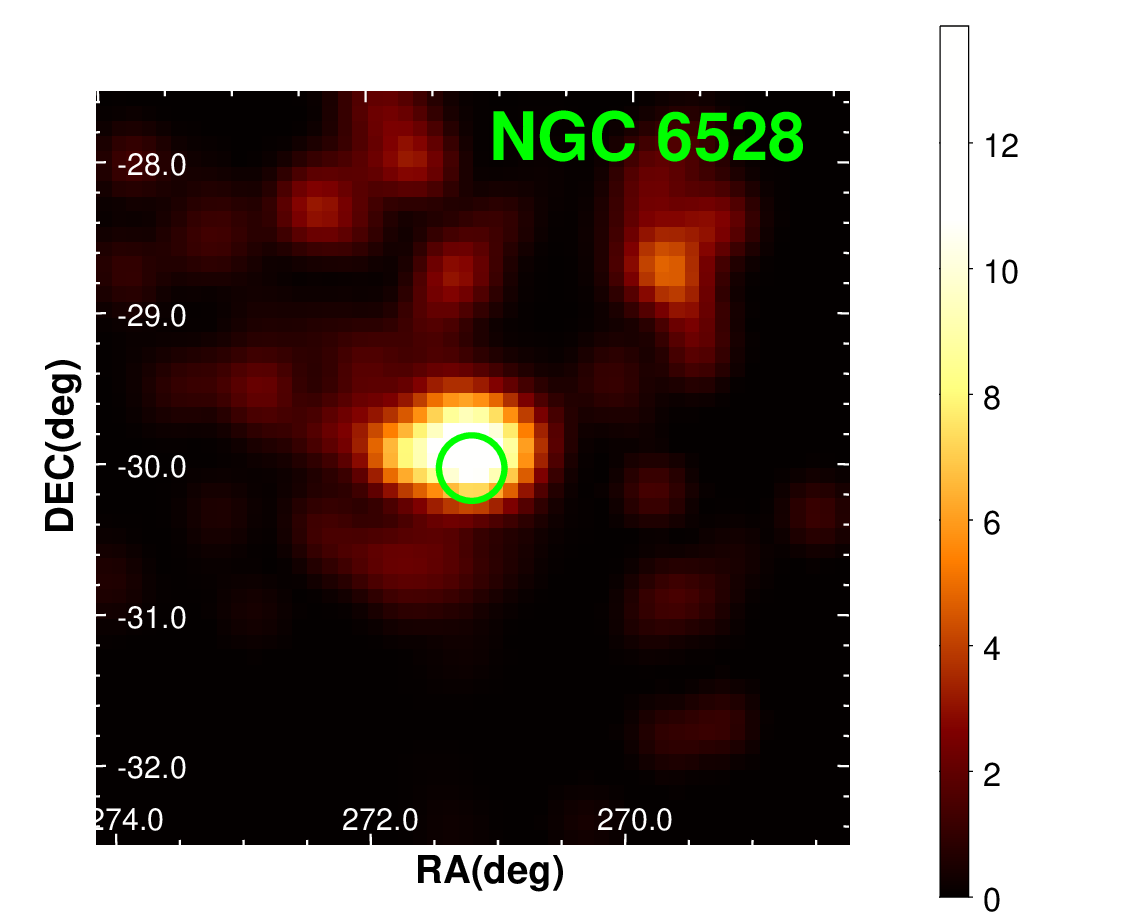}
}
\leftline{
\includegraphics[width=0.25\textwidth]{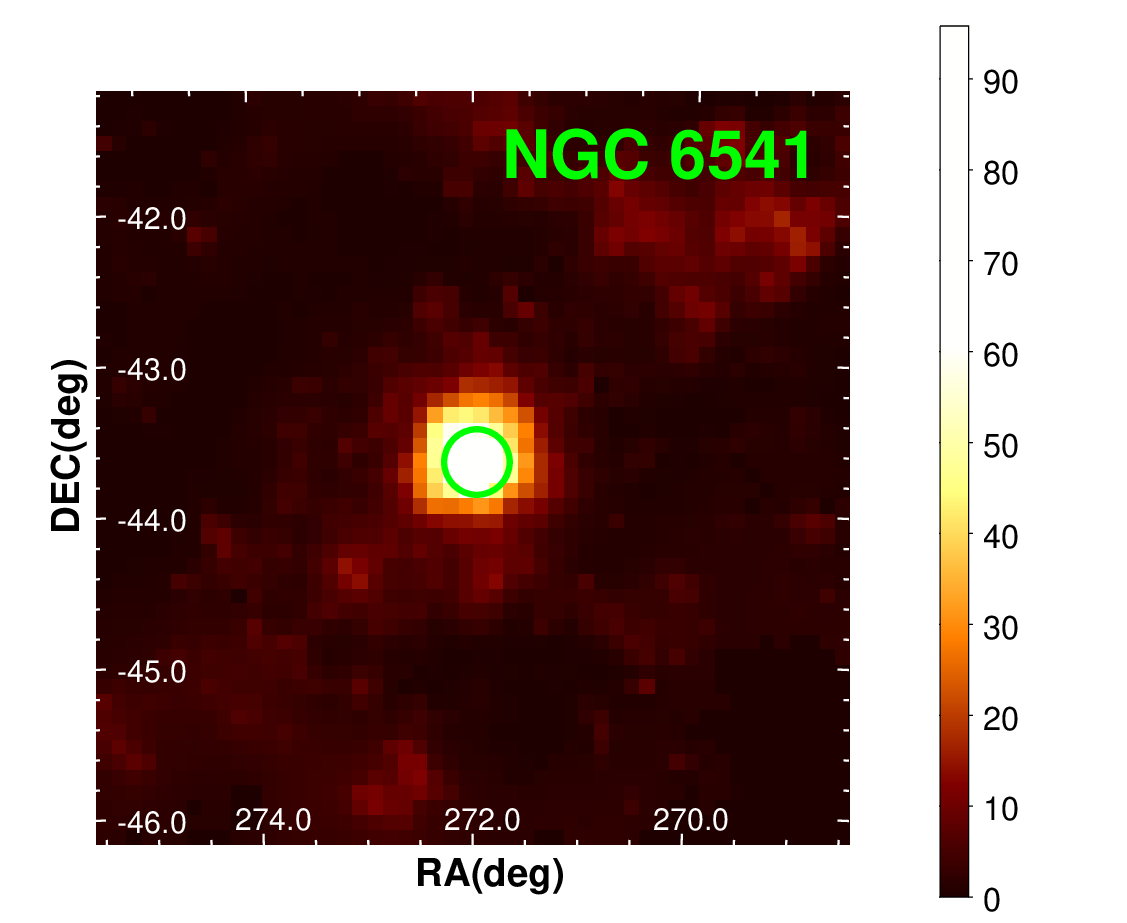}
\includegraphics[width=0.25\textwidth]{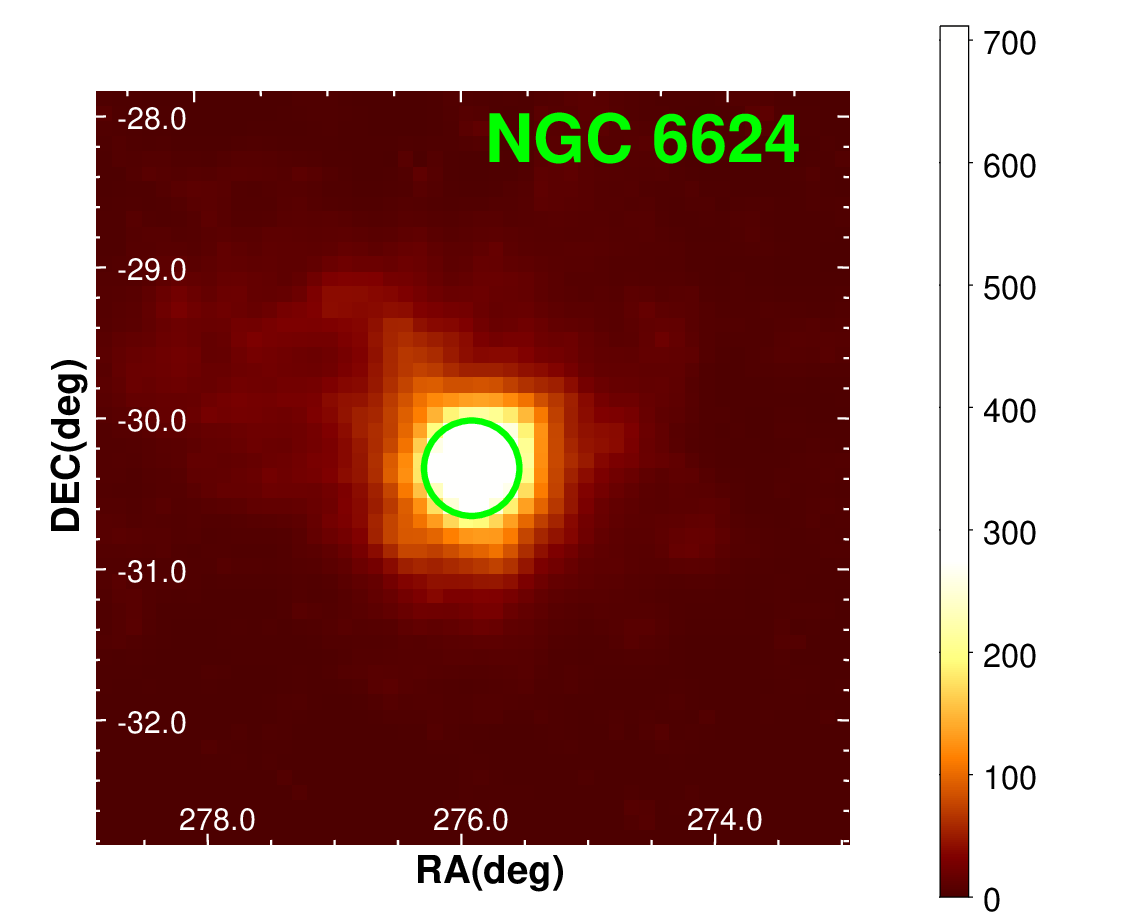}
\includegraphics[width=0.25\textwidth]{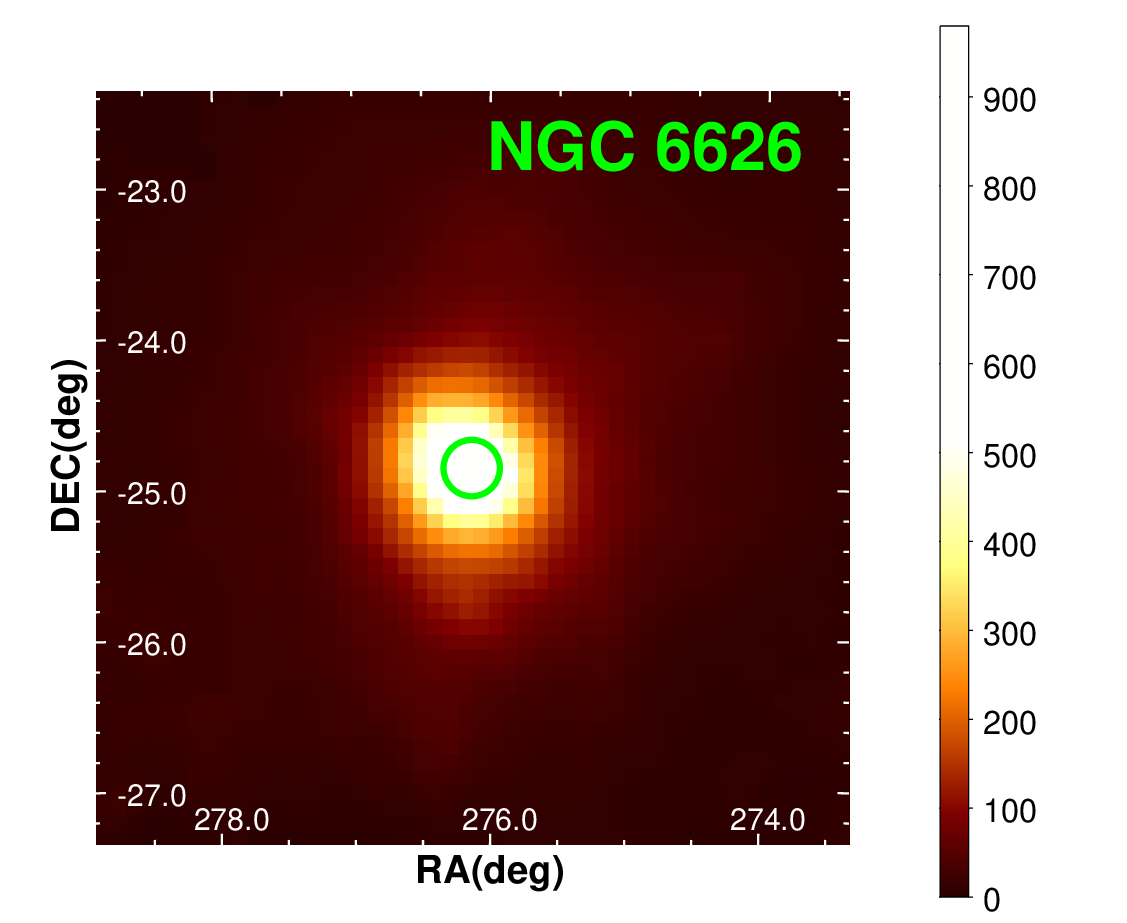}
\includegraphics[width=0.25\textwidth]{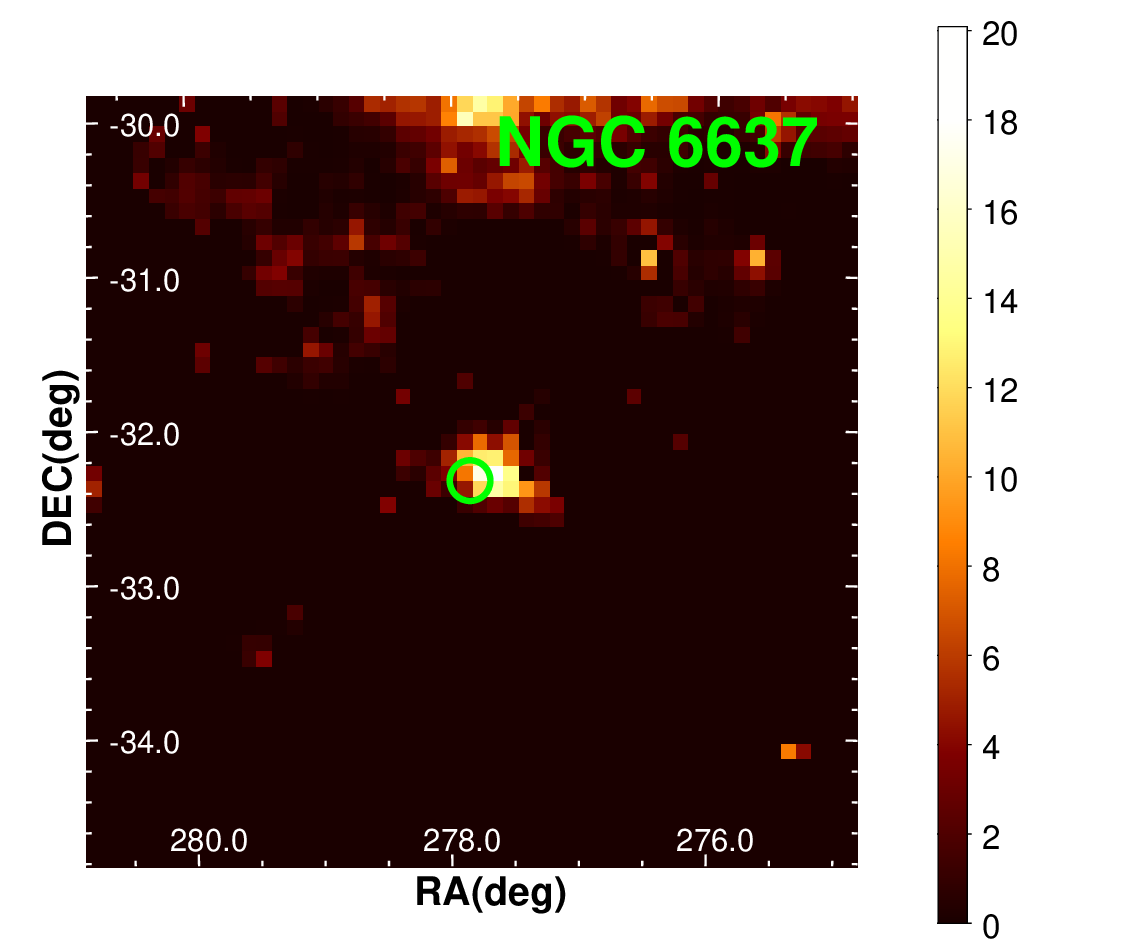}
}
\vspace*{1.5cm}
\caption{$5\degr \times 5\degr$ test statistic maps of the 39 GlCs with TS $>$ 25 in Table-\ref{tab:results}. The GlC name is shown as green text in the upper right of each panel, and the green circle represents the tidal radius of the cluster. The $\gamma$-ray emission are in coincident with the optical centers of the GlCs. The exception is NGC 1904, where the $1\sigma$ error (magenta circle) of the $\gamma$-ray emission peak is in coincident with the $95\%$ error ellipse (cyan) of the 4FGL catalog, which has an offset of $\sim 0.3\degr$ from the cluster center.}
\label{fig:tsmaps}
\end{figure*}

\addtocounter{figure}{-1}
\begin{figure*}
\leftline{
\includegraphics[width=0.25\textwidth]{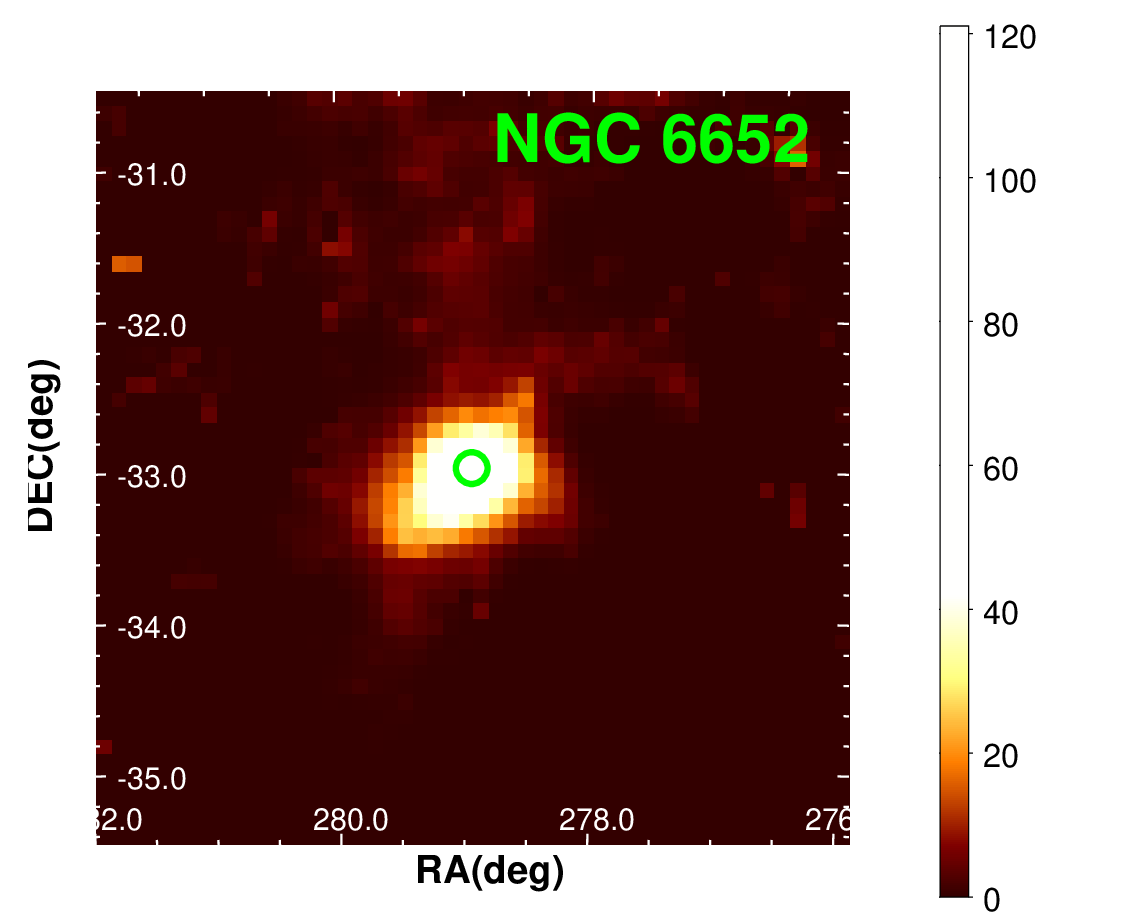}
\includegraphics[width=0.25\textwidth]{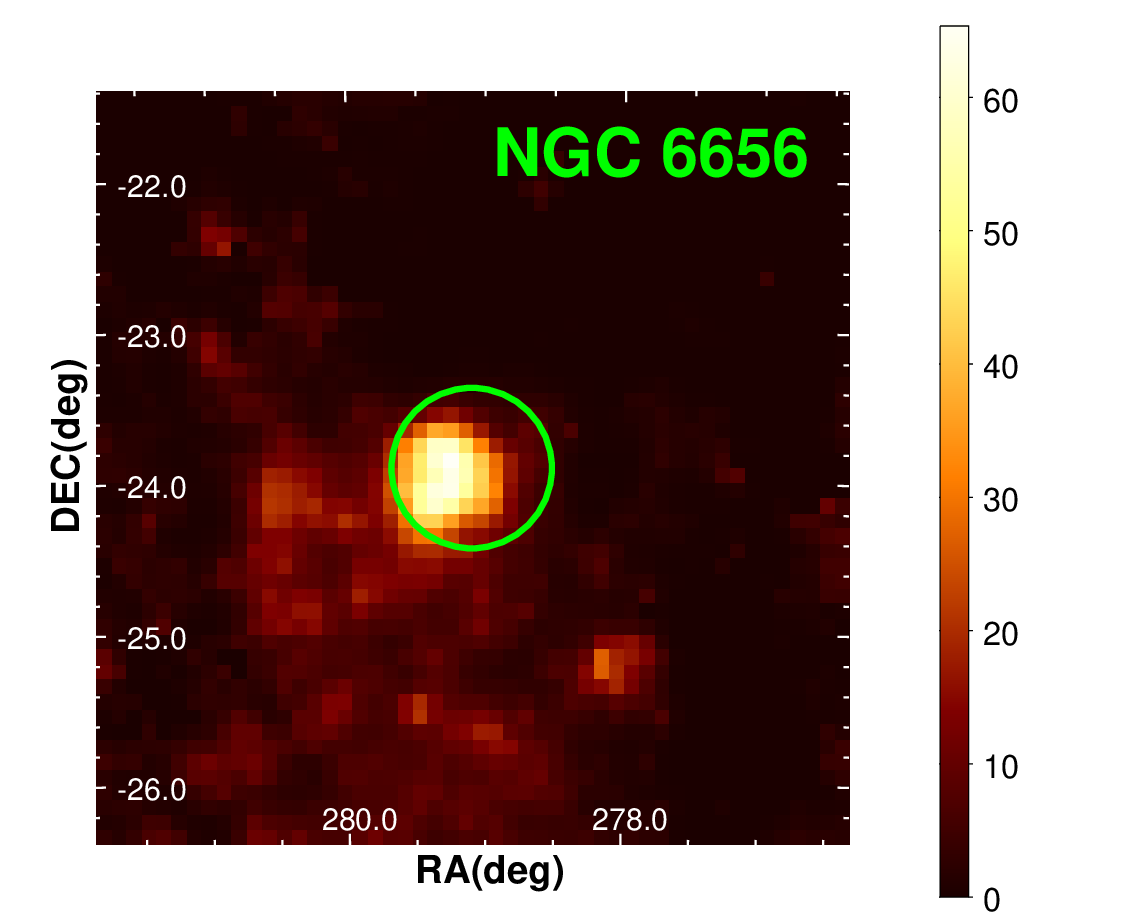}
\includegraphics[width=0.25\textwidth]{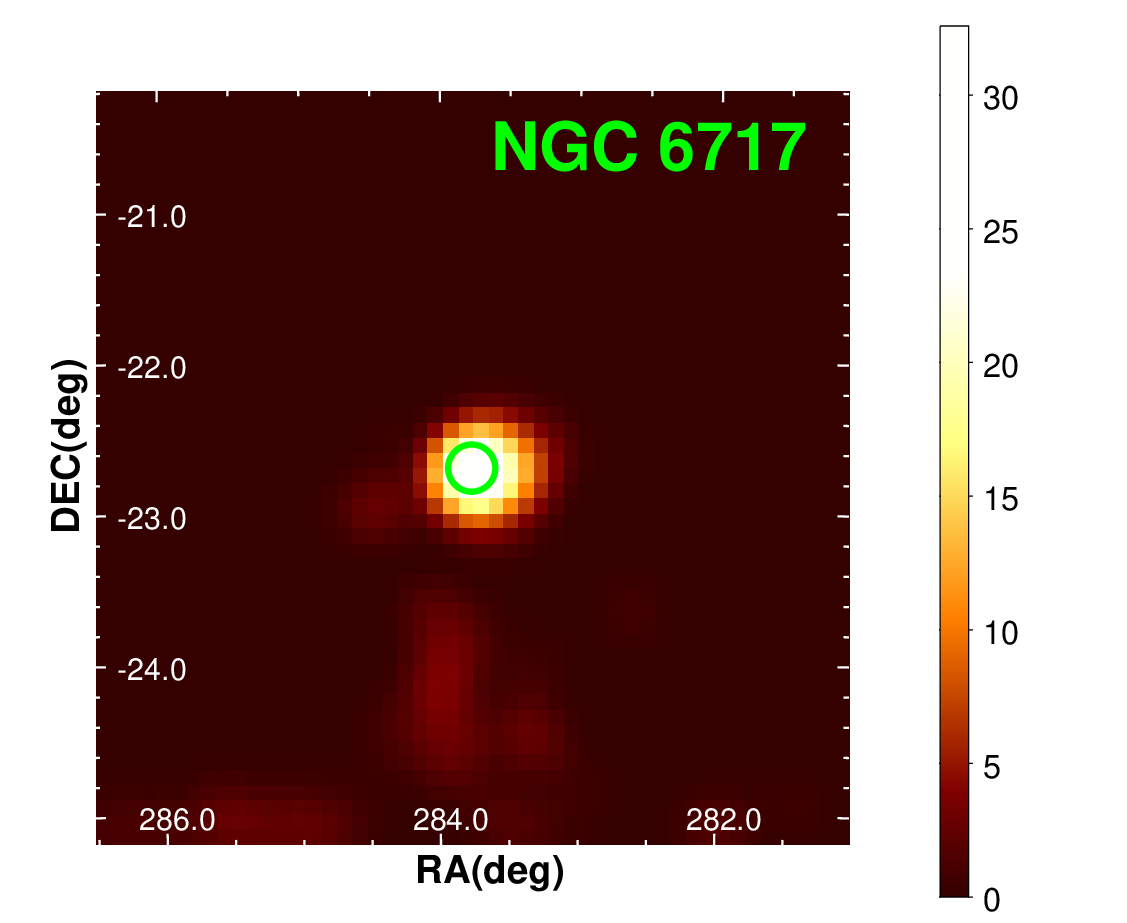}
\includegraphics[width=0.25\textwidth]{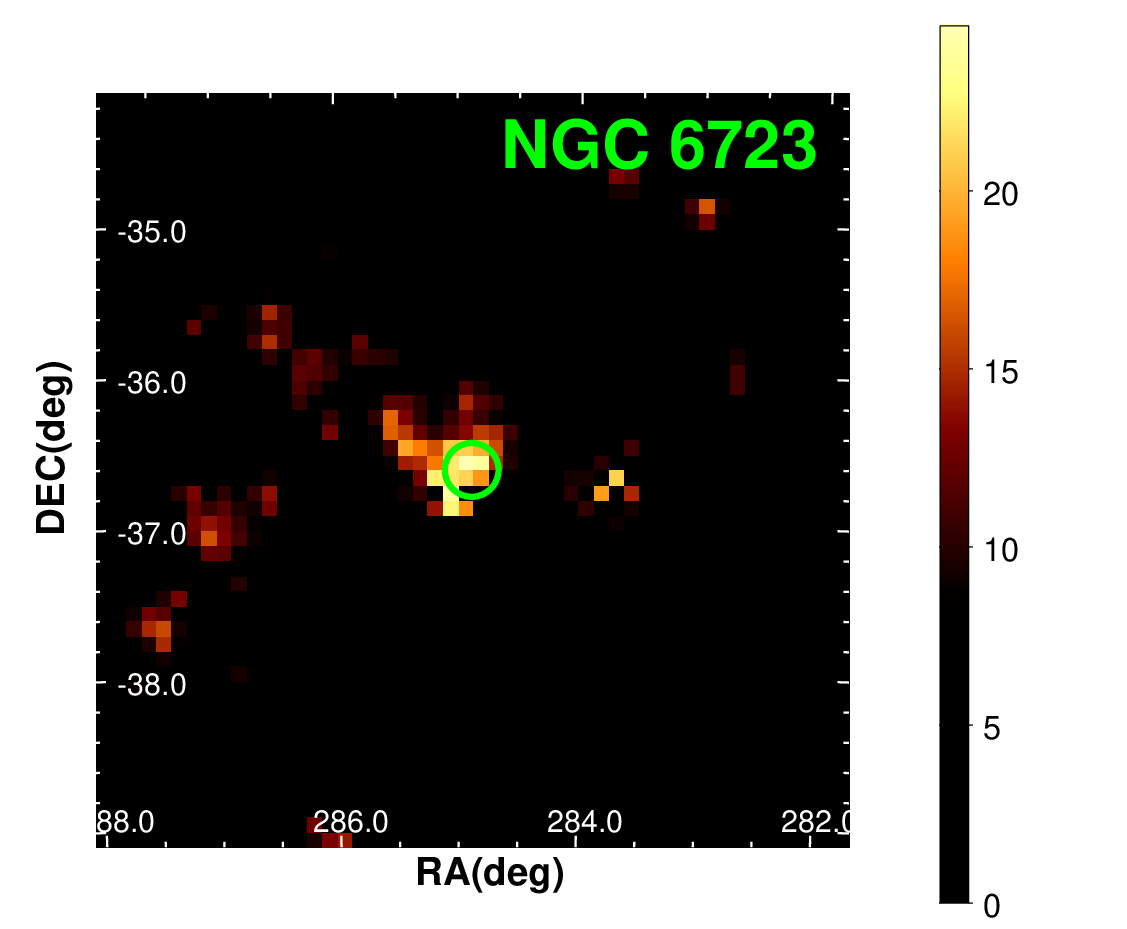}
}
\leftline{
\includegraphics[width=0.25\textwidth]{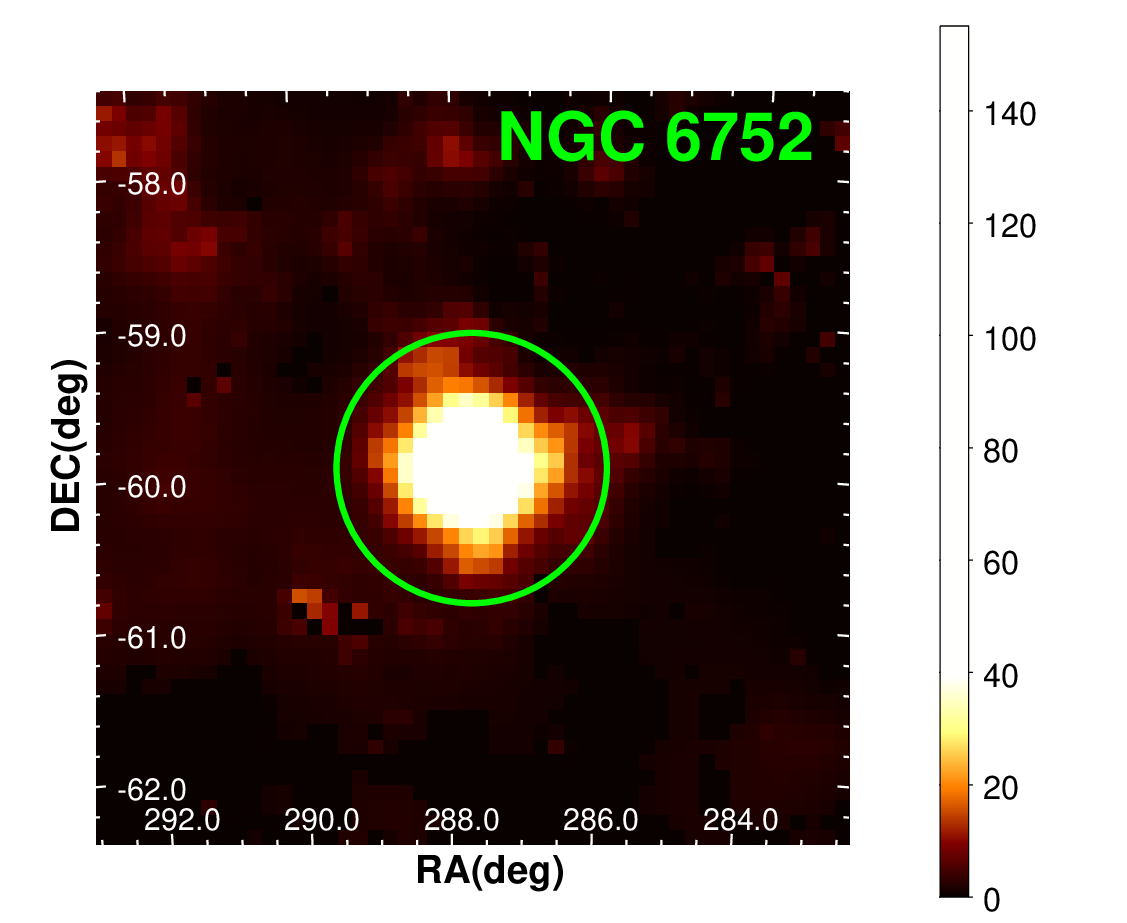}
\includegraphics[width=0.25\textwidth]{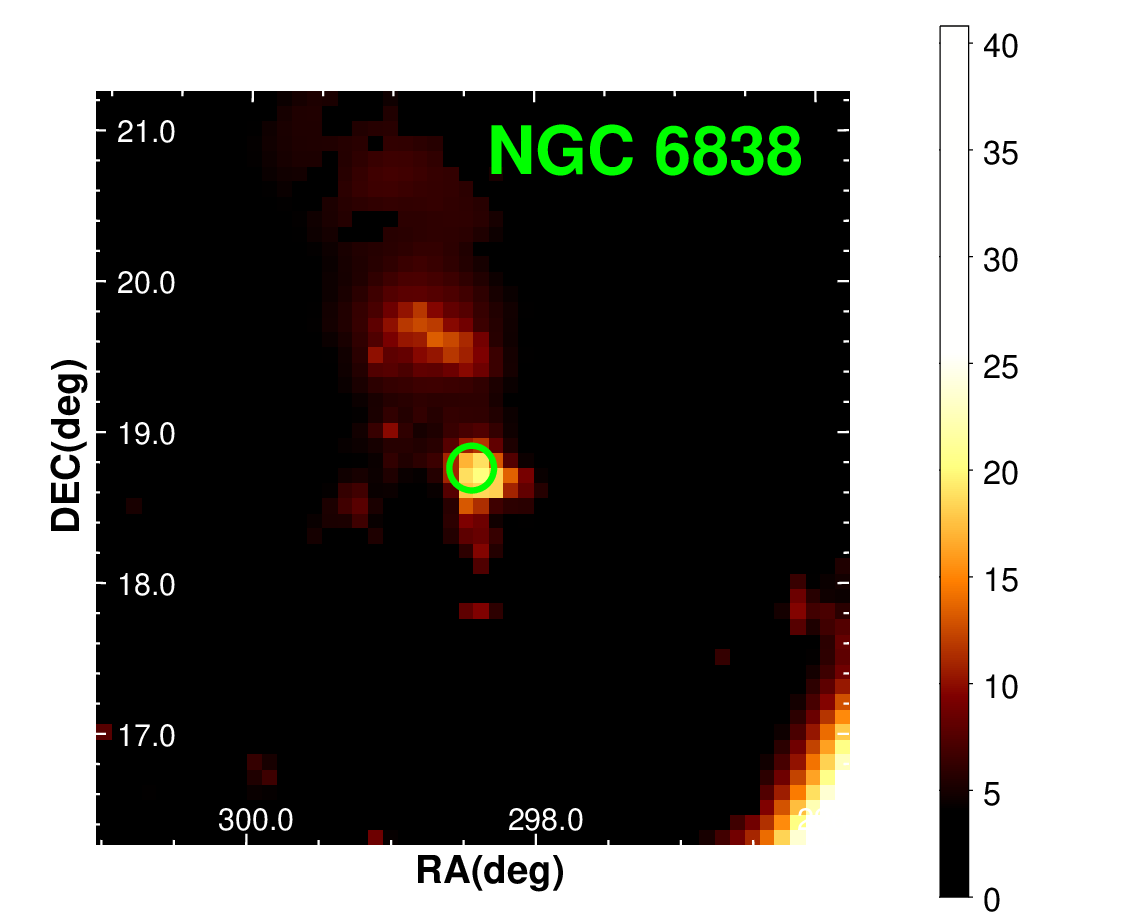}
\includegraphics[width=0.25\textwidth]{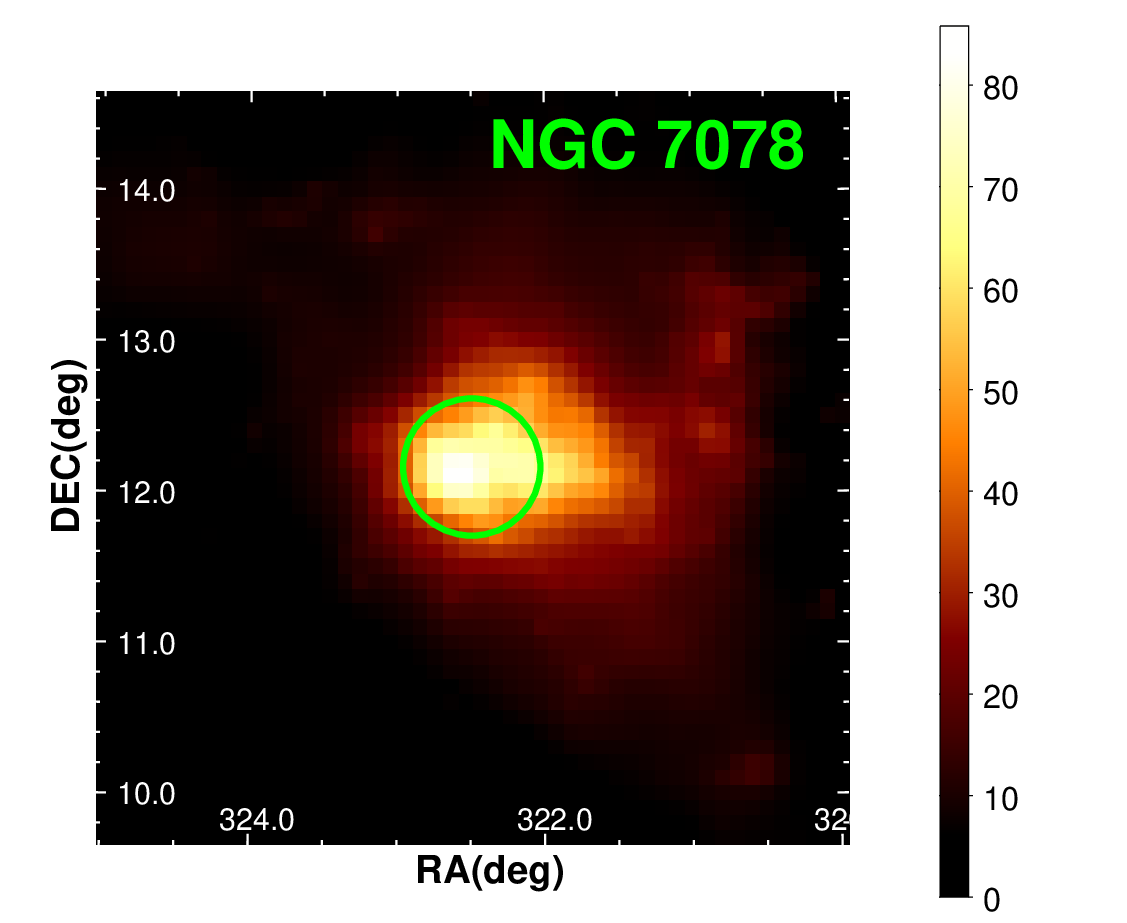}
\includegraphics[width=0.25\textwidth]{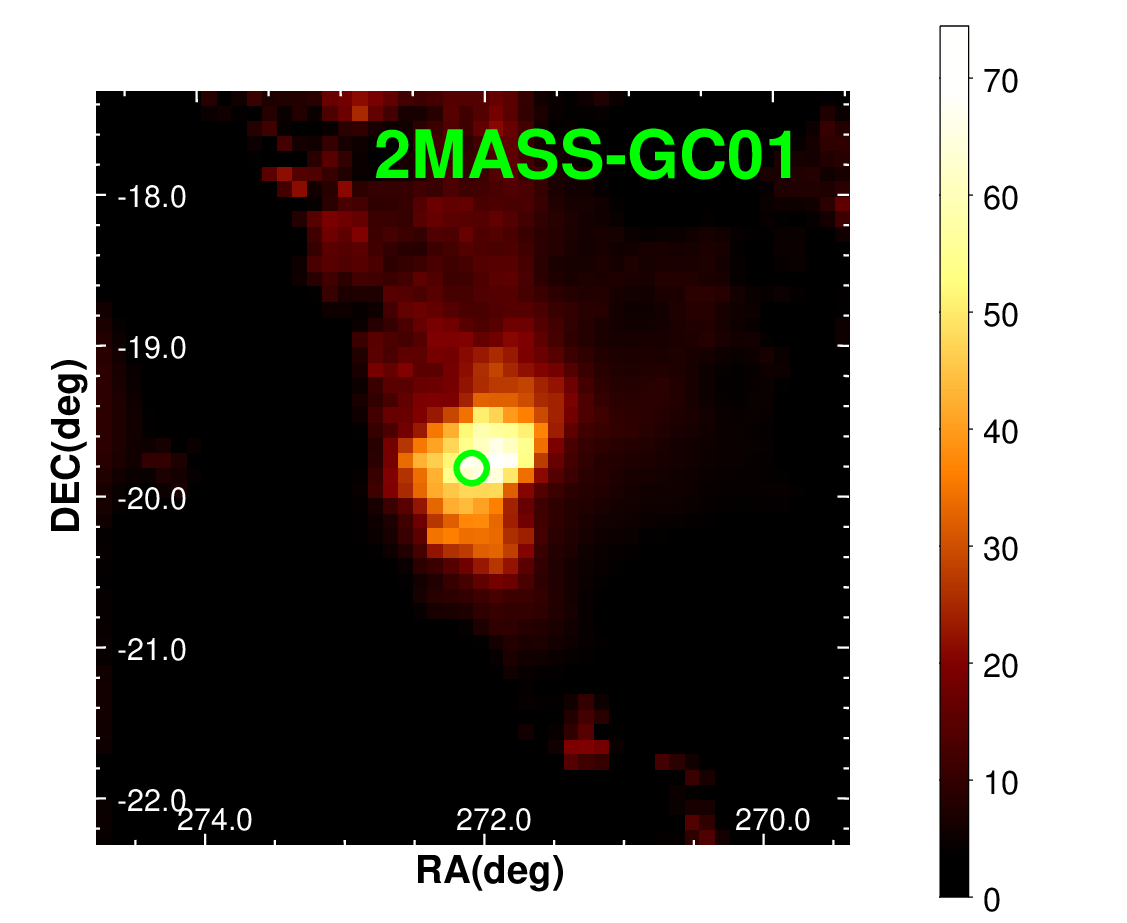}
}
\leftline{
\includegraphics[width=0.25\textwidth]{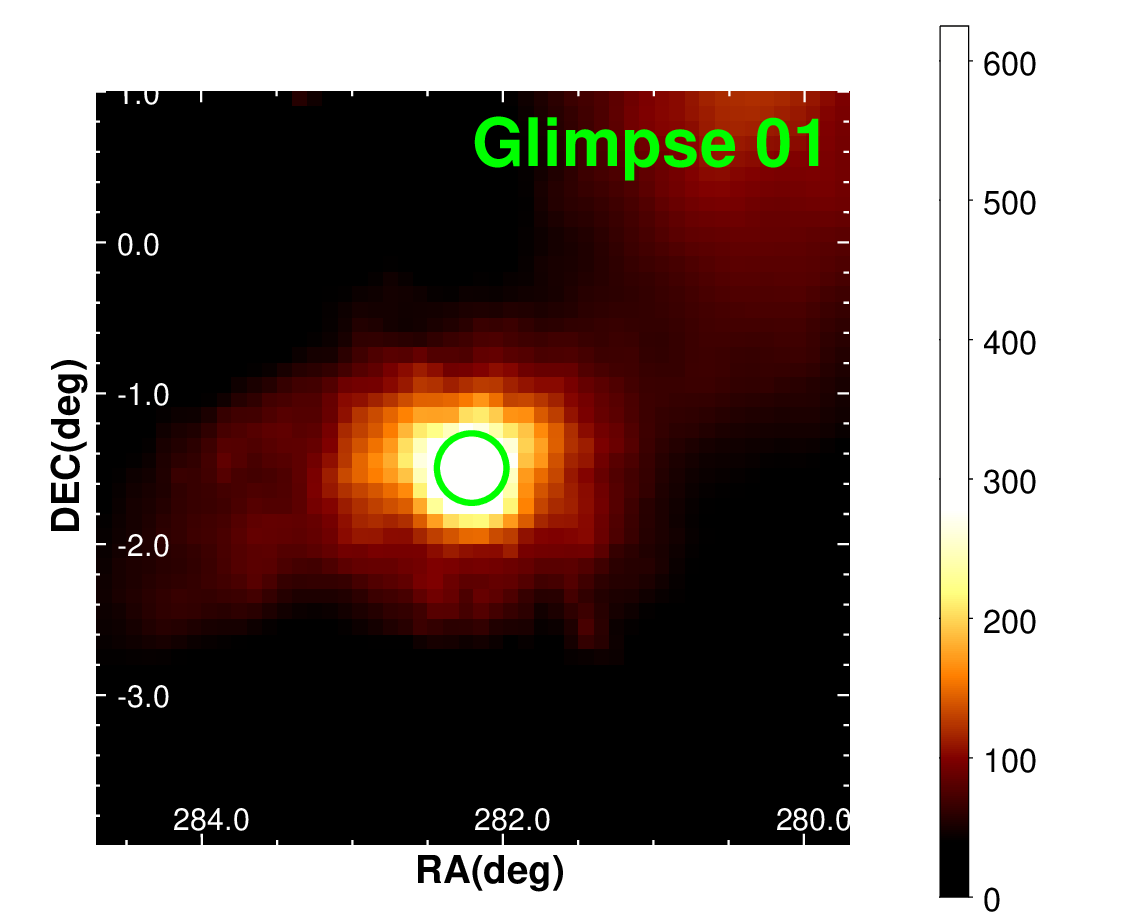}
\includegraphics[width=0.25\textwidth]{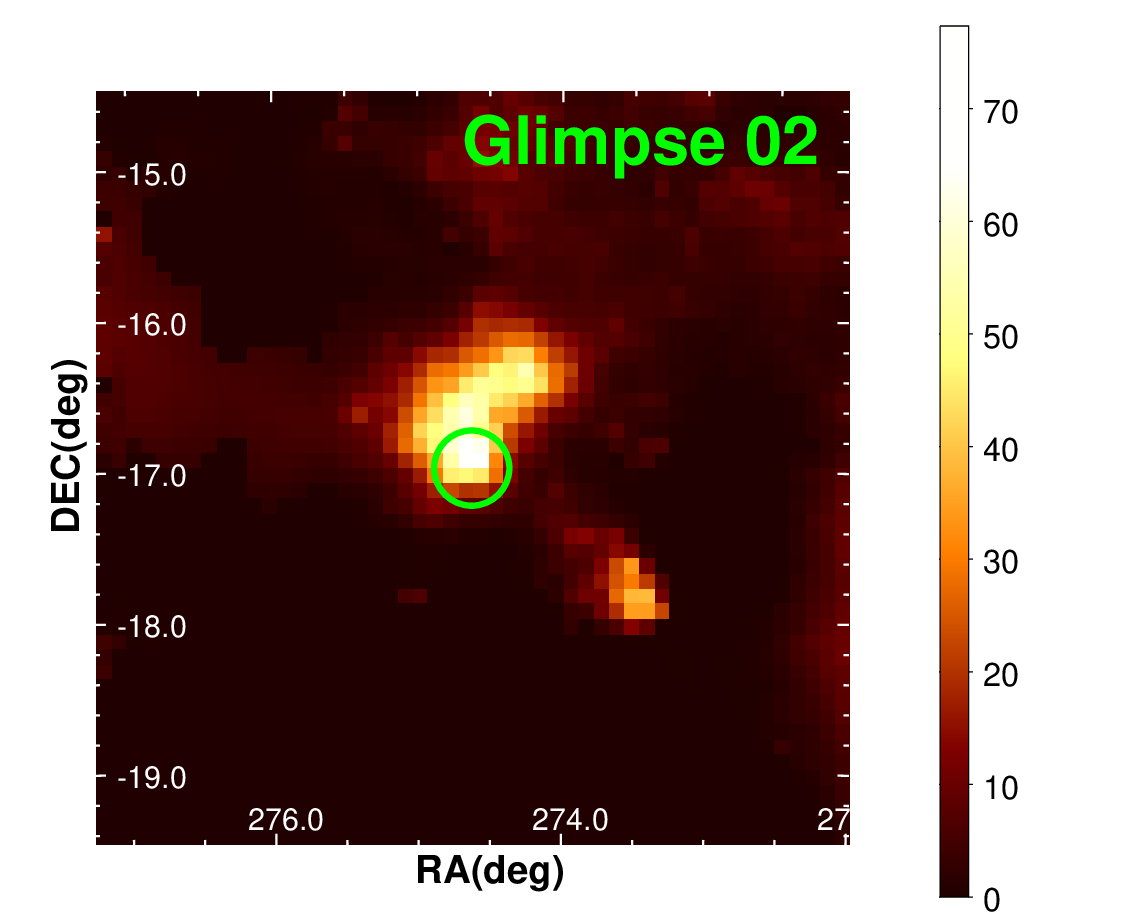}
\includegraphics[width=0.25\textwidth]{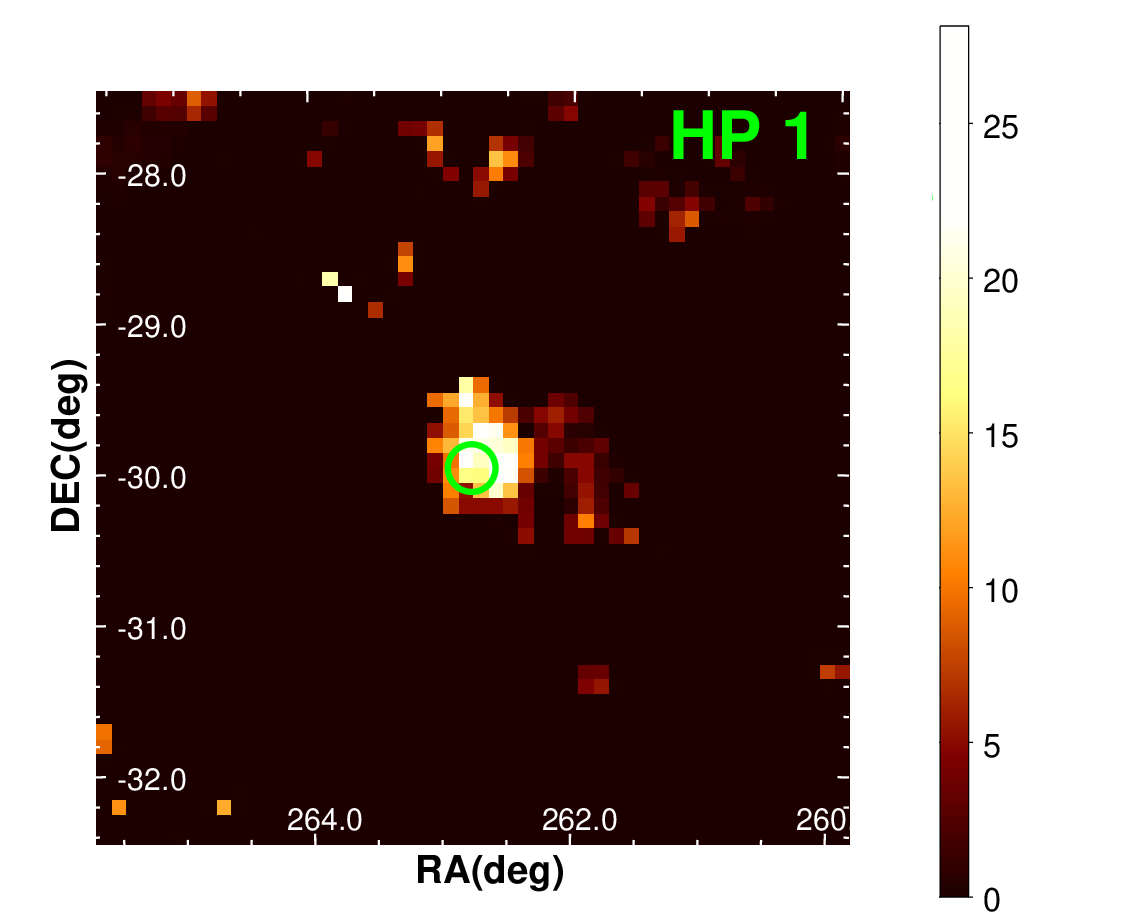}
\includegraphics[width=0.25\textwidth]{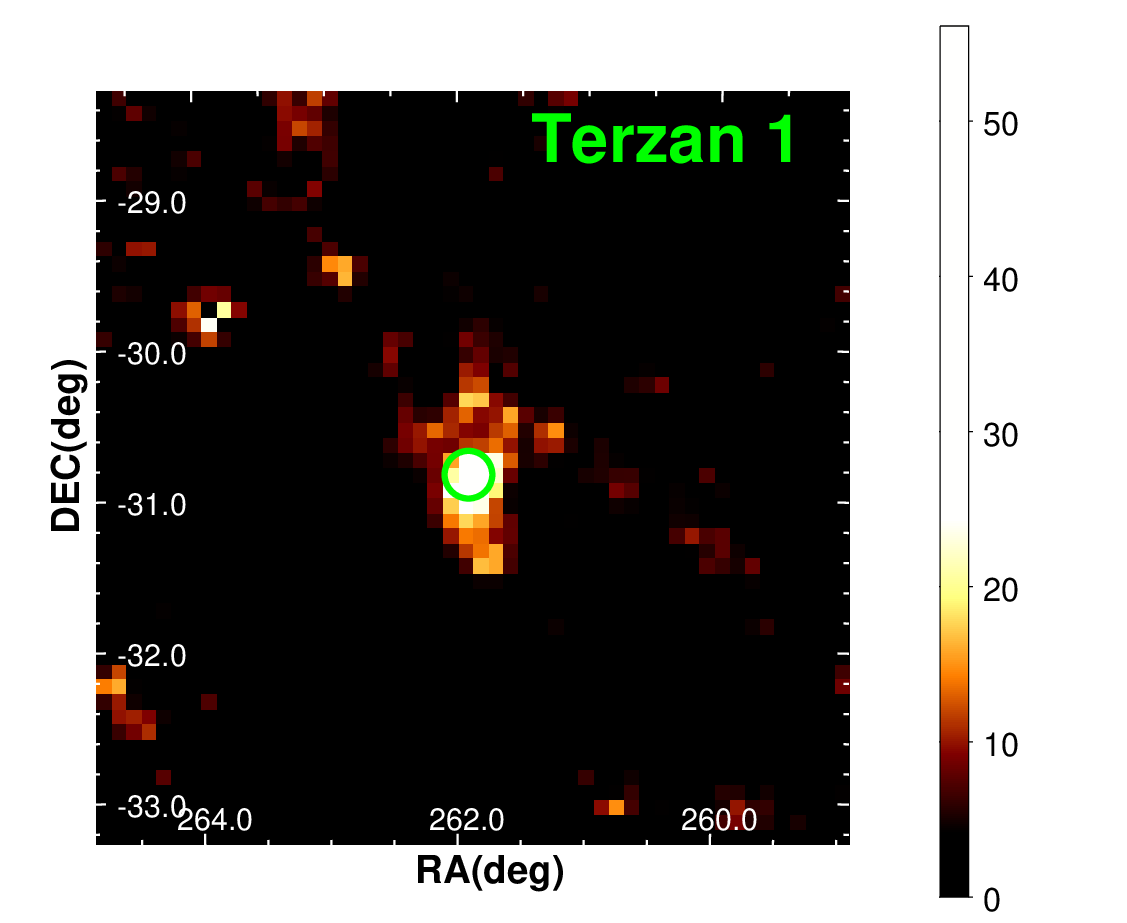}
}
\leftline{
\includegraphics[width=0.25\textwidth]{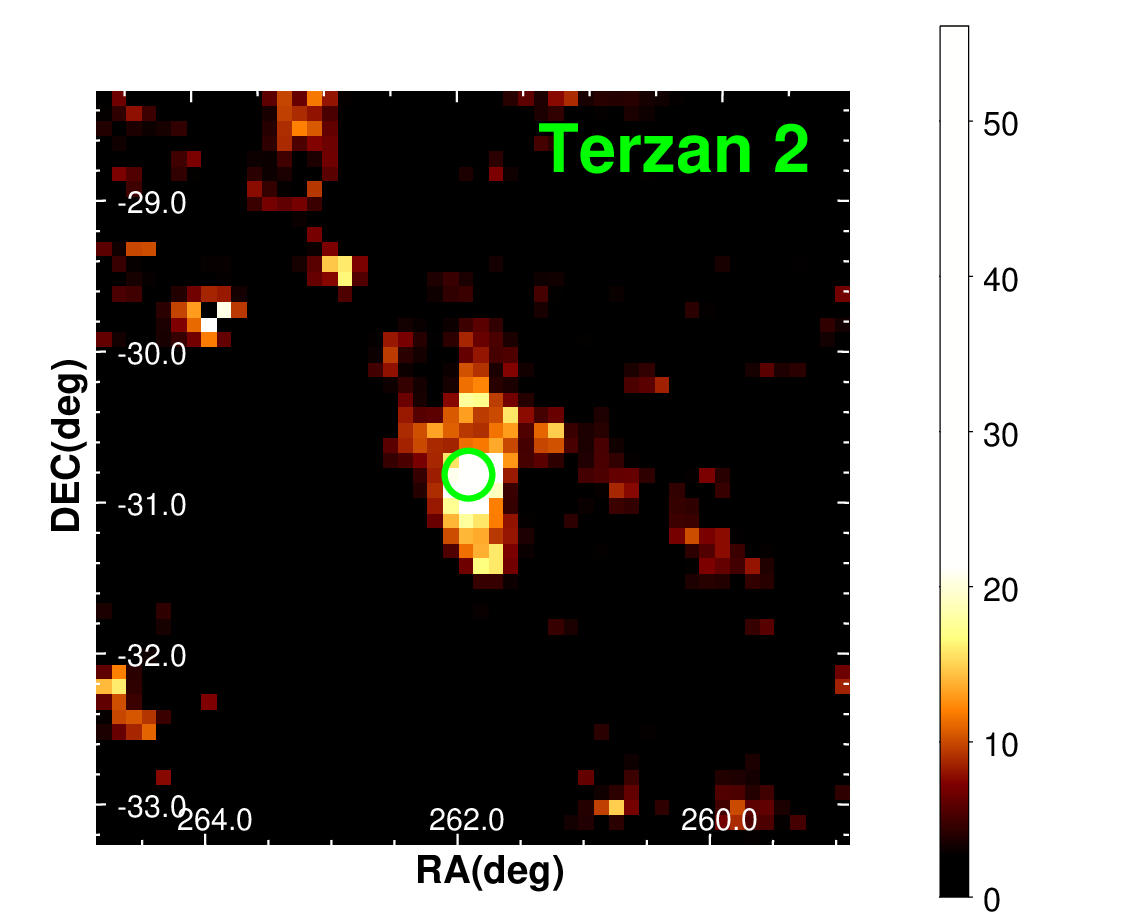}
\includegraphics[width=0.25\textwidth]{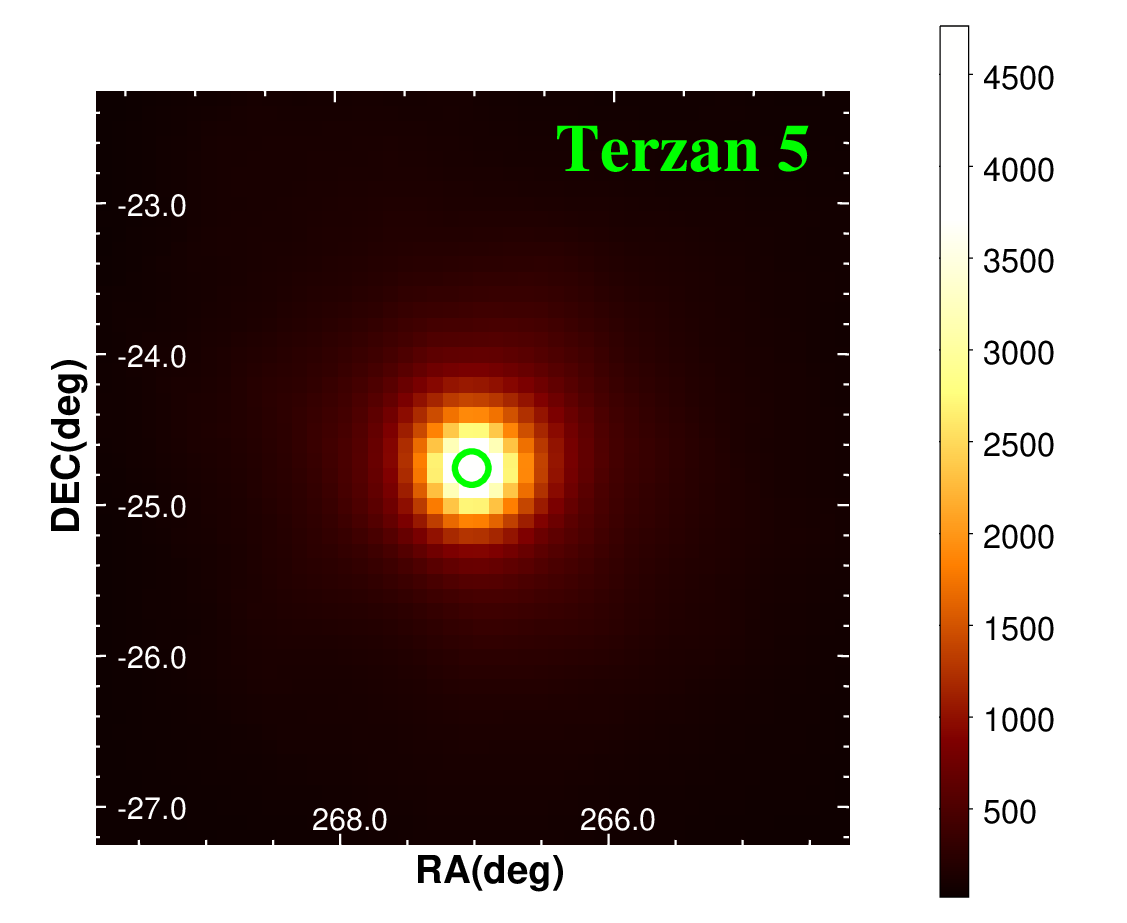}
\includegraphics[width=0.25\textwidth]{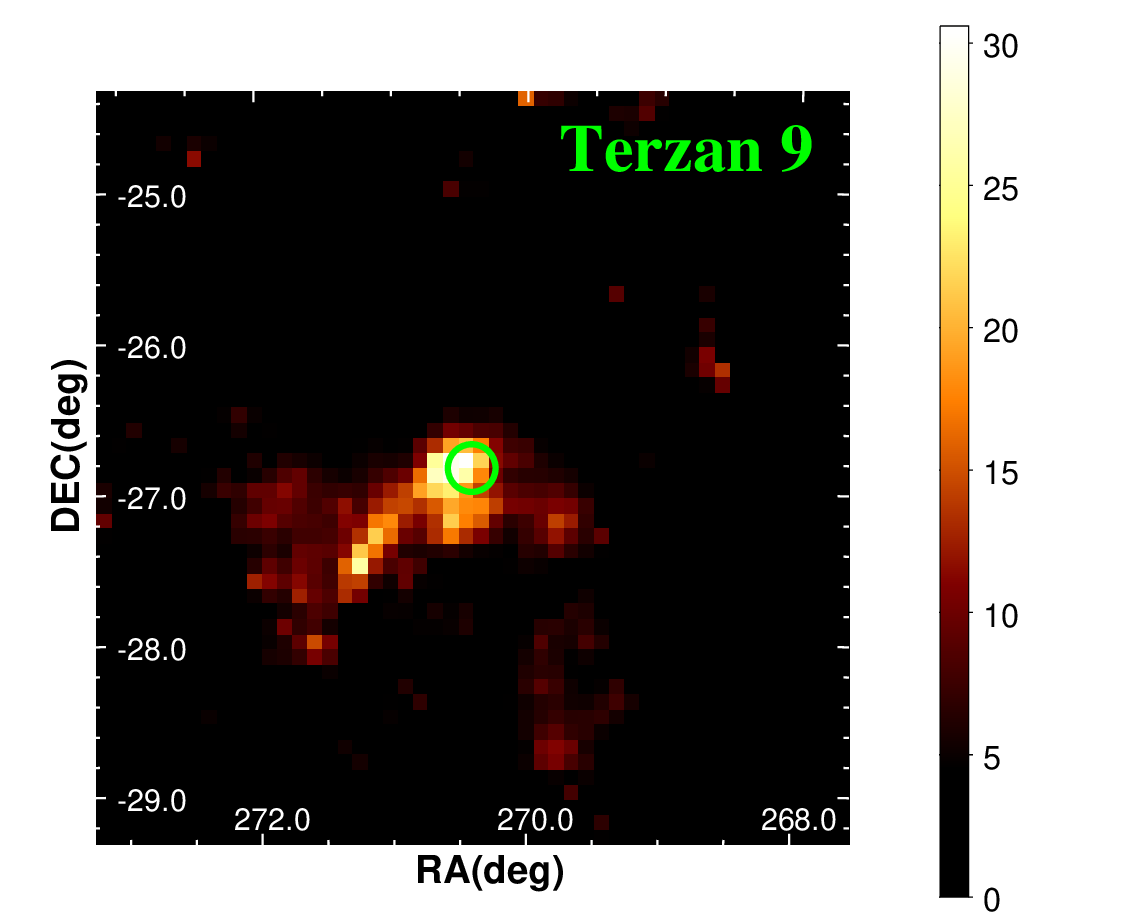}
}
\vspace*{1.5cm}
\caption{Continued test statistic maps of the 39 GlCs presented in Table-\ref{tab:results}.}
\label{fig:tsmaps}
\end{figure*}

\section{Statistic Relations}

As discussed in Section 1, the measured GlC $\gamma$-ray luminosity ($L_{\gamma}$) arises from a collection of MSPs reside in the cluster. Generally speaking, there are two dominant factors may influence the number of MSPs of a given GlC: (i) the population (thus the cluster mass $M$) of neutron stars (NS) hosted by the cluster, and (ii) the stellar dynamical interactions that transfrom NS into LMXBs and MSPs. The second factor also refer as the binary burning process, which is thought to be the essential internal energy source of cluster dynamical evolution. In this regard, the $\gamma$-ray emissivity ($\epsilon_{\gamma}=L_{\gamma}/M$) is better than $L_{\gamma}$ in tracing the formation efficiency of MSPs in GlCs, and both $L_{\gamma}$ and $\epsilon_{\gamma}$ may serve as sensitive probe to study the dynamical evolution of GlCs. With this in mind, we examine the dependence of $L_{\gamma}$ and $\epsilon_{\gamma}$ on various cluster physical parameters in this section. If tidal stripping have played an important effect in the dynamical evolution of GlCs, significant correlations between GlC $\gamma$-ray emission and the Galactic environment parameters are also expected. The GlC parameters are mainly taken from \citet{Baumgardt2018} or \citet{Harris1996}, unless the specific references are quoted. Throughout the paper, we also classified the GlCs into two subgroups (i.e., the dynamically normal and core-collapsed GlCs) according to their labels in \citet{Harris1996}. The core-collapsed GlCs are generally considered to be in the late phase of cluster dynamical evolution, and thus are dynamically more older than the normal GlCs. 

\subsection{Correlations with Cluster Mass}
\begin{figure*}
\centerline{
\includegraphics[width=1.0\textwidth]{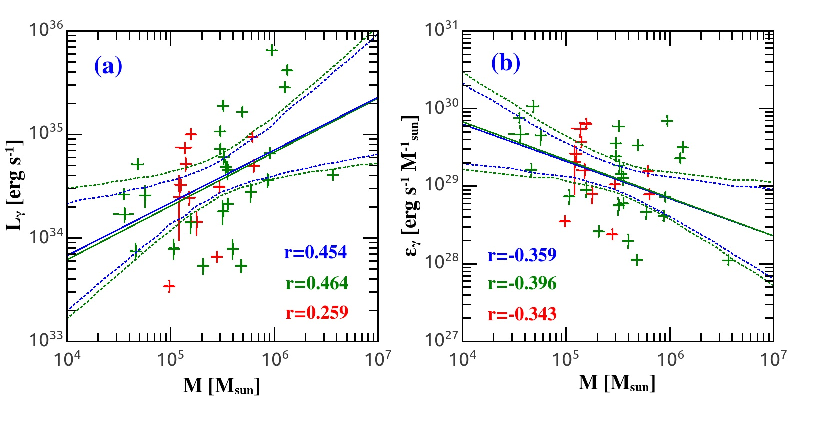}}
\caption{GlC $\gamma$-ray luminosity (a) and $\gamma$-ray emissivity (b) as a function of cluster mass. The olive and red pluses represent the dynamically normal and core-collapsed GlCs separately. The red, olive and blue text donates the Spearman's rank correlation coefficient of the core-collapsed, dynamically normal and the total GlC samples, respectively. The solid lines and associated dotted curves are the best fitting functions and the $95\%$ confidence range: red for core-collapsed GlCs, olive for dynamically normal GlCs, and blue for the total sample.  }
\label{fig:mass}
\end{figure*}

In Figure-\ref{fig:mass} (a), We first explore $L_{\gamma}$ versus $M$ for all GlCs. The GlC mass are adopted from the online catalog\footnote{https://people.smp.uq.edu.au/HolgerBaumgardt/globular/} of \citet{Baumgardt2018}, which are derived by comparing the observed cluster velocity dispersion, surface density, and stellar mass function profiles against N-body simulations. Although the scatter is substantial, a moderate positive correlation between $L_{\gamma}$ and $M$ is observable, from which we find the Spearman's rank coefficient $r=0.454$ and $r=0.464$ for the total and dynamically normal GlCs, with $p=0.004$ and $p=0.017$ for random correlation. We fit a power-law function to the correlations and obtained $L_{\gamma}\propto M^{0.53\pm0.17}$ and $L_{\gamma}\propto M^{0.51\pm0.20}$ for total and the dynamically normal GlCs, respectively. The best fitting functions are shown as the blue and the olive lines in Figure-\ref{fig:mass} (a), and the dotted curves mark the $95\%$ confidence range.   
Compared with the dynamically normal GlCs, the dependence of $L_{\gamma}$ on $M$ is marginal for core-collapsed GlCs, with Spearman's rank coefficient $r=0.259$ and random correlation $p$-value of $p=0.417$. 
The minimum and median $\gamma$-ray luminosity of dynamically normal GlCs is measured to be $L_{\gamma,min}=5.41\times 10^{33}\,{\rm erg\, s^{-1}}$ and $L_{\gamma,med}=4.1\times 10^{34}\,{\rm erg\, s^{-1}}$, which is comparable to that of the core-collapsed GlCs (i.e., $L_{\gamma,min}=3.45\times 10^{33}\,{\rm erg\, s^{-1}}$ and $L_{\gamma,med}=3.22\times 10^{34}\,{\rm erg\, s^{-1}}$). The exception is the most luminous GlCs (i.e., $L_{\gamma}\gtrsim 10^{35}\,{\rm erg\, s^{-1}}$), they are dominated by dynamically normal GlCs and thus are responsible for the much larger scatter of $L_{\gamma}$ in dynamically normal GlCs than the core-collapsed ones. 

In Figure-\ref{fig:mass} (b), we examine the dependence of $\epsilon_{\gamma}$ on $M$. It can be seen that $\epsilon_{\gamma}$ has a substantial scatter ranging from $\sim 10^{28} \, {\rm erg\, s^{-1}}\, M_{\odot}^{-1}$ to $\sim 10^{30} \, {\rm erg\, s^{-1}}\, M_{\odot}^{-1}$. Nevertheless, a marginally significant negative correlation between $\epsilon_{\gamma}$ and $M$ is suggested by the Spearman's rank coefficient, with $r=-0.396$, $r=-0.343$ and $r=-0.359$, and random correlation $p$-value of $p=0.045$, $p=0.276$ and $p=0.027$ for the dynamically normal, core-collapsed and total sample of GlCs, respectively. 
A power-law function fitting gives $\epsilon_{\gamma}\propto M^{-0.49\pm 0.22}$ (olive lines) and $\epsilon_{\gamma}\propto M^{-0.48\pm 0.29}$ (blue lines) for the dynamically normal and total GlCs. These negative correlations are consistent with the sub-linear correlations found in Figure-\ref{fig:mass} (a). 
The average $\gamma$-ray emissivity and standard deviation of all GlCs is measured to be $\langle\epsilon_{\gamma}\rangle=(2.48\pm0.41)\times 10^{29}\, {\rm erg\, s^{-1}}\, M_{\odot}^{-1}$ and $\sigma_{\epsilon_\gamma}=2.52\times 10^{29}\, {\rm erg\, s^{-1}}\, M_{\odot}^{-1}$, and no significant difference between the dynamically normal GlCs ($\langle\epsilon_{\gamma}\rangle=(2.58\pm0.54)\times 10^{29}\, {\rm erg\, s^{-1}}\, M_{\odot}^{-1}$, $\sigma_{\epsilon_\gamma}=2.75\times 10^{29}\, {\rm erg\, s^{-1}}\, M_{\odot}^{-1}$) and core-collapsed GlCs ($\langle\epsilon_{\gamma}\rangle=(2.25\pm0.58)\times 10^{29}\, {\rm erg\, s^{-1}}\, M_{\odot}^{-1}$, $\sigma_{\epsilon_\gamma}=2.02\times 10^{29}\, {\rm erg\, s^{-1}}\, M_{\odot}^{-1}$) are found.

The dependence of $L_{\gamma}$ on $M$ is consistent with the work of \citet{Hooper2016}, where massive GlCs are also found to have larger $\gamma$-ray luminosities. This trend is also in agreement with Figure-\ref{fig:detection} (e), supporting a $\gamma$-ray detection preference for massive GlCs. 
However, the negative $\epsilon_{\gamma}-M$ correlation suggests that the cluster mass is not the decisive factor influencing the abundance (or formation efficiency) of MSPs in GlCs. We note that a similar $\epsilon_{\gamma}-M$ relationship has also been reported by \citet{Fragione2018a} in their figure 3. 

\subsection{Correlations with Stellar Encounter Rates}

Traditionally, there are two parameters in literature to describe the stellar encounter rate of GlCs. The first is the total stellar encounter rate, $\Gamma \propto \int \rho^{2}/\sigma$, an integral of stellar encounters over the cluster volume, with $\rho$ the stellar density and $\sigma$ the cluster velocity dispersion \citep{Verbunt1987, Verbunt2003}. The second is the specific encounter rate, defined as $\Lambda \equiv \Gamma/M$, and measure the chance that a particular star (or binary) undergoes dynamical encounters in GlCs \citep{Pooley2006}. For this reason, $\Gamma$ is thought to be a resonable indicator of the total amount of exotic objects that can be dynamically produced in GlCs \citep{Pooley2003, Cheng2018a}. While $\Lambda$ is considered to be highly correlated with the abundance (or the formation efficiency) of exotic objects in GlCs \citep{Pooley2006, Cheng2018b}. 

\begin{table}
\centering
\caption{Integrated stellar ecounter rate ($\Gamma$) and specific encounter rate ($\Lambda$) of GlCs. The lower and upper limits are given at $1\sigma$ level.}
\label{tab:encounters}
\begin{tabular}{lrrrrrr}
\hline
\multicolumn{1}{c}{Name} & \multicolumn{1}{c}{$\Gamma$} & \multicolumn{1}{c}{$-\delta$} & \multicolumn{1}{c}{$+\delta$} & \multicolumn{1}{c}{$\Lambda$} & \multicolumn{1}{c}{$-\delta$} & \multicolumn{1}{c}{$+\delta$}\\
\hline
\hline
\multicolumn{7}{c}{Dynamically Normal GlCs:} \\
\hline
NGC 104    & 1000 &	134	& 154 &	844	& 113  & 130  \\
NGC 1851   & 1530 &	186	& 198 &	3524& 428  & 456  \\
NGC 2808   & 923  &	83	& 67  &	801	& 72   & 58   \\
NGC 5139   & 90.4 &	20.4& 26.3&	35.2& 7.9  & 10.2 \\
NGC 5286   & 458  &	61	& 58  &	723	& 96   & 92   \\
NGC 5904   & 164  &	30	& 39  &	243	& 45   & 57   \\
NGC 6093   & 532  &	69	& 59  &	1344& 174  & 149  \\
NGC 6139   & 307  &	82	& 95  &	688	& 184  & 214  \\
NGC 6218   & 13.0 &	4.0 & 5.4 &	76.6& 23.8 & 32.1 \\
NGC 6254   & 31.4 &	4.1	& 4.3 &	158	& 21   & 22   \\
NGC 6316   & 77.0 &	14.8& 25.4&	176	& 34   & 58   \\
NGC 6341   & 270  &	29	& 30  &	695	& 75   & 77   \\
NGC 6388   & 899  &	213	& 238 &	766	& 181  & 203  \\
NGC 6402   & 124  &	30	& 32  &	141	& 34   & 36   \\
NGC 6440   & 1400 &	477	& 628 &	2190& 746  & 983  \\
NGC 6441   & 2300 &	635	& 974 &	1600& 442  & 678  \\
NGC 6528   & 278  &	50	& 114 &	3239& 577  & 1328 \\
NGC 6626   & 648  &	91	& 84  &	1746& 245  & 226  \\
NGC 6652   & 700  &	189	& 292 &	7508& 2027 & 3132 \\
NGC 6656   & 77.5 &	25.9& 31.9&	153	& 51   & 63	  \\
NGC 6717   & 39.8 &	13.7& 21.8&	1072& 369  & 587  \\
NGC 6838   & 1.47 &	0.14& 0.15&	41.5& 4.0  & 4.2  \\
2MASS-GC01 & --	  &	--	& --  &	--	& --   & --   \\
Glimpse 01 & 400  &	200	& 200 &	8560& 4280 & 4280 \\
Glimpse 02 & --   &	--  & --  &	--	& --   & --   \\
Terzan 5   & 6800 &	3020& 1040&	3400& 1510 & 520  \\
\hline
\multicolumn{7}{c}{Core-Collapsed GlCs:}\\
\hline
NGC 1904   & 116  &	45  & 68  &	412	& 159  & 240  \\
NGC 6266   & 1670 &	569	& 709 &	1758& 599  & 747  \\
NGC 6397   & 84.1 & 18.3& 18.3&	919	& 200  & 200  \\
NGC 6541   & 386  & 63	& 95  &	746	& 122  & 184  \\
NGC 6624   & 1150 &	178	& 113 &	5742& 889  & 564  \\
NGC 6723   & 11.4 & 4.3 & 8.0 &	41.6& 16.0 & 29.2 \\
NGC 6752   & 401  & 126	& 182 &	1605& 504  & 729  \\
NGC 7078   & 4510 & 986	& 1360&	4705& 1029 & 1419 \\
HP1	       & 0.66 &	0.30& 0.41&	8.51& 3.87 & 5.29 \\
Terzan 1   & 0.29 &	0.17& 0.27&	24.7& 14.5 & 23.0 \\
Terzan 2   & 22.1 &	14.4& 28.6&	486	& 317  & 629  \\
Terzan 9   & 1.71 &	0.96& 1.67&	278	& 156  & 271  \\
\hline 
\end{tabular}
\end{table}

\begin{figure*}
\centerline{
\includegraphics[width=1.0\textwidth]{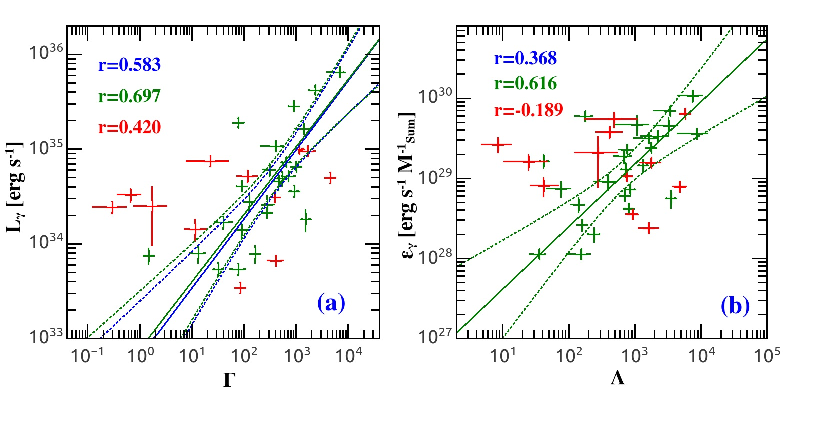}}
\caption{GlC $\gamma$-ray luminosity as a function of the stellar encounter rate $\Gamma$ (a) and emissivity as a function of the specific encounter rate $\Lambda$ (b). The color-coded symbols denate different types of GlCs, as in Figure-\ref{fig:mass}, and the same color coded text gives the Spearman's rank correlation coefficients: red for core-collapsed, olive for dynamically normal, and blue for the total. The olive and blue solid lines mark the best-fitting funtion of the dynamically normal and the total GlCs, with dotted curves represent the $95\%$ fitting confidence range. }
\label{fig:encounterrate}
\end{figure*}

In Figure-\ref{fig:encounterrate} (a), we plot $L_{\gamma}$ versus $\Gamma$ for each GlC. The value of $\Gamma$ are adopted from \citet{Bahramian2013}, which is integrated over the cluster and normalized to a value of 1000 for NGC 104 (Table-\ref{tab:encounters}). 
It is clear that $L_{\gamma}$ is highly correlated with $\Gamma$ for dynamically normal GlCs. The Spearman's correlation coefficient is $r=0.697$, with a random correlation $p$-value of $p=1.1\times10^{-4}$. If core-collapsed GlCs are taken into account, this correlation is still significant, with $r=0.583$ and $p=1.5\times10^{-4}$ for total GlCs. The best-fitting functions can be written as $L_{\gamma}\propto \Gamma^{0.71\pm 0.11}$ (olive curves) and $L_{\gamma}\propto \Gamma^{0.73\pm 0.11}$ (blue curves) for the dynamically normal and total GlCs, respectively. 
These correlations are consistent with the finding of \citet{deMenezes2019}, where the possible number of MSPs ($N$) of GlCs is found to be proportional to $\Gamma$, with $N\propto \Gamma^{0.64\pm 0.15}$ in a sample of 23 GlCs. 

According to the Spearman's rank correlation coefficients, it is clear that $L_{\gamma}$ depends on $\Gamma$ better than $M$, which suggest that stellar dynamical interaction is the fundamental factor that influence the population of MSPs in GlCs. To minimize the dependence on cluster mass, we also examine $\epsilon_{\gamma}$ versus $\Lambda$ in Figure-\ref{fig:encounterrate} (b). The value of $\Lambda$ was calculated with equation $\Lambda=\Gamma/M_{6}$, with $M_{6}$ be the cluster mass in units of $10^{6}M_{\odot}$. Since $\Gamma$ was estimated based on the V-band luminosity \citep{Bahramian2013}, here we calculated $M_{6}$ using the V-band-based magnitude for consistency, following an empirical mass-to-magnitude relation described in \citet{Cheng2018a}.
From Figure-\ref{fig:encounterrate} (b), it is obvious that the dynamically normal GlCs have a larger $\epsilon_{\gamma}$ with increasing $\Lambda$. The Spearman's rank correlation coefficient is $r=0.616$, with $p=1.0\times 10^{-3}$ for random correlation. This correlation becomes less significant when taking the core-collapsed GlCs into acount, with $r=0.368$ and $p=0.025$ for the total GlCs. A power-law fitting yields a function of $\epsilon_{\gamma}\propto \Lambda^{0.78\pm0.15}$ (olive curves) for the dynamically normal GlCs.

Compared with dynamically normal GlCs, the core-collapsed GlCs are found to have much large scatter in Figure-\ref{fig:encounterrate}. 
The difference may arises from the rapid change of cluster structure in these systems. Once GlCs are running out of binary systems and undergo deep core collapse, their structure are not stable any more and the cluster core may suffer from gravothermal oscillations \citep{Fregeau2003}, core-collapsed GlCs therefore may create extremely frequent stellar encounter environments. Indeed, some core-collapsed GlCs (e.g., NGC 6624 and NGC 7078) are proved to be clusters with the highest stellar encounter rates in Table-\ref{tab:encounters}, supporting a currently deep core collapse in these systems. While some clusters (e.g., HP 1 and Terzan 1) are found to have the smallest values of $\Gamma$ and $\Lambda$, which may indicates a core bounce phase after the deep core collapse.  
On the other hand, with extremely large $\Lambda$, it is possible for LMXBs to be dynamically disrupted during deep core collapse \citep{Verbunt2014}, and strong encounters may lead to the ejection of MSPs from the Clusters (Section 4.2). These processes may introduce uncertainties to the $L_{\gamma}-\Gamma$ and $\epsilon_{\gamma}-\Lambda$ correlations for core-collapsed GlCs. %The implication of these effects will be addressed in Section 4.2. %here we continue to investigate the dependence of GlC $\gamma$-ray emission on various cluster parameters as follows.

\subsection{Correlations with Cluster Structure Parameters}

\begin{figure*}%[tbh!]
\centerline{
\includegraphics[width=1.0\textwidth]{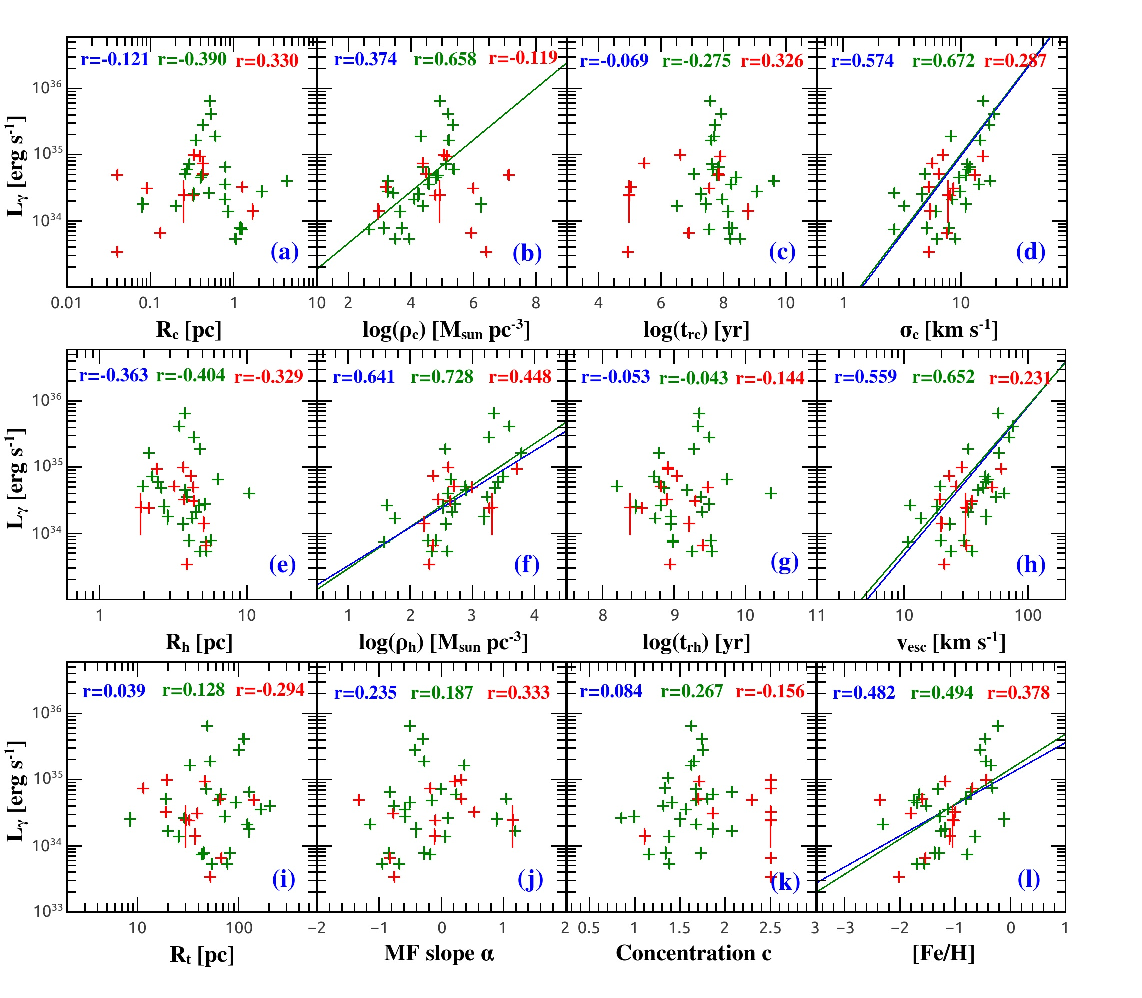}}
\caption{Dependence of the $\gamma$-ray luminosity on various physical properties of GlCs. Top panels, from left to right: cluster core radius $R_{c}$, core density ${\rm log}(\rho_{c})$, core relaxation time ${\rm log}(t_{rc})$, and central velocity dispersion $\sigma_{c}$. Middle panels, from left to right: cluster half-mass radius $R_{h}$, density inside half-mass radius ${\rm log}(\rho_{h})$, half-mass relaxation time ${\rm log}(t_{rh})$, and central escape velocity $v_{esc}$. Bottom panels, from left to right: cluster tidal radius $R_{t}$, global mass function slope $\alpha$, central concentration $c$, and metallicity ${\rm [Fe/H]}$. All these parameters are adopted from the on line catalog of \citet{Baumgardt2018}, except ${\rm log}(t_{rc})$, $c$ and ${\rm [Fe/H]}$ are adopted from \citet{Harris1996}. The color-coded symbols, solid lines and texts have the same meanings as in Figure-\ref{fig:mass}. For simplicity, we have omitted the confidence curves of the best-fitting functions. }
\label{fig:luminosity}
\end{figure*}

\begin{figure*}%[tbh!]
\centerline{
\includegraphics[width=1.0\textwidth]{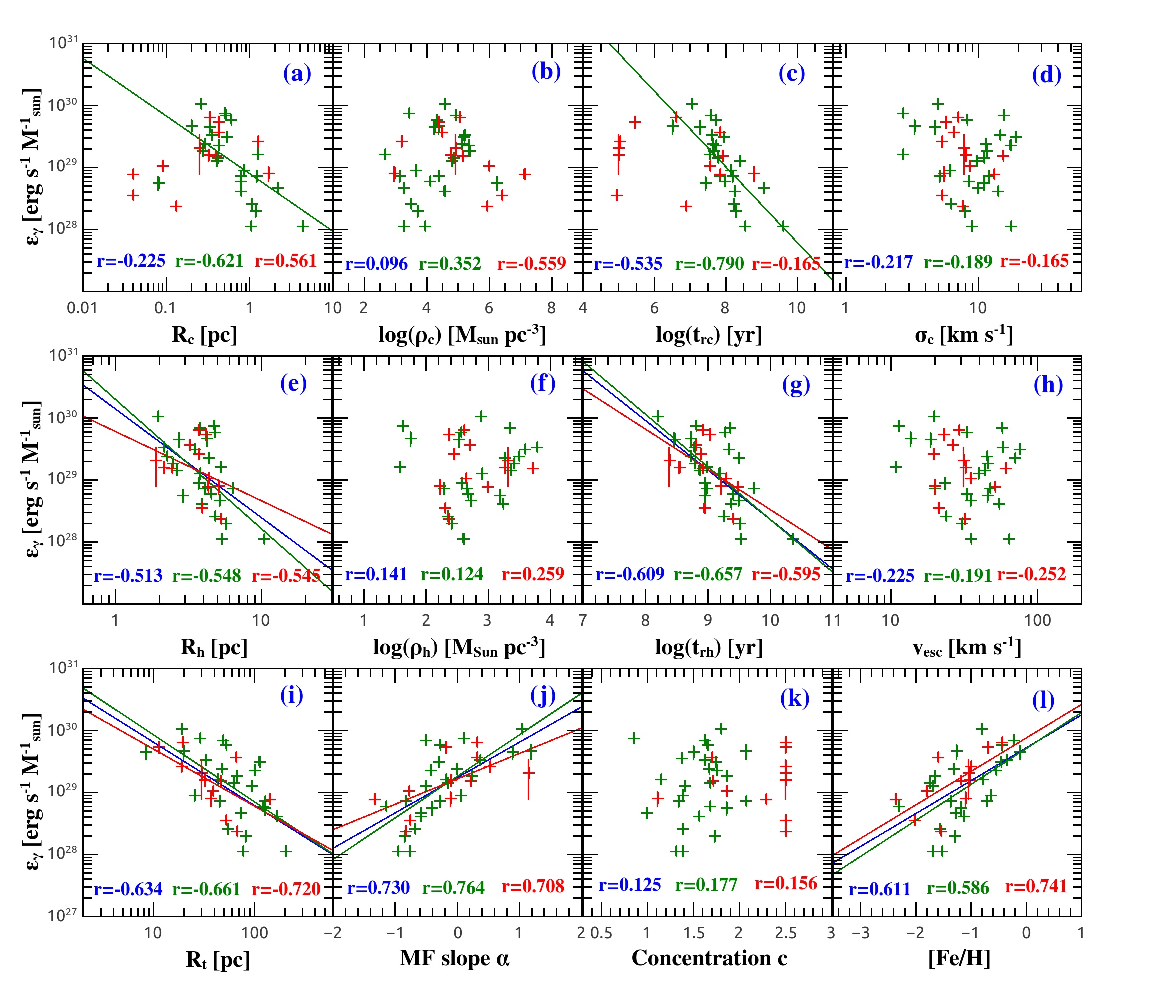}}
\caption{Dependence of the $\gamma$-ray emissivity on various physical properties of GlCs. The layout of the panels, color-coded symbols, lines and comment texts in each panel are same as in Figure-\ref{fig:luminosity}. }
\label{fig:emissivity}
\end{figure*}

To find out the dependence of GlC $\gamma$-ray esmission on GlC dynamical evolution history, we sort the cluster structure parameters and investigate their correlations to $L_{\gamma}$ and $\epsilon_{\gamma}$ in Figure \ref{fig:luminosity} and Figure \ref{fig:emissivity}. 
Generally speaking, MSPs are prone to be dynamically formed in the dense core of GlCs. Therefore, we first plot $L_{\gamma}$ versus GlC core radius $R_{c}$, core density $\rho_{c}$, core relaxation time $t_{rc}$ and central velocity dispersion $\sigma_{c}$ in the upper panels of Figure \ref{fig:luminosity}. No clear correlation exists for the total GlC sample, except that $L_{\gamma}$ is highly correlated with $\sigma_{c}$ ($r=0.574$, $p=2.0\times 10^{-4}$) in Figure \ref{fig:luminosity}(d). %The Spearman's rank correlation coefficients is $r=0.574$, with $p=2.0\times 10^{-4}$ for random correlation. 
When only the dynamically normal GlCs are considered, this correlation is still evident, with Spearman's $r=0.672$ and random correlation $p$-value of $p=2.4\times 10^{-4}$. We fit this correlation with power-law function, which gives $L_{\gamma}\propto \sigma_{c}^{2.36\pm0.44}$ (blue lines) and $L_{\gamma}\propto \sigma_{c}^{2.35\pm0.56}$ (olive lines) for the total and dynamically normal GlCs, respectively. However, it seems that the cluster mass is the more fundamental parameter underlying the $L_{\gamma}- \sigma_{c}$ relation, since the dependence of $\epsilon_{\gamma}$ on $\sigma_{c}$ is less significant in Figure \ref{fig:emissivity}(d), and $\sigma_{c}$ is highly correlated with $M$ in GlCs (Table \ref{tab:spearmans_r}).   

Unlike the linear relations reported Section 3.1 and 3.2, the Spearman's rank correlation coefficients suggest that the dynamically normal GlCs and core-collapsed GlCs are statistically different in Figure \ref{fig:luminosity}(a)-\ref{fig:luminosity}(c). As GlCs evolve to older dynamical age (i.e., with decreasing $t_{rc}$) and become more compact (i.e., with smaller $R_{c}$ and larger $\rho_{c}$), the dynamically normal GlCs are more likely to exhibit larger $L_{\gamma}$. Whereas for core-collapsed GlCs, this trend is reversed. When the influence of cluster mass is eliminated, this tendency still holds in Figure \ref{fig:emissivity}(a)-\ref{fig:emissivity}(c), where the dynamically normal GlCs are measured to have larger $\epsilon_{\gamma}$ with increasing dynamical age, which then reverses to smaller values as clusters evolve into core-collapsed GlCs. We suggest that these features may indicate an episodic ejection of MSPs from GlCs. Namely, during the deep core collapse and subsequent core bounce, significant number of MSPs may have been dynamically ejected from the host clusters. The implication of this effect will be addressed in Section 4.2. Here, we find the significant correlations of dynamically normal GlCs can be expressed as $L_{\gamma}\propto \rho_{c}^{0.39\pm0.14}$, $\epsilon_{\gamma}\propto R_{c}^{-0.93\pm0.27}$ and $\epsilon_{\gamma}\propto t_{rc}^{-0.61\pm0.11}$. They are plot as olive lines in the figures. 

In the middle panels of Figure \ref{fig:luminosity} and Figure \ref{fig:emissivity}, we test the dependence of $L_{\gamma}$ and $\epsilon_{\gamma}$ on the GlC half-mass radius $R_{h}$, stellar density ($\rho_{h}$) inside $R_{h}$, half-mass relaxation time $t_{rh}$ and central escape velocity $v_{esc}$. It is obvious that the dynamically normal GlCs have a larger $L_{\gamma}$ with increasing $\rho_{h}$ and $v_{esc}$, and the best-fitting functions (olive lines) can be written as $L_{\gamma}\propto \rho_{h}^{0.63\pm0.15}$ and $L_{\gamma}\propto v_{esc}^{2.18\pm0.61}$. If core-collapsed GlCs are included, these correlations are still evident, with $L_{\gamma}\propto \rho_{h}^{0.58\pm0.13}$ and $L_{\gamma}\propto v_{esc}^{2.25\pm0.48}$ for the total GlCs (blue lines). However, the dependence of $\epsilon_{\gamma}$ on $R_{h}$ and $v_{esc}$ are not significant in Figure \ref{fig:emissivity}(f) and \ref{fig:emissivity}(h), reflecting again that cluster mass is the more fundamental parameter underlying these relations. On the other hand, GlCs with smaller $R_{h}$ appears to have larger $L_{\gamma}$, as indicated by the mild anti-correlation coefficients in the figures. This trend is still evident in Figure \ref{fig:emissivity}(e), where the best fitting functions between $\epsilon_{\gamma}$ and $R_{h}$ can be expressed as $\epsilon_{\gamma}\propto R_{h}^{-2.09\pm0.61}$, $\epsilon_{\gamma}\propto R_{h}^{-1.12\pm0.91}$ and $\epsilon_{\gamma}\propto R_{h}^{-1.76\pm0.48}$ for the dynamically normal, core-collapsed and total samples of GlCs. Besides, although the dependence of $L_{\gamma}$ on $t_{rh}$ is not apparent in Figure \ref{fig:luminosity}(g), $\epsilon_{\gamma}$ is found to decrease with inceasing $t_{rh}$ in Figure \ref{fig:emissivity}(g), with the best fitting functions can be written as $\epsilon_{\gamma}\propto t_{rh}^{-0.85\pm0.20}$, $\epsilon_{\gamma}\propto t_{rh}^{-0.65\pm0.38}$ and $\epsilon_{\gamma}\propto t_{rh}^{-0.80\pm0.17}$ for dynamically normal, core-collasped and total GlCs, respectively. 

By and large, the dependence of $\gamma$-ray esmission on GlC half-mass paramters is in agreement with that on the core paramters. That is, as GlCs evolve to older dynamical age ($t_{d} \propto t_{rc}^{-1}$ or $t_{d} \propto t_{rh}^{-1}$) and become more compact, GlCs are more likely to exhibit larger $L_{\gamma}$ and $\epsilon_{\gamma}$. The discrepancy may arise from the core collapse, in which the cluster core structure may change dramatically during the gravothermal oscillations. Whereas for the half-mass parameters, their response to core collapse will be more moderate \citep{Fregeau2003}, thus both dynamically normal and core-collapsed GlCs are found to exhibit similar correlations in Figure \ref{fig:luminosity}(e)-\ref{fig:luminosity}(h) and Figure \ref{fig:emissivity}(e)-\ref{fig:emissivity}(h). 

In the bottom panels of Figure \ref{fig:luminosity} and Figure \ref{fig:emissivity}, we examine the dependence of $L_{\gamma}$ and $\epsilon_{\gamma}$ on the cluster tidal radius $R_{t}$, stellar mass function (MF) slope $\alpha$, concentration parameter $c$ and metallicity ${\rm [Fe/H]}$. $L_{\gamma}$ is found to be independent of $R_{t}$, $\alpha$ and $c$ in Figure \ref{fig:luminosity}(i), \ref{fig:luminosity}(j) and \ref{fig:luminosity}(k). However, It is clear that GlCs with smaller $R_{t}$ and larger $\alpha$ are more likely to have larger $\epsilon_{\gamma}$ (Figure \ref{fig:emissivity}(i) and \ref{fig:emissivity}(j)). The power-law fitting yields $\epsilon_{\gamma}\propto R_{t}^{-1.08\pm0.28}$, $\epsilon_{\gamma}\propto R_{t}^{-0.98\pm0.37}$ and $\epsilon_{\gamma}\propto R_{t}^{-1.00\pm0.22}$ for the dynamically normal, core-collapsed and total GlCs, respectively. While for the $\epsilon_{\gamma}$--$\alpha$ relation, which can be expressed as ${\rm log} \epsilon_{\gamma}=a \alpha + b$ in Figure \ref{fig:emissivity}(j), with the best-fitting parameters $a=(0.57\pm0.10)$ and $b=(29.24\pm 0.07)$ for total GlCs, $a=(0.67\pm0.14)$ and $b=(29.26\pm 0.09)$ for dynamically normal GlCs, and $a=(0.41\pm0.15)$ and $b=(29.22\pm 0.11)$ for core-collapsed GlCs, respectively.
%On the other hand, no statistically significant dependence of $\epsilon_{\gamma}$ on $c$ can be found from Figure \ref{fig:emissivity}(k).

\begin{figure*}%[tbh!]
\centerline{
\includegraphics[width=1.0\textwidth]{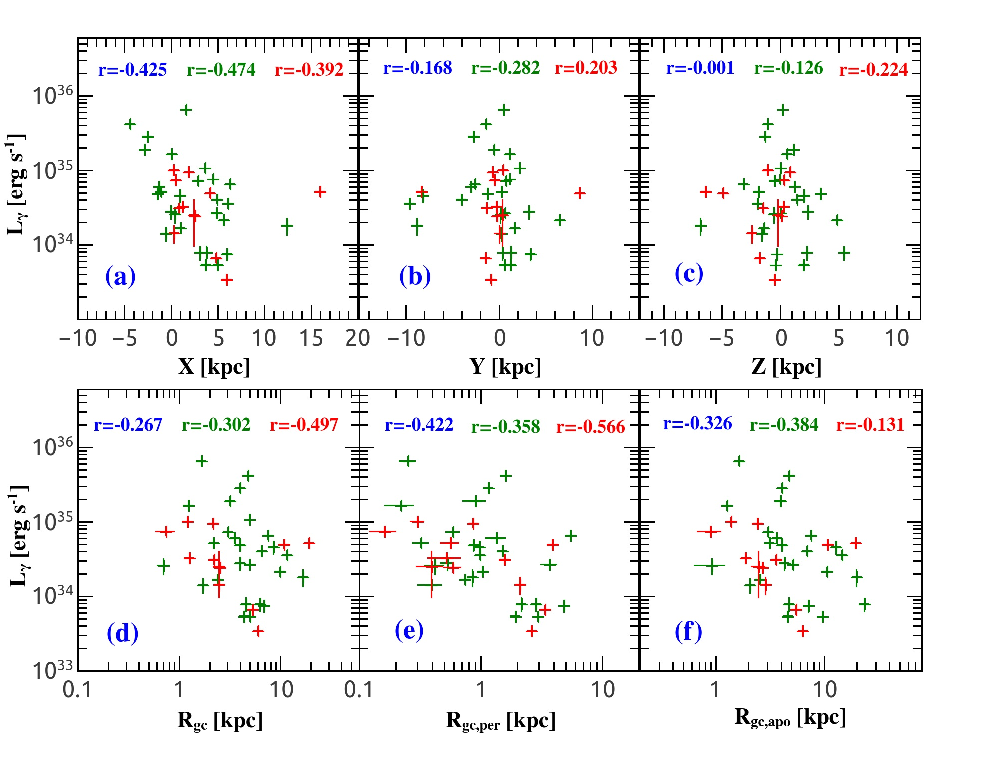}}
\caption{Dependence of GlC $\gamma$-ray luminosity on Galacitic environment parameters. Top panels for cluster coordonates: with $X$ point from GC towards the Sun (a), $Y$ in direction of the Solar motion (b) and $Z$ point towards the North Galactic pole (c). Bottom panels for the cluster orbital parameters in the Milky Way: with $R_{\rm gc}=\sqrt{X^2+Y^2+Z^2}$ represents the current GlC distance to the GC (d), while $R_{\rm gc,per}$ and $R_{\rm gc,apo}$ marks the perigalactic (e) and apogalactic (f) distance of GlC orbit to the GC. The color-coded symbols and texts have the same meanings as in Figure-\ref{fig:mass}. }
\label{fig:lumin_gc}
\end{figure*}

Observationally, many authors argued that LMXBs are more likely to be detected in old, metal-rich GlCs, rather than in young, metal-poor ones \citep{Sivakoff2007, Li2010, Kim2013}. As the offspring of LMXB, Figure \ref{fig:luminosity}(l) also suggests that MSPs are more likely to be found in metal rich GlCs. We fit the $L_{\gamma}$--${\rm [Fe/H]}$ relation with the function ${\rm log}\, L_{\gamma}=a{\rm [Fe/H]}+b$, the best-fitting parameters are $a=(0.53\pm0.18)$ and $b=(35.16\pm 0.21)$ for dynamically normal GlCs, $a=(0.47\pm0.14)$ and $b=(35.09\pm 0.17)$ for total GlCs, respectively. When the influence of cluster mass was eliminated, we find this tendency are still evident in Figure \ref{fig:emissivity}(l). The best-fitting function ${\rm log}\, \epsilon_{\gamma}=a{\rm [Fe/H]}+b$ gives parameters of $a=(0.58\pm0.17)$ and $b=(29.71\pm 0.20)$ for dynamically normal GlCs, $a=(0.54\pm0.18)$ and $b=(29.88\pm 0.26)$ for core-collapsed GlCs, and $a=(0.53\pm0.13)$ and $b=(29.73\pm 0.13)$ for total GlCs, respectively. We note that a similar $L_{\gamma}$--${\rm [Fe/H]}$ relation was also obtained by \citet{deMenezes2019} with 23 GlCs. 

\subsection{Correlations with the Galactic Environment}

\begin{figure*}
\centerline{
\includegraphics[width=1.0\textwidth]{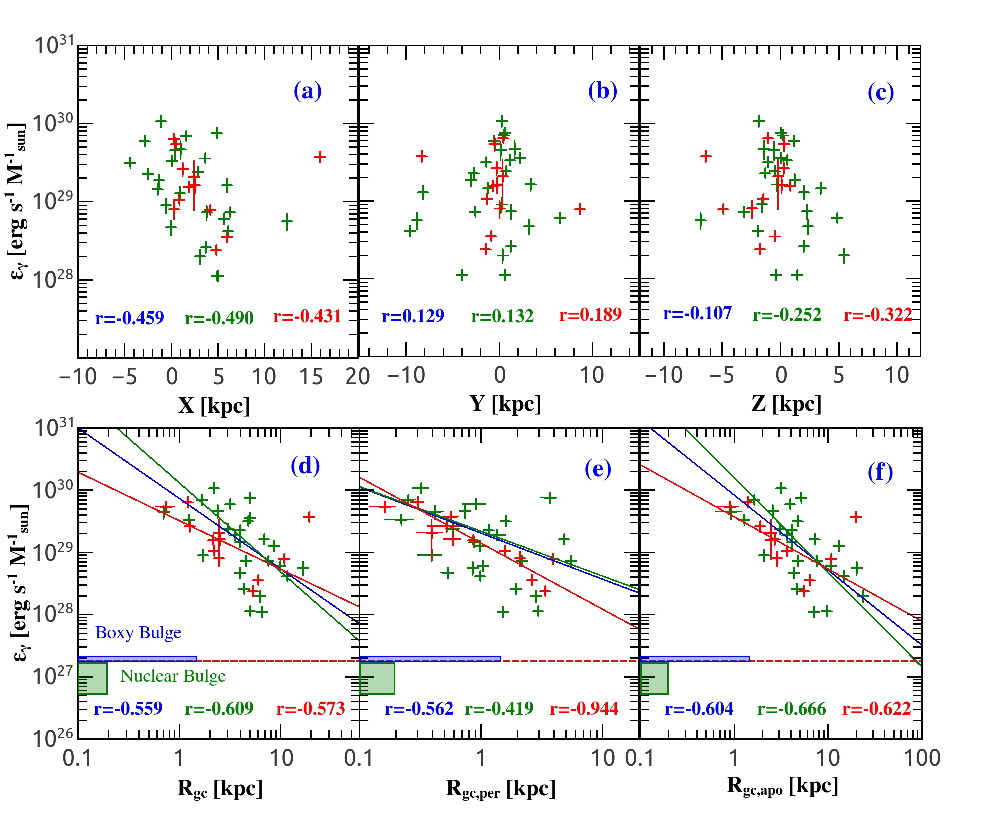}}
\caption{Dependence of GlC $\gamma$-ray esmissivity on Galacitic environment parameters. The layout of the panels are same as in Figure \ref{fig:lumin_gc}. The color-coded symbols and texts have the same meanings as in Figure-\ref{fig:mass}. The measured $\gamma$-ray emissivity of the Galactic boxy bulge and nuclear bulge \citep{Bartels2018} are shown as blue and olive rectangles in the botom panels, while the red dashed horizontal line represents the best fitting $\gamma$-ray esmissivity of the Galactic boxy bulge and nuclear bulge. }
\label{fig:emissi_gc}
\end{figure*}

Contrary to the contraction of the central regions, the outskirts of GlCs are driven to expand to be truncated by the tidal force of the host galaxy. The tidal radius, defined as the boundary where a GlC may lose its stars to the Galaxy, is subject to its coordinates in the Milky Way \citep{Baumgardt2003}: $R_{t}=(GM/2V_{\rm G})^{1/3}R_{gc}^{2/3}$. Here $G$ is the gravitation constant, $V_{\rm G}$ the circular velocity of the Galaxy and $R_{gc}$ the distance of the cluster from the GC. The independence of $L_{\gamma}$ on $R_{t}$ in Figure \ref{fig:luminosity}(i), and the negative $\epsilon_{\gamma}$-$R_{t}$ relation in Figure \ref{fig:emissivity}(j), suggests that tidal stripping effect is another important factor influence the abundance of MSPs in GlCs. Indeed, this tendency also can be confirmed by the positive $\epsilon_{\gamma}$-$\alpha$ correlation, since mass segregation and tidal stripping effects have a preference for the depletion of low-mass stars from GlCs, thus tidally stripped clusters are more likely to exhibit larger stellar MF slope $\alpha$ \citep{Vesperini1997, Baumgardt2003}.  

Therefore, we also investigate the dependence of GlC $\gamma$-ray emission on the Galactic environment parameters in Figure \ref{fig:lumin_gc} and Figure \ref{fig:emissi_gc}. The coordinates of GlCs are adopted from the online catalog of \citet{Baumgardt2019a}, with $X$ point from the Galactic center towards the Sun, $Y$ point in the direction of Galactic rotation at the Solar position, and $Z$ point towards the North Galactic pole, respectively. 
In Figure \ref{fig:lumin_gc}(a) and \ref{fig:emissi_gc}(a), both $L_{\gamma}$ and $\epsilon_{\gamma}$ are found to decrease moderately with increasing $X$. However, these features may reflect an observational selection effect of GlC $\gamma$-ray detection by {\it Fermi}-LAT at the position of Sun (i.e., with $X\approx 8$ kpc, $Y=0$ kpc and $Z=0$ kpc), rather than a physical dependence of GlC $\gamma$-ray emission on $X$.
Nevertheless, it seems that $L_{\gamma}$ tend to reach the maximum values near the GC in $Y$ and $Z$ direction (Figure \ref{fig:lumin_gc}(b) and \ref{fig:lumin_gc}(c)), and this tendency is still evident when the influence of cluster mass was considered in Figure \ref{fig:emissi_gc}(b) and \ref{fig:emissi_gc}(c). 
We further plot in Figure \ref{fig:lumin_gc}(d)-(f) $L_{\gamma}$ as a function of cluster distance to the GC ($R_{\rm gc}$), the perigalactic distance ($R_{\rm gc,per}$) and apogalactic distance ($R_{\rm gc,apo}$) of GlC orbit in the Milky Way. 
Although with large scatter, the mild anti-correlations between $L_{\gamma}$ and the GlC distances to GC is in agreement with the finding of Figure \ref{fig:lumin_gc}(b) and \ref{fig:lumin_gc}(c), suggesting that more MSPs tend to be dynamically formed in GlCs closer to the GC. 
This tendency is even more significant in Figure \ref{fig:emissi_gc}(d)-(f), where the best-fitting functions can be expressed as $\epsilon_{\gamma}\propto R_{\rm gc}^{-1.13\pm0.21}$, $\epsilon_{\gamma}\propto R_{\rm gc,per}^{-0.74\pm0.13}$ and $\epsilon_{\gamma}\propto R_{\rm gc,apo}^{-1.21\pm0.20}$ for total GlCs, $\epsilon_{\gamma}\propto R_{\rm gc}^{-1.43\pm0.28}$, $\epsilon_{\gamma}\propto R_{\rm gc,per}^{-0.73\pm0.16}$ and $\epsilon_{\gamma}\propto R_{\rm gc,apo}^{-1.52\pm0.27}$ for dynamically normal GlCs, and  $\epsilon_{\gamma}\propto R_{\rm gc}^{-0.78\pm0.34}$, $\epsilon_{\gamma}\propto R_{\rm gc,per}^{-1.06\pm0.15}$ and $\epsilon_{\gamma}\propto R_{\rm gc,apo}^{-0.84\pm0.35}$ for core-collapsed GlCs, respectively.

For comparison, it will be interesting to mark in Figure \ref{fig:emissi_gc}(d)-(f) the $\gamma$-ray emissivity of stars in the Inner Galaxy, since tidally disrupted GlCs are thought to have a contribution to at least part of the stars therein. With a total mass of $(1.4\pm0.6)\times 10^{9}\,M_{\odot}$ for nuclear bulge stars and $(0.91\pm0.7)\times 10^{10}M_{\odot}$ for boxy bulge stars, \citet{Bartels2018} showed that the emssion profile of GCE can be better fitted by stellar mass in the boxy and nuclear bulge rather than the dark matter profiles, and the $\gamma$-ray emissivity of the Galactic boxy bulge and nuclear bulge is measured to be $\epsilon_{\gamma}=(1.9\pm0.2)\times 10^{27}\,{\rm erg\,s^{-1}}\,M_{\odot}^{-1}$ and $\epsilon_{\gamma}=(1.1\pm0.6)\times 10^{27}\,{\rm erg\,s^{-1}}\,M_{\odot}^{-1}$ separately. We mark the boxy bulge and nuclear bulge with blue and olive rectangles in Figure \ref{fig:emissi_gc}(d)-(f), with vertical ranges indicate $\gamma$-ray emissivity errors and the the horizontal ranges roughly denote the extent of the boxy bulge ($\sim 10\degr$) and nuclear bulge ($\sim 200$ pc). We also displayed the best fit of GCE emission per unit stellar mass as red dashed horizontal line ($\epsilon_{\gamma}=1.8\times 10^{27}\,{\rm erg\,s^{-1}}\,M_{\odot}^{-1}$, \citealp{Bartels2018}) in the figures. 

\begin{table*}
\centering
\caption{Tested Correlations \& Coefficients. For each correlation, the Spearman's rank cofficient is shown in the upper panel of the grid, while the corresponding random correlation $p$-value is given in the bottom panel.}
\footnotesize
\label{tab:spearmans_r}
\begin{tabular}{cccccccccccccccc}
\hline
\multicolumn{1}{c}{Correlations} & \multicolumn{1}{c}{} & \multicolumn{4}{c}{Total GCs} & \multicolumn{1}{c}{} & \multicolumn{4}{c}{Normal GCs} & \multicolumn{1}{c}{} & \multicolumn{4}{c}{Core-collapsed GCs}\\
\cline{1-1} \cline{3-6} \cline{8-11} \cline{13-16}
\multicolumn{1}{c}{GC param.} &\multicolumn{1}{c}{} & \multicolumn{1}{c}{$L_{\gamma}$} & \multicolumn{1}{c}{$\epsilon_{\gamma}$} & \multicolumn{1}{c}{$M$} & \multicolumn{1}{c}{$R_{\rm gc}$} & \multicolumn{1}{c}{} & \multicolumn{1}{c}{$L_{\gamma}$} & \multicolumn{1}{c}{$\epsilon_{\gamma}$} & \multicolumn{1}{c}{$M$} & \multicolumn{1}{c}{$R_{\rm gc}$} & \multicolumn{1}{c}{} & \multicolumn{1}{c}{$L_{\gamma}$} & \multicolumn{1}{c}{$\epsilon_{\gamma}$} & \multicolumn{1}{c}{$M$} & \multicolumn{1}{c}{$R_{\rm gc}$} \\
\hline
%\startdata
\multirow{2}{*}{$L_{\gamma}$}
& & 1.000  & 0.610  & 0.454  & -0.267 & & 1.000  & 0.544  & 0.464  & -0.302 & & 1.000  & 0.727  & 0.259  &-0.497\\
& & --     & 4.8e-5 & 0.004  & 0.105  & & --     & 0.004  & 0.017  & 0.134  & & 0.007  & 0.007  & 0.417  & 0.101\\
 \cline{3-6} \cline{8-11} \cline{13-16} 
\multirow{2}{*}{$\epsilon_{\gamma}$}                
& & 0.610  & 1.000  & -0.359 & -0.559 & & 0.544  & 1.000  & -0.396 & -0.609 & & 0.727  & 1.000  & -0.343 &-0.573\\             
& & 4.8e-5 & --     & 0.027  & 2.6e-4 & & 0.004  & --     & 0.045  & 9.5e-4 & & 0.007  & --     & 0.276  & 0.051\\ 
 \cline{3-6} \cline{8-11} \cline{13-16}
\multirow{2}{*}{$M$}                
& & 0.454  & -0.359 & 1.000  & -0.193 & & 0.464  & -0.396 & 1.000  & -0.256 & & 0.259  & -0.343 & 1.000  & 0.091\\
& & 0.004  & 0.027  & --     & 0.015  & & 0.017  & 0.045  & --     & 0.003  & & 0.417  & 0.276  & --     & 0.779\\ 
 \cline{3-6} \cline{8-11} \cline{13-16}
\multirow{2}{*}{$\Gamma$} 
& & 0.583  & 0.118  & 0.625  & 0.107  & & 0.697  & 0.270  & 0.529  & -0.004 & & 0.420  & -0.238 & 0.699  & 0.133\\
& & 1.5e-4 & 0.456  & 3.5e-5 & 0.529  & & 1.1e-4 & 0.192  & 0.007  & 0.985  & & 0.175  & 0.457  & 0.011  & 0.681\\ 
 \cline{3-6} \cline{8-11} \cline{13-16}
\multirow{2}{*}{$\Lambda$} 
& & 0.454  & 0.368  & 0.141  & -0.104 & & 0.532  & 0.616  & -0.067 & -0.289 & & 0.364  & -0.189 & 0.517  & 0.028\\
& & 0.005  & 0.025  & 0.406  & 0.539  & & 0.006  & 0.001  & 0.751  & 0.161  & & 0.245  & 0.557  & 0.085  & 0.931\\ 
 \cline{3-6} \cline{8-11} \cline{13-16}
\multirow{2}{*}{$R_{c}$} 
& & -0.121 & -0.255 & -0.286 & 0.553  & & -0.390 & 0.621  & -0.349 & 0.556  & & 0.330  & 0.561  & 0.119  &-0.191\\
& & 0.475  & 0.128  & 2.5e-4 & 4e-14  & & 0.054  & 9.3e-4 & 4.6e-5 & 7e-12  & & 0.295  & 0.058  & 0.538  & 0.320\\ 
 \cline{3-6} \cline{8-11} \cline{13-16} 
\multirow{2}{*}{$\rho_{c}$} 
& & 0.374  & 0.096  & 0.557  & -0.525 & & 0.658  & 0.352  & 0.651  & -0.532 & & -0.119 & -0.559 & 0.152  & 0.236\\
& & 0.023  & 0.572  & 2e-14  & 1e-12  & & 3.5e-4 & 0.085  & 0      & 7e-11  & & 0.713  & 0.059  & 0.431  & 0.217\\ 
 \cline{3-6} \cline{8-11} \cline{13-16}
\multirow{2}{*}{$t_{rc}$} 
& & -0.069 & -0.535 & -0.079 & 0.499  & & -0.275 & -0.790 & -0.202 & 0.413  & & 0.326  & -0.165 & 0.581  & 0.305\\
& & 0.689  & 7.7e-4 & 0.349  & 3e-10  & & 0.194  & 4.3e-6 & 0.033  & 6.1e-6 & & 0.301  & 0.609  & 9.4e-4 & 0.109\\ 
 \cline{3-6} \cline{8-11} \cline{13-16}
\multirow{2}{*}{$\sigma_{c}$} 
& & 0.574  & -0.217 & 0.911  & -0.378 & & 0.672  & -0.189 & 0.942  & -0.411 & & 0.287  & -0.165 & 0.763  & 0.253\\
& & 2.0e-4 & 0.197  & 0      & 9.0e-7 & & 2.4e-4 & 0.367  & 0      & 1.2e-6 & & 0.365  & 0.609  & 1.5e-6 & 0.186\\ 
 \cline{3-6} \cline{8-11} \cline{13-16}
\multirow{2}{*}{$R_{h}$}  
& & -0.363 & -0.513 & -0.226 & 0.666  & & -0.404 & -0.548 & -0.318 & 0.669  & & -0.329 & -0.545 & 0.637  & 0.310\\
& & 0.027  & 0.001  & 0.004  & 0      & & 0.045  & 0.005  & 2.2e-4 & 0      & & 0.297  & 0.067  & 2.0e-4 & 0.102\\ 
 \cline{3-6} \cline{8-11} \cline{13-16}
\multirow{2}{*}{$\rho_{h}$} 
& & 0.641  & 0.141  & 0.665  & -0.559 & & 0.728  & 0.124  & 0.767  & -0.565 & & 0.448  & 0.259  & -0.117 & -0.045\\
& & 2.0e-5 & 0.395  & 0      & 2e-14  & & 3.6e-5 & 0.555  & 0      & 3e-12  & & 0.144  & 0.417  & 0.547  & 0.815\\ 
 \cline{3-6} \cline{8-11} \cline{13-16}
\multirow{2}{*}{$t_{rh}$} 
& & -0.053 & -0.609 & 0.271  & 0.559  & & -0.043 & -0.657 & 0.245  & 0.538  & & -0.144 & -0.595 & 0.797  & 0.346\\
& & 0.756  & 6.3e-5 & 5.5e-4 & 2e-14  & & 0.839  & 3.6e-4 & 0.005  & 4e-11  & & 0.656  & 0.041  & 2.2e-7 & 0.066\\ 
 \cline{3-6} \cline{8-11} \cline{13-16}
\multirow{2}{*}{$v_{esc}$} 
& & 0.559  & -0.225 & 0.893  & -0.389 & & 0.652  & -0.191 & 0.929  & -0.416 & & 0.231  & -0.252 & 0.724  & 0.256\\
& & 3.3e-4 & 0.181  & 0      & 4.0e-7 & & 4.1e-4 & 0.361  & 0      & 8.3e-7 & & 0.471  & 0.430  & 9.1e-6 & 0.180\\
 \cline{3-6} \cline{8-11} \cline{13-16}
\multirow{2}{*}{$R_{t}$} 
& & 0.039  & -0.634 & 0.344  & 0.805  & & 0.128  & 0.661  & 0.351  & 0.759  & & -0.294 & -0.720 & 0.535  & 0.916\\
& & 0.818  & 2.5e-5 & 9.0e-6 & 0      & & 0.543  & 3.2e-4 & 4.2e-5 & 0      & & 0.354  & 0.008  & 0.003  & 3e-12\\
 \cline{3-6} \cline{8-11} \cline{13-16}
\multirow{2}{*}{$\alpha$} 
& & 0.235  & 0.730  & -0.286 & -0.333 & & 0.187  & 0.764  & -0.279 & -0.245 & & 0.333  & 0.708  & -0.534 &-0.398\\
& & 0.161  & 2.9e-7 & 2.6e-4 & 1.8e-5 & & 0.372  & 8.9e-6 &  0.001 & 0.005  & & 0.291  & 0.010  & 0.001  & 0.032\\ 
 \cline{3-6} \cline{8-11} \cline{13-16}
\multirow{2}{*}{$c$} 
& & 0.084  & 0.125  & 0.556  & -0.103 & & 0.267  & 0.177  & 0.607  & -0.098 & & -0.156 & 0.156  & 0.206  & 0.647\\
& & 0.609  & 0.454  & 3e-14  & 0.194  & & 0.179  & 0.386  & 2e-14  & 0.269  & & 0.628  & 0.628  & 0.283  & 1.5e-4\\
 \cline{3-6} \cline{8-11} \cline{13-16}
\multirow{2}{*}{${\rm [Fe/H]}$} 
& & 0.482  & 0.611  & -0.056 & -0.439 & & 0.494  & 0.586  & -0.019 & -0.454 & & 0.378  & 0.741  & -0.163 &-0.508\\
& & 0.003  & 7.7e-5 & 0.497  & 2.1e-8 & & 0.012  & 0.003  & 0.841  & 1.9e-7 & & 0.226  & 0.006  & 0.398  & 0.005\\
 \cline{3-6} \cline{8-11} \cline{13-16}
\multirow{2}{*}{$X$} 
& & -0.425 & -0.459 & -0.068 & 0.680  & & -0.474 & -0.490 & -0.062 & 0.684  & & -0.392 & -0.434 & -0.140 & 0.839\\
& & 0.007  & 0.004  & 0.685  & 2.7e-6 & & 0.012  & 0.011  & 0.764  & 1.2e-4 & & 0.208  & 0.159  & 0.665  &6.4e-4\\
 \cline{3-6} \cline{8-11} \cline{13-16} 
\multirow{2}{*}{$Y$} 
& & -0.169 & 0.129  & -0.267 & -0.198 & & -0.282 & 0.132  & -0.430 & -0.210 & & 0.203  & 0.189  & 0.077  &-0.238\\
& & 0.307  & 0.439  & 0.107  & 0.235  & & 0.154  & 0.519  & 0.028  & 0.304  & & 0.527  & 0.557  & 0.812  & 0.457\\ 
 \cline{3-6} \cline{8-11} \cline{13-16}
\multirow{2}{*}{$Z$} 
& & -0.001 & -0.107 & 0.121  & -0.117 & & -0.126 & -0.252 & 0.094  & 0.037  & & 0.224  & 0.322  & -0.301 &-0.699\\
& & 0.998  & 0.523  & 0.470  & 0.483  & & 0.530  & 0.214  & 0.648  & 0.858  & & 0.484  & 0.308  & 0.342  & 0.011\\ 
 \cline{3-6} \cline{8-11} \cline{13-16}
\multirow{2}{*}{$R_{\rm gc}$} 
& & -0.267 & -0.559 & 0.287  & 1.000  & & -0.302 & -0.609 & 0.249  & 1.000  & & -0.497 & -0.573 & 0.091  & 1.000\\
& & 0.105  & 2.6e-4 & 0.081  & --     & & 0.134  & 9.5e-4 & 0.220  & --     & & 0.101  & 0.051  & 0.779  &   -- \\ 
 \cline{3-6} \cline{8-11} \cline{13-16}
\multirow{2}{*}{$R_{\rm gc,per}$} 
& & -0.433 & -0.562 & 0.143  & 0.662  & & -0.358 & -0.419 & 0.052  & 0.663  & & -0.566 & -0.944 & 0.504  & 0.685\\
& & 0.009  & 3.0e-4 & 0.400  & 7.9e-6 & & 0.079  & 0.037  & 0.807  & 3.0e-4 & & 0.055  & 3.9e-6 & 0.095  & 0.014\\ 
 \cline{3-6} \cline{8-11} \cline{13-16}
\multirow{2}{*}{$R_{\rm gc,apo}$} 
& & -0.326 & -0.604 & 0.280  & 0.974  & & -0.384 & -0.666 & 0.237  & 0.966  & & -0.462 & -0.622 & 0.203  & 0.951\\
& & 0.049  & 7.5e-5 & 0.093  & 0      & & 0.058  & 2.8e-4 & 0.255  & 5e-15  & & 0.131  & 0.031  & 0.527  &2.0e-6\\
\hline
\end{tabular}
\end{table*}

From Figure \ref{fig:emissi_gc}(d)-(f), it can be seen that GlCs have a $\gamma$-ray emissivity $\sim 10-1000$ times of higher than the Galactic boxy bulge and nuclear bulge, which indicates that tidally disrupted GlCs may enhance the $\gamma$-ray emissivity of the Inner Galaxy greatly. 
More importantly, as GlCs inspiral in toward the GC, it seems that normal stars are more likely to be stripped off the clusters by Galactic tidal force, while MSPs are preferentially to be deposited to the Inner Galaxy, thus resulting in a negative dependence of $\epsilon_{\gamma}$ on $R_{\rm gc}$, $R_{\rm gc,per}$ and $R_{\rm gc,apo}$ in GlCs. These features suggest that, in addition to the tidal stripping, mass seggregation also may plays an important role in allocating the objects that can be delivered to the GC. Therefore, an estimation of GlC contribution to the GCE must take these effects into account. We leave the implication of this problem to be addressed in Section 4.4.

We end this section by emphasizing that the GlC parameters are not independent of each other. For example, there is a significant correlation between $\Gamma$ and $M$ in GlCs \citep{Cheng2018a}, and massive GlCs are usually found to have larger central velocity dispersion $\sigma_{c}$, escape velocity $v_{esc}$, core density $\rho_{c}$, and smaller half-mass (core) relaxation time $t_{rh}$ ($t_{rc}$). On the other hand, significant dependencce of cluster paramerters on $R_{c}$ and $R_{\rm gc}$ have been reported in literature \citep{Djorgovski1994}. These mutual relations suggest that the GlC parameters are highly coupled with each other, and they are closely connected with the dynamical evolution of GlCs in the Milky Way tidal field \citep{Meylan1997}. In Table \ref{tab:spearmans_r}, we summarized the Spearman's rank coefficients of $L_{\gamma}$, $\epsilon_{\gamma}$, $M$ and $R_{\rm gc}$ on various GlC parameters, the corresponding random correlation $p$-values are also listed in the table.

\section{Discussion}

By exploring the dependence of $L_{\gamma}$ and $\epsilon_{\gamma}$ on various GlC parameters in Section 3, we have established many correlations and trends between GlC dynamical evolution and their binary burning products. 
In particular, the meaured $\gamma$-ray emission is mainly contributed by the MSPs, which facilitates us to trace the dynamical formation of MSPs within GlCs, their migration with host clusters in the Milky Way Galaxy, and the final settlement and spatial distribution surrounding the GC. As introduced in Section 1, these processes are particularly important for the MSP interpretation of the GCE. Here we discuss our findings and conclusions as follows. 

\subsection{GlC Dynamical Evolution and Formation of MSPs}

As self-gravitating systems, the dynamical evolution of GlCs is balanced by the production of energy in the core and the outflow of energy from the cluster, and two-body relaxation is the fundamental process that dominates the transportation of energy and mass in GlCs. Stars are driven to reach a state of energy equipartition by two-body relaxation, massive stars (or binaries) therefore tend to lose energy and drop to the cluster center (thus influences the types of binary burning products), whereas lower-mass stars tend to gain energy and move faster, and they will migrate outward (thus leading to the energy outflow in GlCs) and drives the cluster envelope to expand \citep{Heggie2003}. 
Binary burning is thought to be the internal energy source that supports the cluster against gravothermal collapse, and the energy production rate of GlCs is sensitive to the encounter rate of the hard binaries \citep{Cheng2018b}:
\begin{equation}
\Gamma_{b}\propto \int f_{b} {\rho}^{2}A_{b}\sigma dV \propto \int \frac{f_{b}{\rho}^{2} a}{\sigma} d V, 
\end{equation}
where $f_{b}$ is the fraction of stars in binary, $A_{b}\propto a/{\sigma}^2$ is the binary encounter cross section, with $a$ being the orbital separation of the binary. In practice, it is hard to determine the distribution of $a$ by observation, while $f_{b}$ is found to vary not too much among GlCs (i.e., $f_{b}\sim 1\%-20\%$; \citealp{Milone2012}). Therefore, $\Gamma_{b}$ is usually simplified as $\Gamma\propto \int \rho^{2}/\sigma$ in literature. %However, this does not weaken the importance of $\Gamma_{b}$ in understanding the dynamical evolution of GCs. 

As binaries are disrupted or been driven to evolve to harder systems by dynamical ecounters, both $f_{b}$ and $A_{b}$ will become smaller and the energy production rate of GlCs may derease according to Equation (1). However, this may not happen in a real cluster. In fact, GlCs contract their cores to increase the stellar density, which enhances the stellar encounter rates (i.e., $\Gamma$ and $\Lambda$) and thus the energy production rate. Accordingly, the cluster two-body relaxation time was adjusted to smaller values, to enhance the energy outflow rate and maintain the balance between the energy production in the cores and the outflow of energy from the systems. As a result, the dynamical evolution of GlCs will become more and more rapid, untill dynamical ejection of close binaries become important (see Section 4.2) and the cluster evolve into a core-collapsed GlC. 

With larger mass than normal stars, NS are expected to concentrate to cluster center and take part in the binary burning process, thus leading to the formation of LMXBs and MSPs in GlCs. 
However, recent N-body simulations suggest that many GlCs may host significant population of primordial stellar mass black holes (BHs), even evolve to an age larger than $\sim 12$ Gyr \citep{Breen2013, Morscher2013, Morscher2015, Wang2016}. Compared with NS, the BHs are expected to drop to cluster center first and form a high-density BH subsystem (BHS), where the BH burning process (i.e., the dynamical hardening of BH binaries, \citealp{Kremer2020a}) may dominates the energy production of GlCs and lead to the formation of gravitational wave sources \citep{Rodriguez2016, Askar2017, Antonini2020, Anagnostou2020}. 
The thermal coupling of BHS with GlCs is sufficient to reshape the structure of the cluster \citep{Merritt2004, Mackey2007, Mackey2008, Giersz2019} and quench or slow down the the mass segregation of less-massive objects (such as NS, binaries and BSS) in GlCs \citep{Weatherford2018, Fragione2018b}. 
Besides, through exchange encounters involving a BH, binaries are preferentially been transformed into BH binaries, BHS therefore may suppress the formation of other binary burning products in GlCs, and clusters with a large number of BHs are more likely to have a lower formation efficiency of LMXBs, MSPs, CVs, ABs, and BSS \citep{Ye2019, Kremer2020b, Fragione2018b}

The mass segregation delay of heavy objects in GlCs has been confirmed by the radial distribution study of X-ray sources, MSPs and BBS in 47 Tuc and Terzan 5 \citep{Cheng2019a, Ferraro2012, Cheng2019b}, where massive objects (i.e., MSPs, Bright X-ray sources or BSS) are found to drop to cluster center earlier and be more centrally concentrated than the less massive ones (i.e., Faint X-ray sources and normal stars). The exception is M28, \citet{Cheng2020a} reported an abnormal deficiency of X-ray sources in the cluster central region with respect to its outskirts, and argued that an early phase of primordial binary disruption by BHS may have happened in M28 \citep{Cheng2020a}. 
On the other hand, significant evidence of BH burning has been reported in GlC $\omega$ Cenaturi, in which the dynamical formation of X-ray sources is found to be highly suppressed, and BHS is the essential internal energy source that supports the cluster against collapse \citep{Cheng2020b}. Indeed, with $\sim4.6\%$ of the cluster mass being invisible, $\omega$ Cenaturi is thought to be the cluster with the heaviest BHS in the Galaxy \citep{Baumgardt2019b}. In Table-1, the $\gamma$-ray emissivity of $\omega$ Cenaturi is also the smallest among the 39 GlCs, suggesting a low formation efficiency of MSPs in this cluster.

The evolution fate of the BHS is beyond the scope of this paper. It is very likely that the majority of the BHs will be ejected from the cluster gradually, and host GlCs tend to contract their cores and increase the stellar density \citep{Arcasedda2018a}. As a consequence, NS starts to concentrate to cluster center and gradually take over the binary burning process, which then results in the formation a large number of LMXBs and MSPs in GlCs. 
Indeed, we have demonstrated that $L_{\gamma}$, the proxy of the population of MSPs in GlCs, is highly correlated with the cluter stellar encounter rate\footnote{According to Equation (1), the dynamical encounters between invisible compact objects are not included in $\Gamma$, thus the energy input of BH burning are also not counted here.}, $\Gamma$, with $L_{\gamma}\propto \Gamma^{0.71\pm0.11}$ in dynamically normal GlCs (Fig \ref{fig:encounterrate}(a)). Besides, the GlC $\gamma$-ray emissivity $\epsilon_{\gamma}$, a proxy of the MSP abundance of GlCs, is highly correlated with the cluster specific encounter rate $\Lambda$, with $\epsilon_{\gamma}\propto \Lambda^{0.78\pm0.15}$ in Fig \ref{fig:encounterrate}(b). These correlations provide strong evidence for the dynamical formation of MSPs in GlCs. 

On the other hand, by examing the dependence of $L_{\gamma}$ and $\epsilon_{\gamma}$ on the cluster structure paramters, we have tested the relationship between MSP formation and the dynamcial evolution of GlCs. For example, both $L_{\gamma}$ and $\epsilon_{\gamma}$ are anti-correlated with GlC core radius $R_{c}$ and half-mass radius $R_{h}$ in dynamically normal GlCs (Fig \ref{fig:luminosity} and Fig \ref{fig:emissivity}). In particular, the $\epsilon_{\gamma}-R_{c}$ and $\epsilon_{\gamma}-R_{h}$ correlations can be expressed as $\epsilon_{\gamma}\propto R_{c}^{-0.93\pm0.27}$ and $\epsilon_{\gamma}\propto R_{h}^{-2.09\pm0.61}$ in Figure \ref{fig:emissivity}(a) and Figure \ref{fig:emissivity}(e), respectively. Furthermore, $\Gamma$ is strongly dependent on the cluster stellar density $\rho$, with $\Gamma \propto \int \rho^{2}/\sigma$. We have also found strong correlations between $L_{\gamma}$ and $\rho_{c}$ and $\rho_{h}$ in dynamically normal GlCs, with $L_{\gamma}\propto \rho_{c}^{0.39\pm0.14}$ and $L_{\gamma}\propto \rho_{h}^{0.63\pm0.15}$ in Figure \ref{fig:luminosity}(b) and \ref{fig:luminosity}(f). 
These trends are in agreement with the dynamical evolution of GlCs, that is, as GlCs contract their central regions to smaller radius and become more dense, more MSPs are expected to be dynamically formed in the clusters.

Finally, it is necessary to emphasize that the dynamical formation of MSPs is far from finished in many dynamically normal GlCs.
As confirmed by the radial distribution studies of X-ray sources \citep{Cheng2019a, Cheng2019b, Cheng2020a} and BSS \citep{Ferraro2012} in GlCs, it is possible that there are considerable numbers of primordial binaries and NS exist in the outskirts of the dynamically young GlCs. Under the effect of mass segregation, these objects tend to drop to the dense core of the cluster, where they could be transformed into LMXBs and MSPs by the binary burning process, thereby increasing the abundance of MSPs in dynamically normal GlCs. 
As a consequence, dynamically normal GlCs with an older dynamical age (i.e., $t_{d} \propto t_{rc}^{-1}$ or $t_{d} \propto t_{rh}^{-1}$) are more likely to exhibit a larger abundance of MSPs, with the $\epsilon_{\gamma}-t_{rc}$ correlation can be written as $\epsilon_{\gamma}\propto t_{rc}^{-0.61\pm0.11}$ in Figure \ref{fig:emissivity}(c), and the $\epsilon_{\gamma}-t_{rh}$ correlation can be expressed as $\epsilon_{\gamma}\propto t_{rh}^{-0.85\pm0.20}$ in Figure \ref{fig:emissivity}(g), respectively.
 
\subsection{Dynamical Disruption of LMXBs and Ejection of MSPs in Core-Collapsed GlCs}
From Equation (1), it can been seen that the evolution of GlCs to core collapse is sensitively depends on the encounter cross section (i.e., $A_{b}\propto a/\sigma^2$) of the binaries. 
With larger $A_{b}$, the primordial binaries are expected to take part in the binary burning process first and been transformed into close binaries, which may lead to the formation of many exotic objects, such as LMXBs (e.g., through exchange encounters involving NS), MSPs, CVs, ABs and BSS in GCs. However, when primordial binaries were exhuasted, GlCs tend to contract their cores and increase the stellar density, thereby increasing the specific encounter rate ($\Lambda$) of stars in the cluster core. The enhancement of $\Lambda$ is helpful for the extraction energy from harder binaries, and even very close binary systems, such as LMXBs, MSPs, CVs and ABs, may suffer strong dynamical encounters in core-collapsed GlCs. %during the evolution of clusters into deep core collapse. 
Nevertheless, it is argured that the ``burning'' of very hard binaries is not sufficient to terminate the deep core collapse of GlCs. The cluster cores are expected to contract further, untill dynamical interactions between single stars take effect and lead to the formation of new binaries in GlCs (i.e., binary formation either via the ``two-body binaries'' or the ``three-body binaries'' channels, \citealp{Heggie2003}). The sudden introduction of a large number of binaries in extremely dense core is sufficient to reverse the cluster core collapse to core bounce, and trigger gravothermal oscillations in GlCs.

Before the discussion of the dynamical evolution fate of LMXBs and MSPs in core-collapsed GlCs, it is necessary to have an intuitive understanding on the dynamic effects of binary burning process of GlCs. 
Accoding to the Hills--Heggie law, through binary--single encounters, hard binaries tend to increase their binding energy ($E_{b}=Gm^{2}/2a$) and become harder. The average increasement of $E_{b}$ per encounter is roughly proportional to its inital value, with $\delta E_{b} \simeq 0.4-0.5E_{b}$ for equal-mass encounters \citep{Hills1975, Heggie1975}. The release of the binding energy is transformed into the kinetic energy of the ejected star, and the remaining binay will receive a recoil velocity as well, with $v_{rec}\propto \sqrt{\delta E_{b}} \propto 1/\sqrt{a}$. Therefore, hard binaries tend to receive larger recoil velocity after strong encounters, which may lead to the recoil of the binaries out of the core into the halo, and sometimes out of the host cluster \citep{Hut1992}. 

With typical central escape velocity of $10\, {\rm km\, s^{-1}}\lesssim v_{esc}\lesssim 100\, {\rm km\, s^{-1}}$, LMXBs and MSPs may receive a large recoil velocity and escape the core-collapsed GlCs after suffering strong encounters. On the other hand, the specific encounter rate $\Lambda$ of deep core collapse GlCs is extremely high, it is possible for LMXBs to be dynamically disrupted (i.e., via the binary-binary encounters) or be greatly modified (i.e., exchange encounters, etc.) before they otherwise could evolve into MSPs \citep{Verbunt2014}. %, and compared with dynamically normal GCs, core-collapsed GCs with large $\Lambda$ are more likely found to harbor isolated pulsars or pulsars with recycling process was interrupted \citep{Verbunt2014}. 
The net effect is that the population of MSPs tend to be gradually exhuasted in core-collapsed GlCs. Indeed, compared with dynamically normal GlCs, the core-collapsed GlCs are found to exhibit smaller $L_{\gamma}$ and $\epsilon_{\gamma}$ with decreasing $R_{c}$, $t_{rc}$, and increasing $\rho_{c}$ in Figure \ref{fig:luminosity}(a)-\ref{fig:luminosity}(c) and Figure \ref{fig:emissivity}(a)-\ref{fig:emissivity}(c). 
And two well-known core-collapsed GCs, NGC 6397 and NGC 6752, are proved to be the clusters with the smallest $L_{\gamma}$ and $\epsilon_{\gamma}$ in Table-\ref{tab:results}.
  
The decreases of MSP populations in core-collapsed GlCs have also been reported by \citet{deMenezes2019}, which was interpreted as the disruption of LMXBs by the extremely large specific encounter rates in the dense cores. However, these authors have neglected the importance of MSP ejection in deep core collapse clusters. The lifetime of MSPs ($\sim 10^{10}$ yrs) is at least one order of magnitudes larger than the cluster core relaxation time ($t_{rc}\sim 10^{5-9}$ yrs), and once MSPs are formed via the recycling process of LMXBs, it is hard to change their properties through direct stellar dynamical interactions. Therefore, the decline of $L_{\gamma}$ and $\epsilon_{\gamma}$ from dynamically normal GlCs to core-collapsed GlCs in Figure \ref{fig:luminosity}(a)-\ref{fig:luminosity}(c) and Figure \ref{fig:emissivity}(a)-\ref{fig:emissivity}(c) may suggests an episodic ejection of MSPs during the deep core collapse phase, and the injection of MSPs into the Milky Way may plays an important role in shaping the spatial extent of the GCE. 
Nevertheless, it is also possible for some MSPs to be retained by the host cluster \citep{Hut1992}, and the retention of these MSPs may responsible for the $\sim 2$ times larger detection rate of $\gamma$-ray emission in core-collapsed GlCs than in dynamically normal GlCs by {\it Fermi}-LAT (Section 2.2).

\subsection{Tidal Stripping and Deposition of MSPs into the Galactic Center}

In addition to the internal effects, the structure of GlCs is also subjected to the external influence of the host galaxy, which may affects the dynamical evolution of GlCs in two ways. First, the expanding envelopes of GlCs are expected to be truncated by the gravitational tidal field of the host galaxy, and compared with clusters in isolation, GlCs in the tidal field may suffer an enhanced rate of energy outflow and mass loss. The tidal stripping therefore has a net effect in accelerating the dynamical evolution of the GlCs. 
Second, through dynamical friction with background stars, GlCs tend to lose energy and spiral in toward the center of the galaxy, where tidal stripping and interactions with the dense nucleus will eventually lead to the dissolution of the clusters \citep{Tremaine1975}. 
In particular, during the passages close to the Galactic bulge or passing through the dense disk, GlCs may suffer gravitational shocks and these two effects could be enhanced greatly. The bulge shocking and disk shocking tend to heat up the GlC outskirts and increase evaporation of the cluster \citep{Gnedin1997}, which in turn accelerate the core collapse and shorten the destruction time of GlCs \citep{Gnedin1999}.
Taking these aspects together, it is suggested that GlCs are the ancient building blocks of our Galaxy, and the inner galactic structures, such as the nuclear star cluster and the nuclear bulge, are assembled at least in part, by the merger of tidally disrupted GlCs that spiraled into the deep gravitational well of the GC \citep{Antonini2012, Antonini2013, Gnedin2014}.

\begin{figure*}
\centerline{
\includegraphics[width=1.0\textwidth]{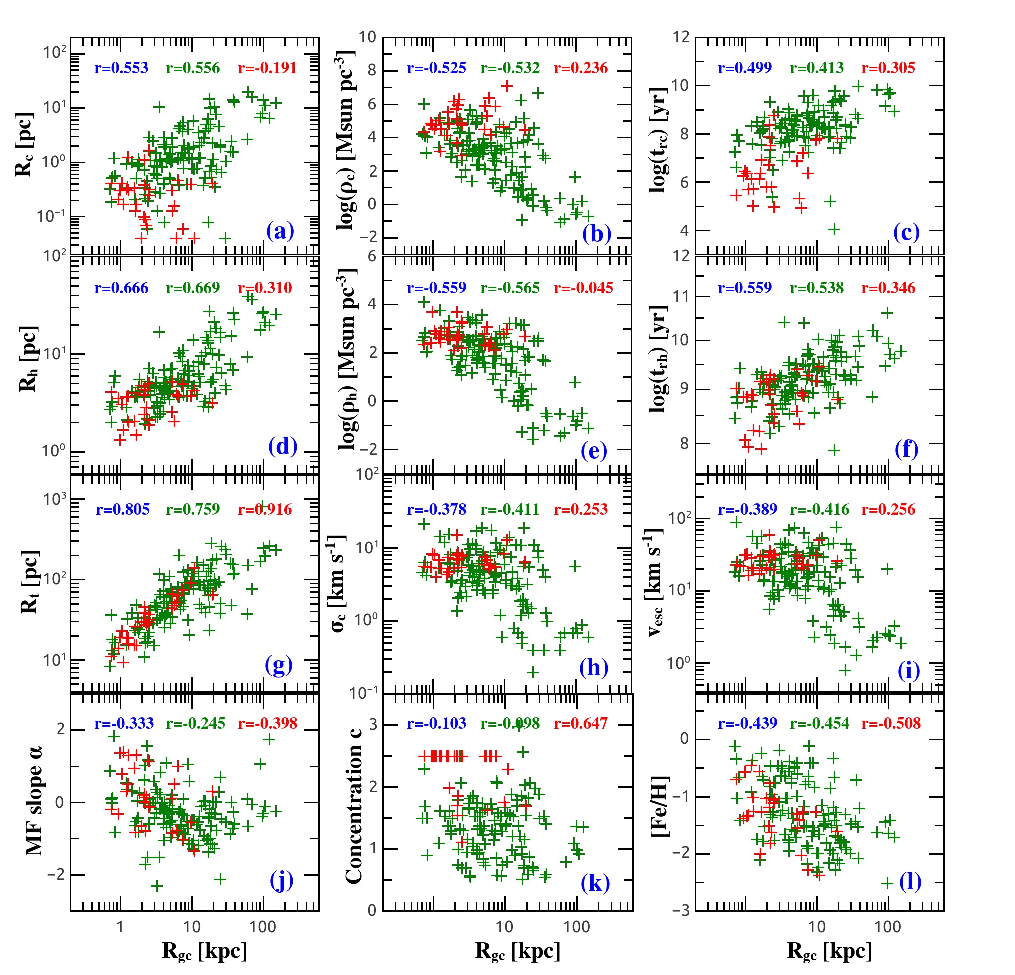}}
\caption{GlC parameters as a function of cluster distance from GC. All the 157 GlCs are shown in this figure, with olive and red pluses represent the dynamically normal and core-collapsed GlCs separately. The red, olive and blue text donates the Spearman's rank correlation coeficient of the core-collapsed, dynamically normal and the total GlC samples, respectively. }
\label{fig:Rgc}
\end{figure*}

Observationally, the dynamical evolution of GlCs and its coupling with host galaxy has been widely confirmed in literature, either via the stellar streams of tidal stripping \citep{Leon2000, Sollima2020}, or the dependence of cluster structure parameters on GlC positions within the Galaxy \citep{Djorgovski1994}. As shown in Figure-\ref{fig:Rgc}, with decreasing distance from the GC, GlCs are found to be more concentrated and have smaller and denser cores. These trends are consistent with the dynamical evolution of GlCs in the Galactic environment, namley, as GlCs udergo orbital decay and spiral in toward the GC, the clusters tend to suffer stronger tidal stripping and will be more dynamically evolved. Besides, it is proposed that mass segregation and tidal stripping effects have a preference in evaporating the low-mass stars from GlCs, which is strong enough to turn cluster initially increasing MF into MF that decrease towards the low-mass end \citep{Vesperini1997, Lamers2013, Baumgardt2003}, and the global MF slope $\alpha$ is strongly anti-correlated with the half-mass relaxation time $t_{rh}$ in GlCs \citep{Sollima2017, Baumgardt2017}.

The relationship of GlC dynamical evolution and its coupling with the Galactic environment can also be confirmed by our $\gamma$-ray data. For example, we showed that there is a mild anti-correlation between GlC $\gamma$-ray luminosity $L_{\gamma}$ and their current distances $R_{gc}$, perigalactic distances $R_{gc,per}$ and apogalactic distances $R_{gc,apo}$ from the GC in Figure \ref{fig:lumin_gc}. Compared with the strong dependence of $L_{\gamma}$ on $\Gamma$ in Figure \ref{fig:encounterrate}(a), these correlations suggest that tidal stripping is a secondary factor affecting the binary burning process of GlCs and the populations of MSPs formed therein. On the other hand, tidal stripping plays a crucial role in diminishing the mass of the GlCs and leading to the final dissolution of the clusters, which can be further confirmed by the much stronger anti-correlations between GlC $\gamma$-ray emissivity $\epsilon_{\gamma}$ and $R_{gc}$, $R_{gc,per}$ and $R_{gc,apo}$ in Figure \ref{fig:emissi_gc}. 
In fact, we also showed that there is no evident dependence of $L_{\gamma}$ on GlC tidal radius $R_{t}$ in Figure \ref{fig:luminosity}(i), but $\epsilon_{\gamma}$ is found to strongly anti-correlated with $R_{t}$ in Figure \ref{fig:emissivity}(i). Therefore, the loss of cluster mass is the most plausible reason for these anti-correlations.

Finally, it is necessary to emphasize the importance of mass segregation in affecting the tidal stripping and dissolution of GlCs. Unlike the low-mass normal stars that tend to migrate outward in GlCs and will be stripped off the clusters preferentially, LMXBs and MSPs are more likely to segregate to GlC center and could be retained by the clusters for much longer time. As a result, GlCs with increasing MS slope $\alpha$ are found to exhibit larger $\epsilon_{\gamma}$ in Figure \ref{fig:emissivity}(j), and as GlCs udergo orbital decay and spiral in toward the Inner Galaxy, LMXBs and MSPs are more likely to be delivered to the GC, leading to a strong anti-correlations between $\epsilon_{\gamma}$ and $R_{gc}$, $R_{gc,per}$ and $R_{gc,apo}$ in Figure \ref{fig:emissi_gc}. 

\subsection{Contribution of GlC MSPs to the GCE}

As introduced in Section 1, it is proposed that the GCE may arising from a large number ($\sim 10^3-10^4$) of unresolved MSPs residing in the GC \citep{Abazajian2011, Yuan2014, Bartels2016}. The generation of MSPs could be arise from the {\it in situ} star formation and evolution at GC \citep{Yuan2014, Eckner2018, Gautam2022}, or inherited from GlCs that were brought to the Inner Galaxy by dynamical friction and assembled the Galactic nuclear star cluster and bulge \citep{Bednarek2013, Brandt2015, Abbate2018, Fragione2018a, Arcasedda2018b}. Compared with the dynamical channel within GlCs, the {\it in situ} formation of MSPs at GC is disfavoured because of LMXBs, the progenitors of MSPs, are observed to be quite rare in the bulge of our Galaxy than the prediction of the MSP interpretation \citep{Cholis2015b, Haggard2017}. 
Besides, \citet{Boodram2022} argued that the natal velocity kicks received by newly formed NS during supernova may lead to a lower abundance of MSPs (thus a lower synthetic $\gamma$-ray surface brightness) in the central degrees ($R\lesssim 150$ pc) of the Galatic bulge with respect to its outskirts, and which is inconsistent with the measured $\gamma$-ray surface brightness profile of the GCE \citep{Boodram2022}.

On the other hand, we have demonstrated that the dynamical channel of MSPs are very efficient during cluster dynamical evolution, and as GlCs losing kinetic energy and sipiral in toward the GC, MSPs are preferentially to be deposited into the the deep gravitational potential well of the Galaxy\footnote{In fact, it can be seen from Figure \ref{fig:emissi_gc}(e) that the GlC MSPs could be effectively delivered to a distance of $R\sim 100$ pc from the Galactic center.}, which may avoid the problems reported in the work of \citet{Boodram2022}.
The spatial distribution of GlCs are found to be spherically distributed around the GC \citep{Baumgardt2018}, and their velocity dispersion is highly isotropic within the central Galaxy (i.e., $R\lesssim 10$ kpc; \citealp{Vasiliev2019}).  
Therefore, the tidal stripping and ejection of MSPs from GlCs to the Milky Way are also expected to create a spherically distributed MSP population in the Inner Galaxy, and with radial distribution profile resamble that of the GCE \citep{Brandt2015, Abbate2018, Fragione2018a, Arcasedda2018b}. 
Moreover, considering the lifetime ($\sim10^{10}$ yr) of MSPs is much longer than that ($\sim10^{8}$ yr) of LMXBs,
the very small LMXBs to MSPs ratio observed in the Galatic bulge can be interpreted as the dynamical disruption of LMXBs during the deep core collapse phase of cluster evolution (Section 4.2), and the sudden interruption of the dynamical formation of LMXBs since host GlCs were tidally disrupted \citep{Brandt2015}. 

To estimate of the contribution of Galactic GlC MSPs to the GCE, the knowledge of the luminosity function (LF) of MSPs is particularly important. By utilizing the cluster stellar encounter rate $\Gamma$ to estimate the formation rate of MSPs in GlCs, \citet{Hooper2016} constrain the $\gamma$-ray LF of MSPs in GlCs and found that they are as luminous as the Galactic field MSPs, with log-normal LF of $L_{\rm 0}\simeq 8.8 \times 10^{33}\, {\rm erg\ s^{-1}}$ and $\sigma \simeq 0.62$. Since MSPs would spin down quickly and fade away in $\gamma$-ray luminosity, \citet{Hooper2016} therefore argued that the GlC MSPs can account for only a few percent or less of the observed GCE. Nevertheless, the LF of \citet{Hooper2016} also predict that most of the GlCs will be dominated by only one or two MSPs in $\gamma$-ray luminosity, which is inconsistent with the radio survey result of MSPs in GlCs, where many clusters detected by {\it Fermi} are found to host far more than one MSPs\footnote{http://www.naic.edu/~pfreire/GCpsr.html}. 
On the other hand, by using the cluster $\epsilon_{\gamma}$ to scale the stellar mass deposited by GlCs at GC, \citet{Brandt2015} showed that a reasonable disrupted GlC mass, such as that calculated in \citet{Gnedin2014}, may account for the total $\gamma$-ray emission of the GCE. 
 
However, both the work of \citet{Hooper2016} and \citep{Brandt2015} have neglected the influence of cluster dynamical evolution on the formation rate of MSPs in GlCs. As dicussed in Section 4.1 and Section 4.2, the dynamical formation of MSPs is far from finished in dynamically young GlCs, while the population of MSPs in core-collapsed GlCs is underestimated, since strong encounters may have lead to the ejection of many MSPs from these systems. These effects can also be confirmed by the large variance of measured $\epsilon_{\gamma}$ in GlCs.   
From Figure \ref{fig:emissi_gc}(e)-\ref{fig:emissi_gc}(f), it can been seen that $\epsilon_{\gamma}$ of GlCs is 1-3 orders of magnitude larger than that of the Galactic bulge stars, the large variance of $\epsilon_{\gamma}$ suggests that the final population of MSPs     created by individual GlCs could be more than tens times higher than that currently hosted in the cluster. 
Indeed, by assuming a cluster $\epsilon_{\gamma}$ of ${\rm log}\, \epsilon_{\gamma}=32.66\pm0.06 -(0.63\pm0.11){\rm log}\, M$ for GlCs, \citet{Fragione2018a} simulating the spirial in and tidal stripping of GlCs in the Milky Way Galaxy. Although their model has neglected the essential features of cluster dynamical evolution and stellar dynamics (i.e., internal energy sources, stellar mass segregation, core collapse, etc.) within GlCs, the empirical $\epsilon_{\gamma}-M$ relation of \citet{Fragione2018a} is consistent with our results in Figure \ref{fig:mass}(b), and as GlCs spiral in toward GC and get less massive, the $\epsilon_{\gamma}$ of remaining cluster debris will become larger and larger, as displayed in Figure \ref{fig:emissi_gc}(e)-\ref{fig:emissi_gc}(f).  
More importantly, \citet{Fragione2018a} showed that the cluster $\epsilon_{\gamma}$ derived $\gamma$-ray luminosity is about one order of magnitude larger than the observed GCE, and spin down of MSPs will reproduce a $\gamma$-ray luminosity consistent with the GCE.

\section{Summary }
We present a $\gamma$-ray study of the 157 Milky Way Globular Clusters (GlCs), based on the archival {\it Fermi}-LAT data with a time span of $\sim 12$ years. By examing the dependence of GlC $\gamma$-ray luminosity ($L_{\gamma}$) and emissivity ($\epsilon_{\gamma}=L_{\gamma}/M$) on various cluster parameters, our main findings are as follows.

1. 39 GlCs are found to be $\gamma$-ray bright (i.e., with $TS>25$) in our work, which corresponds to about $25\%$ (39/157) of the total Milky Way GlCs. The detection rate of core-collapsed GlCs (12/29) is $\sim2$ times higher than the dynamically normal GlCs (27/128). 

2. Compared with cluster mass ($M$), the GlC $\gamma$-ray emission is highly correlated with the stellar encounter rate ($\Gamma$) and the specific encounter rate ($\Lambda=\Gamma/M$), with $L_{\gamma}\propto \Gamma^{0.71\pm0.11}$ and $\epsilon_{\gamma}\propto \Lambda^{0.78\pm0.15}$ in dynamically normal GlCs. These correlations provide strong evidence for the dynamical formation of MSPs in GlCs. 

%3. The GlC $\gamma$-ray emission is highly correlated with their structure parameters. With increasing core/half-mass density, or decreasing core/half-mass radius and relaxtion time, the dynamically normal GlCs are more likely to exhibit larger $L_{\gamma}$ and $\epsilon_{\gamma}$. %in Figure-\ref{fig:luminosity} and Figure-\ref{fig:emissivity}. These trends suggest that the formation of MSPs is highly dependent on the dynamical evolution history of the host cluster, and as GlCs evolve to older dynamcial age and become more compact, more MSPs are expected to be produced in the cluster.
3. The formation of MSPs is also highly dependent on the dynamical evolution history of the host GlCs. As GlCs evolve to older dynamcial age (i.e., $t_{d} \propto t_{rc}^{-1}$ or $t_{d} \propto t_{rh}^{-1}$) and become more compact (i.e., with smaller $R_{c}$ or $R_{h}$, and larger $\rho_{c}$ or $\rho_{h}$), the dynamically normal GlCs are expected to produce more MSPs, thereby exhibit larger $L_{\gamma}$ and $\epsilon_{\gamma}$ in Figure-\ref{fig:luminosity} and Figure-\ref{fig:emissivity}. 

4. However, compared with dynamically normal GlCs, the core-collapsed GlCs are found to exhibit decreasing $L_{\gamma}$ and $\epsilon_{\gamma}$ as their cores undergo deep core collapse. This feature may implies that even LMXBs could be dynamically disrupted or be greatly modified in extremely dense cluster cores, and strong encounters may lead to the ejection of MSPs from core-collapsed GlCs.

5. With decreasing tidal radius ($R_t$) and distance ($R_{gc}$) from the Galactic Center (GC), GlCs are found to have increasing $\gamma$-ray emissivity, with $\epsilon_{\gamma}\propto R_t^{-1.0\pm0.22}$ and $\epsilon_{\gamma}\propto R_{gc}^{-1.13\pm0.21}$. Besides, $\epsilon_{\gamma}$ is positively correlated with GlC stellar mass function slope $\alpha$, with $\epsilon_{\gamma}\propto 10^{(0.57\pm0.1)\alpha}$. These correlations suggest that both tidal stripping and mass segregation effects are improtant factors influencing the abundance of MSPs in GlCs, and as GlCs undergo orbital decay and spiral in toward the GC, MSPs are more likely to be deposited into the GC rather than normal stars.   

6. We gauge the cluster $\epsilon_{\gamma}$ is about 1-3 orders of magnitude larger than that of the Galactic bulge stars, which implies that GlCs may enhance the $\gamma$-ray emissivity of the Galactic bulge greatly. The large variance of cluster $\epsilon_{\gamma}$ may arise from the ongoing dynamical formation (or ejection) of MSPs in GlCs, and the different degree of tidal stripping among the clusters. More importantly, the $\epsilon_{\gamma}$ relations derived in our paper is in agreement with the empirical relation adopted in the simulation of \citet{Fragione2018a}, which states that tidally disrupted GlCs may provide a natural astrophysical explanation to the observed GCE.

\section*{Acknowledgments}
This work is supported by the Youth Program of the National Natural Science Foundation of China No. 12003017.

%%%%%%%%%%%%Begin the Reference%%%%%%%%%%%%%%%%%%%%%%%%%%\

\end{document}